\documentclass[%
 reprint,
superscriptaddress,
 amsmath,amssymb,
 aps,
nofootinbib
]{revtex4-1}

\usepackage{placeins}
\usepackage{float}
\usepackage{rotating}
\usepackage{multirow}
\usepackage{amsmath}
\usepackage{booktabs}
\usepackage{tasks}
\usepackage{float}
\usepackage{siunitx}
\usepackage{graphicx}
\usepackage{tabularx}% Include figure files
\usepackage{dcolumn}% Align table columns on decimal point
\usepackage{bm}% bold math
\usepackage{hyperref}% add hypertext capabilities
\usepackage[mathlines]{lineno}% Enable numbering of text and display math
\AtBeginDocument{%
  \heavyrulewidth=.08em
  \lightrulewidth=.05em
  \cmidrulewidth=.03em
  \belowrulesep=.65ex
  \belowbottomsep=0pt
  \aboverulesep=.4ex
  \abovetopsep=0pt
  \cmidrulesep=\doublerulesep
  \cmidrulekern=.5em
  \defaultaddspace=.5em
}

\usepackage{xcolor}

\graphicspath{{./}{Figures/}}

\begin{document}

%%%%%%%%%%%%%%%%%%%%%%%%%%%%%%%%%%%%%%%%%%%%%%%%%%%%%%%%%%%
\title{Search for GeV-scale Dark Matter from the Galactic Center with IceCube-DeepCore}

%%%%%%%%%%%%%%%%%%%%%%%%%%%%%%%%%%%%%%%%%%%%%%%%%%%%%%%%%%%
\affiliation{III. Physikalisches Institut, RWTH Aachen University, D-52056 Aachen, Germany}
\affiliation{Department of Physics, University of Adelaide, Adelaide, 5005, Australia}
\affiliation{Dept. of Physics and Astronomy, University of Alaska Anchorage, 3211 Providence Dr., Anchorage, AK 99508, USA}
\affiliation{School of Physics and Center for Relativistic Astrophysics, Georgia Institute of Technology, Atlanta, GA 30332, USA}
\affiliation{Dept. of Physics, Southern University, Baton Rouge, LA 70813, USA}
\affiliation{Dept. of Physics, University of California, Berkeley, CA 94720, USA}
\affiliation{Lawrence Berkeley National Laboratory, Berkeley, CA 94720, USA}
\affiliation{Institut f{\"u}r Physik, Humboldt-Universit{\"a}t zu Berlin, D-12489 Berlin, Germany}
\affiliation{Fakult{\"a}t f{\"u}r Physik {\&} Astronomie, Ruhr-Universit{\"a}t Bochum, D-44780 Bochum, Germany}
\affiliation{Universit{\'e} Libre de Bruxelles, Science Faculty CP230, B-1050 Brussels, Belgium}
\affiliation{Vrije Universiteit Brussel (VUB), Dienst ELEM, B-1050 Brussels, Belgium}
\affiliation{Dept. of Physics, Simon Fraser University, Burnaby, BC V5A 1S6, Canada}
\affiliation{Department of Physics and Laboratory for Particle Physics and Cosmology, Harvard University, Cambridge, MA 02138, USA}
\affiliation{Dept. of Physics, Massachusetts Institute of Technology, Cambridge, MA 02139, USA}
\affiliation{Dept. of Physics and The International Center for Hadron Astrophysics, Chiba University, Chiba 263-8522, Japan}
\affiliation{Department of Physics, Loyola University Chicago, Chicago, IL 60660, USA}
\affiliation{Dept. of Physics and Astronomy, University of Canterbury, Private Bag 4800, Christchurch, New Zealand}
\affiliation{Dept. of Physics, University of Maryland, College Park, MD 20742, USA}
\affiliation{Dept. of Astronomy, Ohio State University, Columbus, OH 43210, USA}
\affiliation{Dept. of Physics and Center for Cosmology and Astro-Particle Physics, Ohio State University, Columbus, OH 43210, USA}
\affiliation{Niels Bohr Institute, University of Copenhagen, DK-2100 Copenhagen, Denmark}
\affiliation{Dept. of Physics, TU Dortmund University, D-44221 Dortmund, Germany}
\affiliation{Dept. of Physics and Astronomy, Michigan State University, East Lansing, MI 48824, USA}
\affiliation{Dept. of Physics, University of Alberta, Edmonton, Alberta, T6G 2E1, Canada}
\affiliation{Erlangen Centre for Astroparticle Physics, Friedrich-Alexander-Universit{\"a}t Erlangen-N{\"u}rnberg, D-91058 Erlangen, Germany}
\affiliation{Physik-department, Technische Universit{\"a}t M{\"u}nchen, D-85748 Garching, Germany}
\affiliation{D{\'e}partement de physique nucl{\'e}aire et corpusculaire, Universit{\'e} de Gen{\`e}ve, CH-1211 Gen{\`e}ve, Switzerland}
\affiliation{Dept. of Physics and Astronomy, University of Gent, B-9000 Gent, Belgium}
\affiliation{Dept. of Physics and Astronomy, University of California, Irvine, CA 92697, USA}
\affiliation{Karlsruhe Institute of Technology, Institute for Astroparticle Physics, D-76021 Karlsruhe, Germany}
\affiliation{Karlsruhe Institute of Technology, Institute of Experimental Particle Physics, D-76021 Karlsruhe, Germany}
\affiliation{Dept. of Physics, Engineering Physics, and Astronomy, Queen's University, Kingston, ON K7L 3N6, Canada}
\affiliation{Department of Physics {\&} Astronomy, University of Nevada, Las Vegas, NV 89154, USA}
\affiliation{Nevada Center for Astrophysics, University of Nevada, Las Vegas, NV 89154, USA}
\affiliation{Dept. of Physics and Astronomy, University of Kansas, Lawrence, KS 66045, USA}
\affiliation{Centre for Cosmology, Particle Physics and Phenomenology - CP3, Universit{\'e} catholique de Louvain, Louvain-la-Neuve, Belgium}
\affiliation{Department of Physics, Mercer University, Macon, GA 31207-0001, USA}
\affiliation{Dept. of Astronomy, University of Wisconsin{\textemdash}Madison, Madison, WI 53706, USA}
\affiliation{Dept. of Physics and Wisconsin IceCube Particle Astrophysics Center, University of Wisconsin{\textemdash}Madison, Madison, WI 53706, USA}
\affiliation{Institute of Physics, University of Mainz, Staudinger Weg 7, D-55099 Mainz, Germany}
\affiliation{Department of Physics, Marquette University, Milwaukee, WI 53201, USA}
\affiliation{Institut f{\"u}r Kernphysik, Universit{\"a}t M{\"u}nster, D-48149 M{\"u}nster, Germany}
\affiliation{Bartol Research Institute and Dept. of Physics and Astronomy, University of Delaware, Newark, DE 19716, USA}
\affiliation{Dept. of Physics, Yale University, New Haven, CT 06520, USA}
\affiliation{Columbia Astrophysics and Nevis Laboratories, Columbia University, New York, NY 10027, USA}
\affiliation{Dept. of Physics, University of Oxford, Parks Road, Oxford OX1 3PU, United Kingdom}
\affiliation{Dipartimento di Fisica e Astronomia Galileo Galilei, Universit{\`a} Degli Studi di Padova, I-35122 Padova PD, Italy}
\affiliation{Dept. of Physics, Drexel University, 3141 Chestnut Street, Philadelphia, PA 19104, USA}
\affiliation{Physics Department, South Dakota School of Mines and Technology, Rapid City, SD 57701, USA}
\affiliation{Dept. of Physics, University of Wisconsin, River Falls, WI 54022, USA}
\affiliation{Dept. of Physics and Astronomy, University of Rochester, Rochester, NY 14627, USA}
\affiliation{Department of Physics and Astronomy, University of Utah, Salt Lake City, UT 84112, USA}
\affiliation{Dept. of Physics, Chung-Ang University, Seoul 06974, Republic of Korea}
\affiliation{Oskar Klein Centre and Dept. of Physics, Stockholm University, SE-10691 Stockholm, Sweden}
\affiliation{Dept. of Physics and Astronomy, Stony Brook University, Stony Brook, NY 11794-3800, USA}
\affiliation{Dept. of Physics, Sungkyunkwan University, Suwon 16419, Republic of Korea}
\affiliation{Institute of Physics, Academia Sinica, Taipei, 11529, Taiwan}
\affiliation{Dept. of Physics and Astronomy, University of Alabama, Tuscaloosa, AL 35487, USA}
\affiliation{Dept. of Astronomy and Astrophysics, Pennsylvania State University, University Park, PA 16802, USA}
\affiliation{Dept. of Physics, Pennsylvania State University, University Park, PA 16802, USA}
\affiliation{Dept. of Physics and Astronomy, Uppsala University, Box 516, SE-75120 Uppsala, Sweden}
\affiliation{Dept. of Physics, University of Wuppertal, D-42119 Wuppertal, Germany}
\affiliation{Deutsches Elektronen-Synchrotron DESY, Platanenallee 6, D-15738 Zeuthen, Germany}

\author{R. Abbasi}
\affiliation{Department of Physics, Loyola University Chicago, Chicago, IL 60660, USA}
\author{M. Ackermann}
\affiliation{Deutsches Elektronen-Synchrotron DESY, Platanenallee 6, D-15738 Zeuthen, Germany}
\author{J. Adams}
\affiliation{Dept. of Physics and Astronomy, University of Canterbury, Private Bag 4800, Christchurch, New Zealand}
\author{S. K. Agarwalla}
\thanks{also at Institute of Physics, Sachivalaya Marg, Sainik School Post, Bhubaneswar 751005, India}
\affiliation{Dept. of Physics and Wisconsin IceCube Particle Astrophysics Center, University of Wisconsin{\textemdash}Madison, Madison, WI 53706, USA}
\author{J. A. Aguilar}
\affiliation{Universit{\'e} Libre de Bruxelles, Science Faculty CP230, B-1050 Brussels, Belgium}
\author{M. Ahlers}
\affiliation{Niels Bohr Institute, University of Copenhagen, DK-2100 Copenhagen, Denmark}
\author{J.M. Alameddine}
\affiliation{Dept. of Physics, TU Dortmund University, D-44221 Dortmund, Germany}
\author{S. Ali}
\affiliation{Dept. of Physics and Astronomy, University of Kansas, Lawrence, KS 66045, USA}
\author{N. M. Amin}
\affiliation{Bartol Research Institute and Dept. of Physics and Astronomy, University of Delaware, Newark, DE 19716, USA}
\author{K. Andeen}
\affiliation{Department of Physics, Marquette University, Milwaukee, WI 53201, USA}
% ...
% \date{\today}

%%%%%%%%%%%%%%%%%%%%%%%%%%%%%%%%%%%%%%%%%%%%%%%%%%%%%%%%%%%
\begin{abstract}
Models describing dark matter as a novel particle often predict that its annihilation or decay into Standard Model particles could produce a detectable neutrino flux in regions of high dark matter density, such as the Galactic Center. In this work, we search for these neutrinos using $\sim$9 years of IceCube-DeepCore data with an event selection optimized for energies between 15 GeV to 200 GeV. 
We considered several annihilation and decay channels and dark matter masses ranging from 15 GeV up to 8 TeV. No significant deviation from the background expectation from atmospheric neutrinos and muons was found. The most significant result was found for a dark matter mass of 201.6 GeV annihilating into a pair of $b\bar{b}$ quarks assuming the Navarro-Frenk-White halo profile with a post-trial significance of $1.08 \;\sigma$. We present upper limits on the thermally-averaged annihilation cross-section of the order of  $10^{-24}~\mathrm{cm}^3 \mathrm{s}^{-1}$, as well as lower limits on the dark matter decay lifetime up to $10^{26}~\mathrm{s}$ for dark matter masses between 5 GeV up to 8 TeV. These results strengthen the current IceCube limits on dark matter masses above 20 GeV and provide an order of magnitude improvement at lower masses. In addition, they represent the strongest constraints from any neutrino telescope on GeV-scale dark matter and are among the world-leading limits for several dark matter scenarios.
\end{abstract}

\newpage

%%%%%%%%%%%%%%%%%%%%%%%%%%%%%%%%%%%%%%%%%%%%%%%%%%%%%%%%%%%
\maketitle
%%%%%%%%%%%%%%%%%%%%%%%%%%%%%%%%%%%%%%%%%%%%%%%%%%%%%%%%%%%
%%%%%%%%%%%%%%%%%%%%%%%%%%%%%%%%%%%%%%%%%%%%%%%%%%%%%%%%%%%
\section{Introduction}\label{sec:Introduction}
Multiple cosmological observations provide strong evidence for the existence of \textit{Dark Matter} (DM) \cite{Persic:1995ru, Rubin:1970zza, Zwicky:1937zza, Jee:2007nx, Jee:2008qj}, an elusive component that constitutes approximately $22 \%$ of today's energy density of the Universe and $85 \%$ of its present matter content \cite{Planck:2013pxb}. Despite an extensive body of evidence, the fundamental nature of DM remains unknown. Numerous experimental and theoretical efforts are currently underway to better understand this fundamental component of the Universe. A detailed review on DM can be found in Refs.~\cite{RevModPhys.90.045002, Cirelli:2024ssz}.

The difficulty in understanding the properties of DM stems from the lack of direct evidence for its interactions with ordinary matter, aside from its gravitational effects on Standard Model (SM) particles. However, various \textit{Beyond the Standard Model} (BSM) extensions predict a weak coupling between DM and SM particles. These theories propose DM candidates as new particles that interact feebly with known SM particles~\cite{Bertone:2004pz}. These candidates are usually assumed to have been produced in the early Universe and to be stable, or long-lived enough, to explain the presently inferred DM abundance, which is influenced by the thermal history of the Universe. A widely popular candidate is the \textit{Weakly Interacting Massive Particle} (WIMP), with masses ranging from a~GeV up to $\mathcal{O}(100) \; \mathrm{TeV}$ \cite{Arcadi:2017kky}.

Indirect detection of DM relies on identifying anomalous fluxes of SM particles resulting from DM decay or annihilation in regions with an expected over-density of DM particles \cite{PerezdelosHeros:2020qyt, Gaskins:2016cha}. Among the possible target regions, the \textit{Galactic Center} is particularly promising due to its large DM content and relative proximity to Earth. Notably, the Fermi-LAT telescope has detected a gamma-ray excess in the Galactic Center (the so-called GeV excess), an observation that has been the subject of study for over a decade \cite{Daylan:2014rsa, Fermi-LAT:2015sau, Fermi-LAT:2017opo, Hooper:2022bec}. This excess has sparked a long debate over its possible origin, whether it arises from the annihilation of thermal relic DM particles with GeV-scale masses, or from astrophysical sources such as a population of unresolved millisecond pulsars \cite{Murgia:2020dzu,Cholis:2021rpp, Caron:2022akb, Holst:2024fvb}. While the origin of the GeV excess remains unexplained, complementary searches using different astrophysical sources and detection messengers, such as neutrinos \cite{Beacom:2006tt,Yuksel:2007ac, Arguelles:2022nbl, IceCube:2023ies, Super-Kamiokande:2020sgt, ANTARES:2019svn, KM3NeT:2024xca}, can provide additional insights into the nature of DM.

In this article, we present an indirect search for DM through neutrino signals from DM annihilation or decay from the Galactic Center using data from the IceCube neutrino telescope \cite{Aartsen:2016nxy}. We consider DM interactions that produce two SM particles in the final state, referred to as \textit{annihilation or decay channels}. Specifically, we study benchmark cases where DM annihilates or decays into $b\bar{b}, \; W^{+}W^{-}, \; \tau \bar{\tau}, \; \mu^+ \mu^-$ and $\nu \bar{\nu}$ of all three standard neutrino flavors assuming a 100\% branching ratio into each of these channels separately. While the $\nu \bar{\nu}$ channel can yield a direct neutrino signal from primary production, the other SM channels undergo further decays or hadronization, producing secondary neutrinos that contribute to the detectable flux.

These benchmark channels are chosen to represent a broad range of possible neutrino energy spectra: from soft spectra (e.g., $b\bar{b}$) to harder spectra (e.g., $W^+W^-$, $\tau^+\tau^-$, $\mu^+\mu^-$), and including sharply peaked spectra from direct neutrino production ($\nu\bar{\nu}$). The latter is commonly referred to as the \textit{neutrino line channel}, due to its distinctive monochromatic peak in the energy spectrum at an energy corresponding to the DM mass (for annihilation) or half the DM mass (for decay). While these benchmarks do not cover other possibilities of two-body or multi-body final states, they provide representative spectral shapes. Any potential signal with a similar spectral feature would likely be captured within the search of one of these benchmark cases.

IceCube has previously conducted DM analyses (see for example Refs.~\cite{ IceCube:2016oqp,IceCube:2017rdn,IceCube:2023ies,IceCube:2011kcp,IceCube:2018tkk}). While earlier studies \cite{IceCube:2016oqp,IceCube:2017rdn}  used solely directional information, the more recent work in Ref.~\cite{IceCube:2023ies} incorporated both direction and energy, along with an optimized selection targeting the neutrino line channel.

Since then, IceCube has not only accumulated more data, but also witnessed significant improvements in detector understanding, including enhanced calibration of individual optical modules using in-situ data \cite{IceCube:2020nwx} and advancements in ice modeling \cite{Abbasi:2024wpf}. In this work, we adopt the most recent IceCube-DeepCore dataset, optimized for detecting sub-TeV neutrinos. This dataset was originally developed and used for oscillation studies \cite{IceCube:2023ewe, IceCube:2024xjj}. Furthermore, we employ a state-of-the-art reconstruction algorithm designed for GeV-scale low-energy events \cite{IceCube:2022kff}. As a result, our analysis is expected to enhance IceCube sensitivity to DM with GeV-scale masses.

This article is structured as follows: Section \ref{sec:Icecube} describes the IceCube detector design and the event sample used in this analysis. Section \ref{sec:signal_expectation} presents the computation of the expected DM signal, while Section \ref{sec:analysis_method} outlines the statistical method employed. We also discuss the impact of systematic uncertainties in Section \ref{sec:systematics}. The results are presented and discussed in Section \ref{sec:results}. Finally, the summary and conclusion are given in Section \ref{sec:conclusion}.

%%%%%%%%%%%%%%%%%%%%%%%%%%%%%%%%%%%%%%%%%%%%%%%%%%%%%%%%%%%
\section{IceCube and Event Selection}\label{sec:Icecube}
The IceCube Neutrino Observatory \cite{Aartsen:2016nxy} is a Cherenkov neutrino telescope located at the South Pole, embedded in the ice at depths ranging from 1.5 km to 2.5 km. The detector comprises a total of 5,160 photomultiplier tubes (PMTs) \cite{Abbasi_2010} arranged in a three-dimensional array covering approximately 1 $\mathrm{km}^3$. Each PMT, along with its digital electronics, is hosted within a Digital Optical Module (DOM) \cite{IceCube:2008qbc}. These DOMs are then connected by cables to form 86 vertical \textit{strings}.

Throughout most of the IceCube array, the average spacing between DOMs is about 125 m horizontally and 17 m vertically. This arrangement is optimized for the detection of astrophysical neutrinos with energies ranging from a few TeV up to several PeV. At the center-bottom of IceCube, where the ice is clearest, a denser sub-array referred to as DeepCore is located \cite{IceCube:2011ucd}. DeepCore features a higher concentration of DOMs, with a horizontal spacing of 42 m to 72 m and a vertical spacing of 7 m. The DeepCore DOMs also obtain higher quantum efficiency PMTs compared to the rest of IceCube. The design of DeepCore extends the IceCube detection capability to energies as low as $\sim$10 GeV.

IceCube detects neutrino signals through Cherenkov light induced by relativistic charged particles produced from the interaction between neutrinos and surrounding matter. The charged-current (CC) interaction of $\nu_\mu/\bar{\nu}_{\mu}$ creates an energetic muon, leading to elongated light patterns of \textit{track} event topology. On the other hand, $\nu_e/\overline{\nu}_{e}$ CC and the neutral current (NC) interaction of all neutrino flavors induce a spherical light emission, referred to as \textit{cascade} events, due to the production of electron and hadronic showers. For $\nu_\tau/\bar{\nu}_{\tau}$ CC interactions, below $1$ PeV, the tau decay length is short enough compared to the average DOM separation in IceCube such that these interactions also mostly appear as cascades. In 17.8\% of the cases, $\nu_\tau/\bar{\nu}_{\tau}$ CC interactions contribute to track events due to the muonic decay of the tau lepton.

This work uses the recent DeepCore data sample which benefits from numerous improvements in the calibration, simulation, and understanding of the ice properties \cite{IceCube:2020nwx, Abbasi:2024wpf}. For events originating from the Northern Hemisphere, Earth serves as a natural shield, significantly reducing atmospheric muon contamination. For sources in the Southern Sky, such as the Galactic Center, the rest of the IceCube detector can serve as an active veto to suppress the contribution of atmospheric muons. The event selection is designed to minimize this background and the detector noise while retaining a large statistics sample of pure and well-reconstructed neutrinos at GeV energies. A comprehensive description of the novel DeepCore sample is presented in Ref.~\cite{IceCube:2023ewe}. Both track and cascade topologies are included in this analysis, while also employing a cutting-edge likelihood-based reconstruction developed specifically for low energy events in IceCube-DeepCore \cite{IceCube:2022kff}.

\begin{figure}
    \centering    \includegraphics[width=0.5\textwidth]{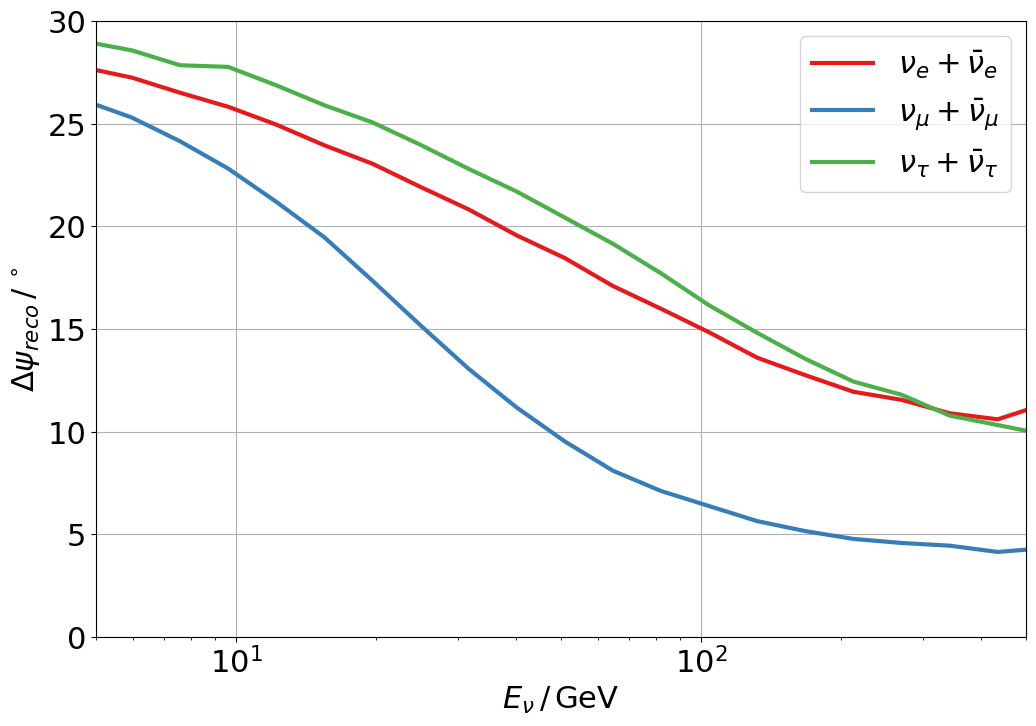}
    \caption{Angular resolution of each neutrino flavor in terms of the opening angle from the Galactic Center as a function of neutrino energy.}
    \label{fig:angular_resolution}
\end{figure}

\begin{figure}
    \centering    \includegraphics[width=0.5\textwidth]{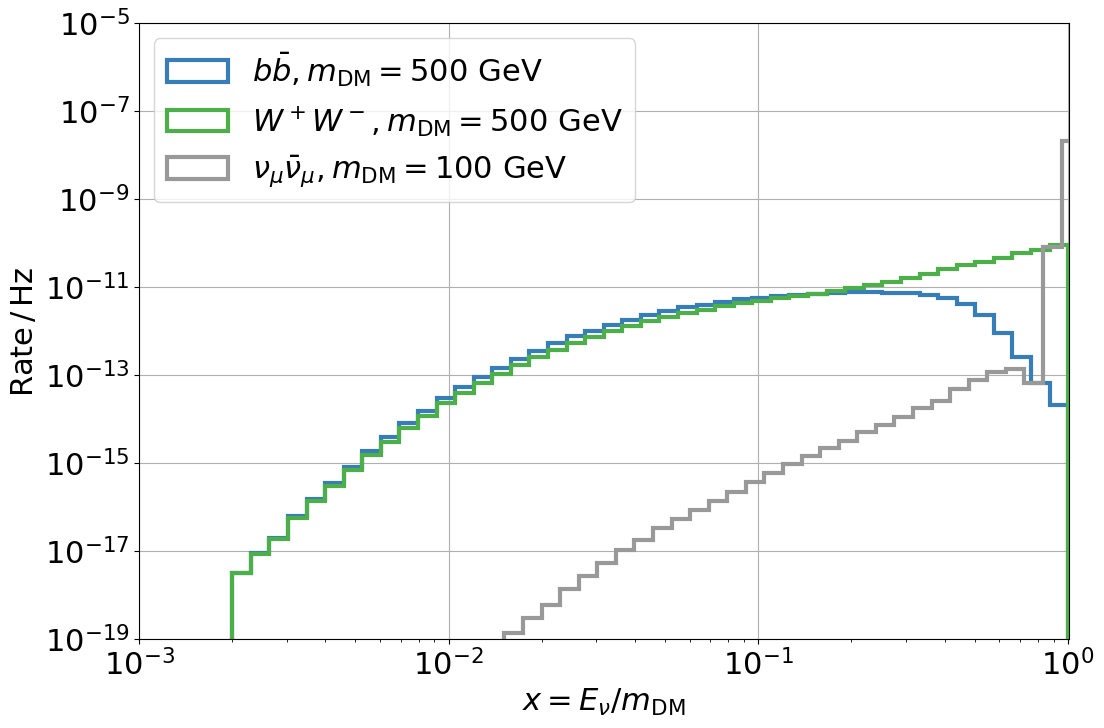}
    \caption{Predicted event rate as the function of true neutrino energy for different dark matter masses and channels, assuming Navarro-Frenk-White (NFW) profile and annihilation with $\langle \sigma v \rangle = 10^{-26}~\mathrm{cm^3 \mathrm{s}^{-1}}$.}
    \label{fig:signal_acceptance}
\end{figure}

Fig.~\ref{fig:angular_resolution} and Fig.~\ref{fig:signal_acceptance} illustrate the performance of the DeepCore sample in the search for DM in the Galactic Center. Fig.~\ref{fig:angular_resolution} depicts the angular resolution, represented by the median error in the reconstructed opening angle from the Galactic Center, as a function of the neutrino energy. At 10 GeV, the resolution can achieve approximately 30 degrees for the median error. The angular resolution improves with increasing energy down to 5-15 degrees at 100 GeV, with muon neutrinos achieving the best resolution due to their long track signature in the detector. Fig.~\ref{fig:signal_acceptance} presents the expected rates of DM signals without accounting for resolution effects. The neutrino line channel, $\nu_\mu \overline{\nu}_\mu$, exhibits a sharp peak at the DM particle mass. The expected rate for the $b \overline{b}$ channel decreases at higher energies due to its softer energy spectrum, whereas the $W^+ W^-$ channel shows an increasing trend owing to its harder spectrum.

At the reconstructed energy range considered in this analysis (1-1000 GeV), the background is dominated by atmospheric neutrinos, with approximately $10\%$ of the sample attributed to atmospheric muons. The recent discovery by IceCube of high-energy neutrino diffuse emission in the Galactic Plane (GP) \cite{IceCube:2023ame} produced by the interaction of cosmic rays with the Milky Way interstellar medium introduces an additional background component for DM searches. Given the absence of measurements of this flux below 1 TeV, we estimate its contribution by extrapolating the best-fit and theoretically predicted spectra presented in Ref.~\cite{IceCube:2023ame} down to GeV energies, assumed to be due to $pp$ interaction with the same power-law behavior and an unchanged spatial distribution. This estimation indicates that the GP neutrino flux contributes between 0.01\% and 0.05\% to the total event count, depending on the model assumptions. An evaluation of the impact of this astrophysical flux on our analysis is presented in Section~\ref{sec:systematics}.

%%%%%%%%%%%%%%%%%%%%%%%%%%%%%%%%%%%%%%%%%%%%%%%%%%%%%%%%%%%
\section{Signal Expectation}\label{sec:signal_expectation}
In this analysis, both DM self-annihilation and decay in the Galactic Center are considered. The corresponding differential neutrino flux from such processes with DM particles of mass $m_\mathrm{DM}$ can be computed as a function of neutrino energy $E_\nu$ and the opening angle from the Galactic Center $\Psi$ as follows:
\begin{align}
     \frac{\mathrm{d}\phi_{\nu}^{\mathrm{ann.}}}{\mathrm{d}E_{\nu}} (E_\nu, \Psi) &=  \frac{\langle \sigma \upsilon \rangle}{4\pi\mathit{k} m_{\mathrm{DM}}^2} \frac{\mathrm{d}N_{\nu}}{\mathrm{d}E_{\nu}} \int_{\Omega} \int_{\mathrm{l.o.s}} \rho^2 \left(r(l, \Psi) \right) \mathrm{d} l  \; \mathrm{d} \Omega \; , \nonumber \\
  \frac{\mathrm{d}\phi_{\nu}^{\mathrm{dec.}}}{\mathrm{d}E_{\nu}} (E_\nu, \Psi) &= \frac{1}{4\pi m_{\mathrm{DM}} \tau} \frac{\mathrm{d}N_{\nu}}{\mathrm{d}E_{\nu}} \int_{\Omega} \int_{\mathrm{l.o.s}} \rho \left(r(l, \Psi) \right) \mathrm{d} l  \; \mathrm{d} \Omega \; ,
   \label{eq:signal-flux}
\end{align} 
where $\langle \sigma \upsilon \rangle$ is the thermally-averaged self-annihilation cross-section, and $\tau$ is the decay lifetime of dark matter. In this work, we focus on the case of fermionic DM where the factor $k$ depends on the nature of the DM particles: $k=4$ for Dirac particles and $k=2$ for Majorana ones. We assume that DM is a Majorana particle \cite{Queiroz:2016sxf}, corresponding to the more optimistic signal flux. The constraints on $\langle \sigma \upsilon \rangle$ derived from this work can be translated to the Dirac case by scaling them up by a factor of 2.

The last integral encodes the astrophysical information of the DM halo in the Milky Way and is referred to as the J-factor for the DM annihilation and D-factor for the DM decay. These terms govern the spatial dependence of the dark matter signal and can be expressed as the three-dimensional integration along the line-of-sight (l.o.s) and over the solid angle $\Omega$:
\begin{align}
    J (\Psi) &= \int_{\Omega} \int_{\mathrm{l.o.s}} \rho^2 \left(r(l, \Psi) \right) \; \mathrm{d} l  \; \mathrm{d} \Omega, \nonumber\\
    D (\Psi) &=\int_{\Omega} \int_{\mathrm{l.o.s}} \rho \left(r(l, \Psi) \right) \; \mathrm{d} l  \; \mathrm{d} \Omega,
    \label{eq:Jfactor}
\end{align}
where $r(l, \psi)$ is the galactocentric distance which can be expressed as a function of opening angle $\Psi$ between the direction of the Galactic Center and l.o.s, and the distance $l$ along the l.o.s. The DM density $\rho$ is assumed to be spherically symmetric. 

Understanding the morphology of DM halos remains a central challenge and constitutes a major source of astrophysical uncertainty~\cite{Benito_2019}. In the case of the Milky Way, observational constraints on the DM distribution are robust only in the Solar neighborhood, where the local DM density is inferred to be $\rho_\odot \simeq 0.4~\mathrm{GeV\;cm^{-3}}$ from stellar kinematics~\cite{Nesti:2013uwa, de_Salas_2019, Ou:2023adg, S_ding_2025}. In contrast, the DM density and spatial profile in the inner Galactic region, which are most relevant for indirect searches targeting the Galactic Center, remain largely unconstrained by observations.

This uncertainty is closely related to the so-called \textit{core--cusp problem}~\cite{deBlok:2009sp}, originally identified in studies of dwarf and low-surface-brightness galaxies. Collisionless $N$-body simulations of cold DM predict a steeply rising DM density toward the halo center, commonly referred to as a cuspy profile and commonly parametrized by the Navarro--Frenk--White (NFW) profile~\cite{NFW_1996}. However, observations of galactic rotation curves in dwarf galaxies often favor a nearly constant-density core, motivating cored halo profiles such as the Burkert profile~\cite{Burkert_1995}.

More recent hydrodynamical simulations that include baryonic physics suggest a more nuanced picture. Feedback processes associated with star formation and supernovae can redistribute DM and flatten the central density, an effect expected to be particularly efficient in low-mass systems such as dwarf galaxies. Conversely, in MW–like galaxies, the large gravitational potential of baryonic matter in the central region can lead to an adiabatic contraction of the DM halo, resulting in a steeper, cusp-like density profile~\cite{DiCintio:2013qxa, Pontzen:2014lma, Chan:2015tna}.

Given these uncertainties, and in order to bracket a conservative range of possible DM distributions in the Milky Way, this analysis adopts two benchmark halo models: the Burkert profile, representing an extreme cored scenario, and the NFW profile, representing a cuspy distribution. Both DM density profiles can be expressed as a function of distance $r$ to the Galactic Center, following a unified parameterization taken from Ref.~\cite{An:2012pv}:
\begin{equation}
    \rho(r) = \frac{\rho_0}{(\delta + \frac{r}{r_s})^\gamma \left[ 1 + \left(\frac{r}{r_s}\right)^\alpha \right]^{(\beta-\gamma)/\alpha}},
\end{equation}
where $(\alpha, \beta, \delta, \gamma)$
are equal to $(2,3,1,1)$ for the Burkert profile and $(1,3,0,1)$ for the NFW profile.
The values of normalization density $\rho_0$ and scale radius $r_s$ are taken from Ref.~\cite{Nesti:2013uwa}, where they are derived from kinematic measurements in the MW. The corresponding $J$- and $D$-factors are computed by integrating the DM density (or its square) along the line of sight and over the solid angle of interest using the {\sc Clumpy} package~\cite{Hutten:2018aix}.

The energy spectra of DM signal depend on the term $\mathrm{d}N_{\nu}/\mathrm{d}E_{\nu}$ which is the differential number of neutrinos per energy and annihilation (decay). This quantity is determined by the mass of DM particles as well as the primary SM products of the annihilation or decay. As mentioned in Section~\ref{sec:Introduction}, we consider the following \textit{channels} where DM annihilates/decays into: $b\bar{b}, \; W^{+}W^{-}, \; \tau \bar{\tau}, \; \mu^+ \mu^-$ and $\nu \bar{\nu}$ of all standard 3 neutrino flavors, with the assumption of a 100\% branching ratio into each individual channel. 

The neutrino spectra $\mathrm{d}N_{\nu}/\mathrm{d}E_{\nu}$ are generated using $\chi\mathrm{aro}\nu$ \cite{Liu_2020}, a framework for the computation of neutrino fluxes from WIMP annihilation and decay. $\chi\mathrm{aro}\nu$ uses PYTHIA-8.2 to simulate neutrino production for DM masses below the electroweak (EW) scale (set at 100 GeV). For DM masses above this scale, the framework adopts the recent spectra calculation \cite{Bauer:2020jay}, which includes the state-of-the-art EW corrections. The resulting neutrino flux is then propagated from the Galactic Center to the Earth, assuming oscillation-averaged neutrino spectra propagation over the large travel distance, with oscillation parameters taken from a global fit of neutrino oscillation measurements \cite{Esteban:2020cvm}.

The signal expectation in the two-dimensional space of reconstructed energy and opening angle is then computed by convolving the true incoming flux in equation \eqref{eq:signal-flux} with \textit{the response matrix}. This response matrix, built from Monte-Carlo (MC) simulation, encodes the probability of a neutrino with given true properties (neutrino/antineutrino, flavor, energy, and direction) interacting, passing the selection, and being reconstructed. A kernel density estimation \cite{KDEpy} is applied to obtain a smooth response matrix and reduce the effect of MC errors caused by limited statistics of the MC sample.

%%%%%%%%%%%%%%%%%%%%%%%%%%%%%%%%%%%%%%%%%%%%%%%%%%%%%%%%%%%
\section{Analysis Method}\label{sec:analysis_method}

This analysis adopts a binned Poisson likelihood method on two observables: the reconstructed energy ($E_{reco}$) and the reconstructed opening angle from the Galactic Center ($\Psi_{reco}$) of the events. All cascade-like and track-like events are included to account for all neutrino flavors. The likelihood function is defined as the product of the Poissonian probabilities $\mathcal{P}$ of observing $n^{obs}$ events given the predicted number of events $\mu$ in each observable bin:
\begin{equation}
     \mathcal{L}(\xi) = \prod\limits_{i} \mathcal{P} (n^{obs}_{i}, \mu_i) \,.
     \label{eq:likelihood}
\end{equation}
The expected number of events in bin $i$ is given by $\mu_i = n_{tot}^{obs} f(i, \xi)$ where $n_{tot}^{obs} = \sum_i n_i^{obs}$ is the total number of observed events, which is fixed by the data. The fraction $f(i, \xi)$ is the fraction of events in bin $i$ under a given hypothesis and defined as:
\begin{equation}
     f(i \, ;  \, \xi) = \xi \, \mathcal{S}_i \, + \, (1 - \xi) \mathcal{B}_i \,,
     \label{eq:event-fraction}
\end{equation}
where $\mathcal{S}$ and $\mathcal{B}$ denote the \textit{probability density functions} (PDFs) for the signal and the background distribution respectively and $\xi$ is the overall signal fraction.

\begin{figure*}
    % \centering
    \begin{minipage}{0.48\textwidth}
        \includegraphics[width=\linewidth]{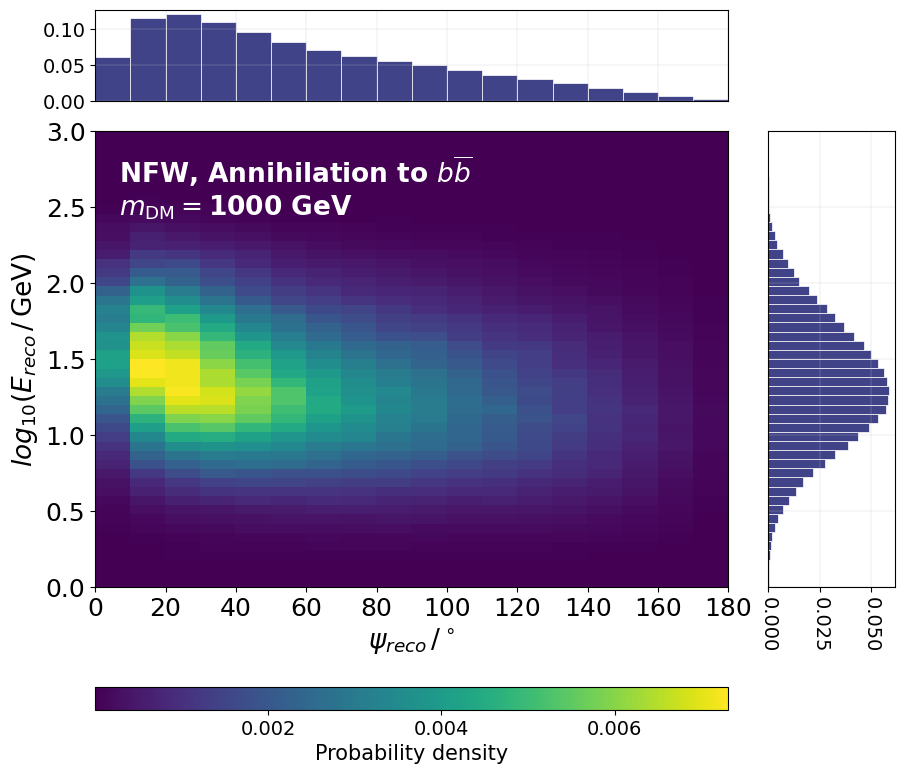}
    \end{minipage}
    \hfill
    \begin{minipage}{0.48\textwidth}
        \includegraphics[width=\linewidth]{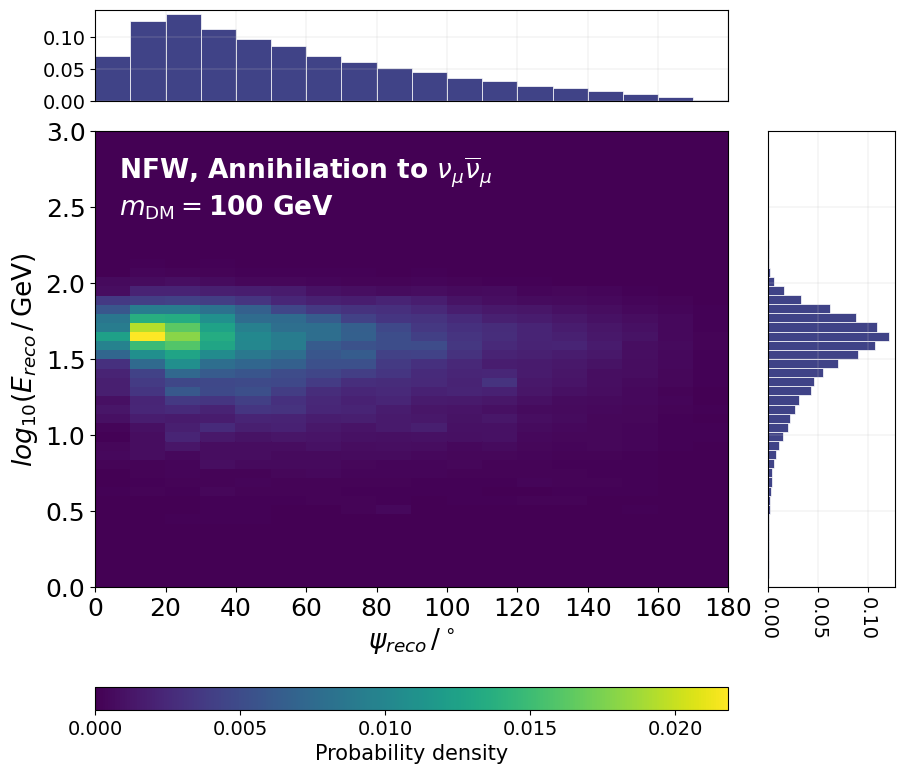}
    \end{minipage}
\caption{Two-dimensional PDFs as a function of reconstructed energy ($E_{reco}$) and reconstructed opening angle relative to the Galactic Center ($\psi_{reco}$) for a DM signal with mass 1000 GeV annihilating into $b\overline{b}$ \textit{(left)} and with mass 100 GeV annihilating into $\nu_\mu \overline{\nu}_{\mu}$  \textit{(right)}, both assuming NFW profile.}
\label{fig:signal_pdfs}
\end{figure*}

Fig.~\ref{fig:signal_pdfs} presents examples of signal PDFs for the case of DM with masses of 1000 GeV and 100 GeV annihilating into pairs of $b \bar{b}$ (left panel) and $\nu_{\mu} \bar{\nu}_{\mu}$ (right panel) assuming the NFW profile of the Galactic halo. As expected, the neutrino line channel exhibits a sharp energy peak near the DM mass, with the smearing of the monochromatic line attributed to the energy resolution of the detector. On the other hand, the expected energy spectrum of the $b \bar{b}$ channel exhibits an excess extending across a broader range of energies below the DM mass. This feature results from the interplay between the power-law spectra of the secondary neutrinos and the efficiency of the detector. In both cases, the directional distribution shows a pronounced excess toward the Galactic Center, reflecting the spatial dependence of the J-factor in the assumed halo profile.

\begin{figure}
    \centering
    \includegraphics[width=.49\textwidth]{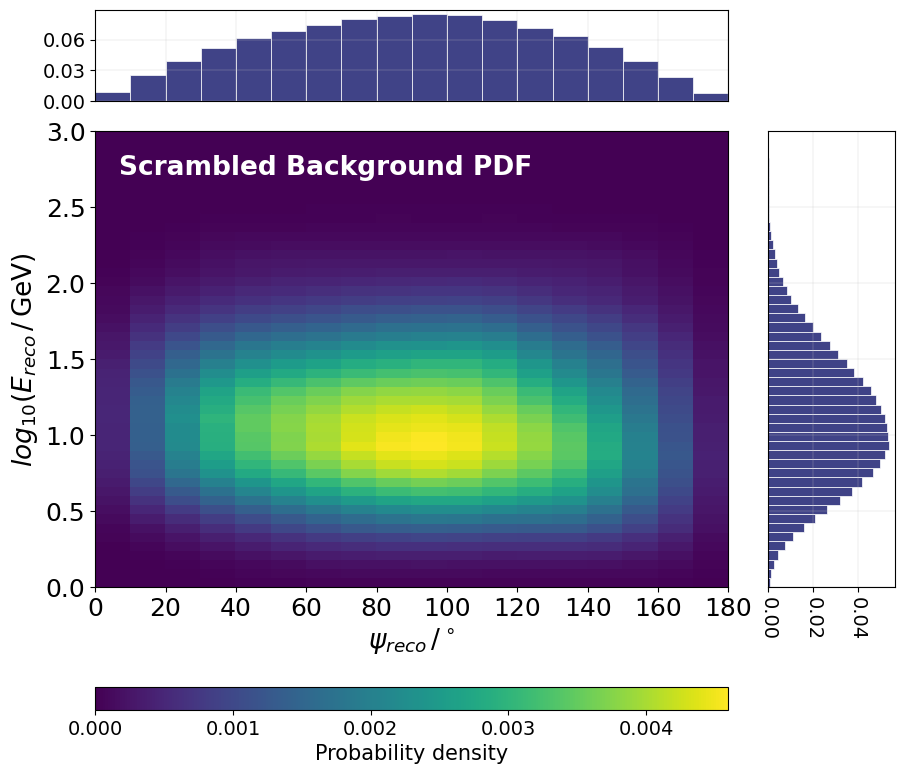}
    \caption{Two-dimensional PDF of background estimated as RA-scrambled data.}
    \label{fig:background_pdf}
\end{figure}

The background PDF is estimated using a data-driven method which was also adopted in the previous IceCube analysis for the DM signal from the Galactic Center \cite{IceCube:2023ies}. This background estimation is performed as the average of 100 data samples where each data event is assigned a random right-ascension (RA) coordinate which is equivalent to time-scrambling of data in IceCube's local coordinate system. The resulting background PDF, dubbed here as the scrambled background $\mathcal{B}^{scr}$, is shown in Fig.~\ref{fig:background_pdf}. This method is valid and commonly used in neutrino telescopes thanks to their active livetime of over $98 \%$ and the rotation of the Earth which causes the detected atmospheric neutrino and muon backgrounds to be uniform in the right ascension. The scrambling technique allows the PDFs to be marginally dependent on MC (through only signal PDF), thus making the analysis robust to the systematics effects. In this approach, a potential signal in the data could contribute to the scrambled background. To account for this effect, a background-only hypothesis, as described in equation \eqref{eq:event-fraction}, can be computed as:
\begin{equation}
    \mathcal{B}_i = \frac{1}{1-\xi} (\mathcal{B}_i^{\mathrm{scr}} - \xi \mathcal{S}^{\mathrm{scr}}_i) \, ,
    \label{eq:bkg-with-signalsub}
\end{equation}
where $\mathcal{S}^{\mathrm{scr}}_i$ is the right ascension scrambled signal PDF. The corresponding likelihood is referred to as the \textit{signal-subtraction likelihood}. With this adjustment, the testing hypothesis PDF in equation \eqref{eq:event-fraction} now becomes:

\begin{equation}
    f(i; \xi) = \xi \mathcal{S}_i + \mathcal{B}_i^\mathrm{scr} - \xi \mathcal{S}_i^\mathrm{scr} \; .
    \label{eq:fraction-with-signalsub}
\end{equation}

In this analysis, the signal fraction $\xi$ is the only parameter to be fitted, so that the best-fit signal fraction $\hat{\xi}$ maximizes the likelihood in equation \eqref{eq:likelihood}. As shown in equation \eqref{eq:signal-flux}, the total number of signal events is proportional to the thermally-averaged cross-section $\langle \sigma \upsilon \rangle$ or the inverse of the DM lifetime $\tau$. Therefore, the best-fit signal fraction can be translated to the corresponding values of these two physical parameters. 

To quantify the statistical significance of the best-fit result $\mathcal{H}(\hat{\xi})$ relative to the null hypothesis $\mathcal{H}(\xi=0)$, we employ the \textit{discovery test statistic} as defined in Ref.~\cite{Cowan:2010js}:
\begin{equation}
    q  = \left\{
        \begin{array}{ll}
            -2 \ln \dfrac{\mathcal{L}(0)}{\mathcal{L}(\hat{\xi})}  & \mbox{if } \hat{\xi} \geq 0 \;, \\
            0 & \mbox{otherwise.}
        \end{array}
 \right. 
 \label{eq:discovery_ts}
\end{equation}
This test statistic ensures that the null hypothesis can only be rejected in case a positive signal fraction is preferred by the data. 

In this article, we test 30 logarithmically spaced mass values for each combination of DM annihilation or decay, the NFW or Burkert profile, and the considered production channels:  $b\bar{b}, \; W^{+}W^{-}, \; \tau \bar{\tau}, \; \mu^+ \mu^-$, $\nu_e \bar{\nu}_e$, $\nu_\mu \bar{\nu}_\mu$, $\nu_\tau \bar{\nu}_\tau$. The lowest and highest tested masses are chosen so that the median of the expected signal rate remains within 95\% containment of MC sample energy distribution. This criterion ensures the reliability of the signal PDFs by avoiding regions with low MC statistics.

%%%%%%%%%%%%%%%%%%%%%%%%%%%%%%%%%%%%%%%%%%%%%%%%%%%%%%%%%%%%%%%%%
\section{Systematic Uncertainties}\label{sec:systematics}
% Effects of systematic uncertainties
Since the background PDF is derived using a data-driven method, the systematic effects are accounted by the scrambling of data. Nevertheless, sources of systematic uncertainties can still have a notable impact on the signal PDFs as they are built from the MC simulation.

To evaluate the effects of each source of systematics, we first compute the signal PDFs using dedicated MC simulations where the systematic parameters are varied within their uncertainties.  The fit results from these varied cases are then compared to the nominal scenario, where all systematics are set to their default values. In the nominal case, statistical fluctuations are also accounted for by performing the fit on multiple trials generated as Poissonian fluctuations of the null hypothesis. This approach allows us to quantify the effect of systematic uncertainties in relation to statistical fluctuations. Specifically, the $1\sigma$ statistical fluctuations correspond to variations in the fitted sensitivity between about $0.6$ and $2$ times the median expectation, while the $2\sigma$ fluctuations range from about $0.3$ up to a factor of $3$.

The first set of systematics studied in this work is the detector-related uncertainties. They include the quantum efficiency of the DOMs, the ice properties, and the Deep Inelastic Scattering (DIS) cross-section which is the dominant neutrino interaction in IceCube. The impact of these uncertainties on the sensitivity can range from a few percent up to a maximum of $50\%$ depending on the specific DM scenario and systematic variation considered. However, they remain smaller than the statistical-only uncertainty and are well within its $1~\sigma$ range.

Another source of systematic uncertainty considered is the diffuse astrophysical neutrino emission from the Galactic Plane (GP), recently discovered by the IceCube Collaboration~\cite{IceCube:2023ame}, which acts as an additional background in DM searches targeting the Galactic Center. Unlike atmospheric and extragalactic backgrounds, the GP diffuse flux is not homogeneous in RA and therefore cannot be accounted for by the RA scrambling method in the background PDF estimation. To evaluate the effects of GP astrophysical flux, the hypothesis PDF in equation \eqref{eq:fraction-with-signalsub} can be modified as follows:
\begin{equation}
    f(i; \xi) = \xi \mathcal{S}_i + \mathcal{B}_i^\mathrm{scr} - \xi \mathcal{S}_i^\mathrm{scr} + g \mathcal{G}_i - g \mathcal{G}_i^\mathrm{scr},
    \label{eq:fraction-withGP}
\end{equation}
where the GP expectation is now included with the fraction $g$ and the PDF $\mathcal{G}$. Since the RA-scrambled data can contain the GP emission, the subtraction of the RA-scrambled GP events is also accounted via the RA-scrambled PDF $\mathcal{G}^\mathrm{scr}$. The total contribution of the DM signal and the GP component is restricted to $0 \leq \xi + g \leq 1$, ensuring a valid interpretation of the event fractions.

Given the large uncertainties on the prediction of GP neutrino emission and the lack of measurements of this flux in the energy range of this analysis, various model variations are tested. For the PDFs, we use two different templates: the $\pi^0$ model and the KRA-$\gamma$ model as those used in Ref.~\cite{IceCube:2023ame}. Both templates account for the diffuse $\gamma$-ray emission measured by the Fermi-LAT telescope~\cite{Fermi-LAT:2012edv}.
The $\pi^0$ template is constructed assuming that the diffuse $\gamma$-ray emission is dominated by the $\pi_0$ decay from cosmic-ray interactions with the interstellar medium. It adopts a spatially uniform spectral index of $\pi_0$ component across the Galactic disk, with a single power-law behavior of $E^{-2.7}$ in photon energy.
In contrast, the KRA-$\gamma$ model~\cite{Gaggero:2014xla,Gaggero:2015xza} is based on a cosmic-ray transport framework with spatially dependent diffusion coefficients and Galactic winds. Within this framework, the model parameters are adjusted to be consistent with the observed diffuse $\gamma$-ray from the Milky Way. As a consequence, the predicted diffuse neutrino spectrum is spatially varying, with a harder average spectral index of approximately $E^{-2.5}$.

The event fraction $g$ is determined through the normalization of the GP emission. Different treatments of this parameter are considered: setting it to the model prediction or to the IceCube measurement corresponding to each template (extrapolated down to GeV energies under the same power-law assumption), as discussed in Section~\ref{sec:Icecube}; in addition, we adopt an alternative treatment where $g$ is left as a free nuisance parameter. Furthermore, we have tested pseudo-data with injected GP models under both the template assumptions in order to test a possible mismatch in the signal modeling of the GP emission. 

Different combinations of GP injection and hypothesis assumptions result in sensitivities that deviate by up to 30\%, which is significantly smaller than the statistical fluctuations. This can be expected due to the very small fraction of GP emission compared to the atmospheric background at GeV energies, making the analysis insensitive to its contribution. In addition, the expected GP emissions are also orders of magnitude lower than the DM signals at the $90\%$ confidence level (C.L.) sensitivity (see Appendix~\ref{Appendix A} for more details). Therefore, in the results presented in the next section, the GP component is not considered as part of the hypothesis.

In summary, all the tested systematic uncertainties have a minor impact compared to statistical fluctuations. The dominant source of uncertainty remains the astrophysical profile of the DM halo, since the DM density in the Galactic Center could vary by up to three orders of magnitude depending on the assumed distribution \cite{Nesti:2013uwa, IceCube:2023ies}. The effect is expected to be more pronounced for annihilation signals, where the J-factor scales with $\rho^2$, than for decay signals, where the D-factor is only proportional to $\rho$. As a result, we follow the approach used in previous work \cite{IceCube:2023ies}, where all detector uncertainties are fixed to their nominal values, and the results are presented for the two halo profiles, NFW and Burkert, as these two models can be considered the two extreme scenarios of DM signal flux.

%%%%%%%%%%%%%%%%%%%%%%%%%%%%%%%%%%%%%%%%%%%%%%%%%%%%%%%%%%%
\section{Results}\label{sec:results}
A likelihood maximization is performed for each DM scenario and the corresponding test statistic, $q$, is computed. The value of $q$ is converted into a significance (z-score) assuming Wilk's theorem with a $\chi^2$ distribution of one degree of freedom \cite{Wilks:1938dza, 10.1214/aoms/1177728725}. This assumption is validated by using $10\,000$ background-only pseudo-samples as RA-scrambled data samples. Across all tested DM scenarios, no significant excess above $3 \sigma$ is observed. 

In the absence of significant excess, the upper limit on the thermally-averaged cross-section and the lower limit on the DM lifetime are presented for both NFW and Burkert halo profiles. These limits are derived by using the likelihood interval method \cite{Cowan:2010js}, following the same approach of the previous IceCube DM search \cite{IceCube:2023ies}, with the \textit{test statistic for upper limits} on $\xi$ defined as:
\begin{equation}
    q_{\xi}  = \left\{
        \begin{array}{ll}
            -2 \ln \dfrac{\mathcal{L}(\xi)}{\mathcal{L}(\hat{\xi})}  & \mbox{if } \hat{\xi} \leq \xi \;,\\
            0 & \mbox{otherwise.}
        \end{array}
 \right. 
 \label{eq:limit_ts}
\end{equation}
The $90\%$ C.L. upper limit on the signal fraction $\xi_{90}$ is defined as the value of $\xi$ that yields a $10 \%$ significance under the repeated trials of the hypothesis $\mathcal{H}(\xi_{90})$ \cite{Cowan:1998ji, Cowan:2010js}. After validating that $q_{\xi_{90}}$ asymptotically follows a $\chi^2$ distribution with one degree of freedom normalized to $1/2$ of the samples, consistent with a generalization \cite{10.1214/aoms/1177728725} of Wilk's theorem \cite{Wilks:1938dza}, $q_{\xi} = 1.64$ is chosen for the computation of $\xi_{90}$. 

In the following subsections, we first discuss the limits on DM thermally-averaged cross-section and lifetime, given the absence of a significant excess. Then we also illustrate the best-fit hypothesis and its compatibility with the data.

\subsection{Limits}
\begin{figure*}[tb!]
    % \centering
    \begin{minipage}{0.48\textwidth}
        \includegraphics[width=\linewidth]{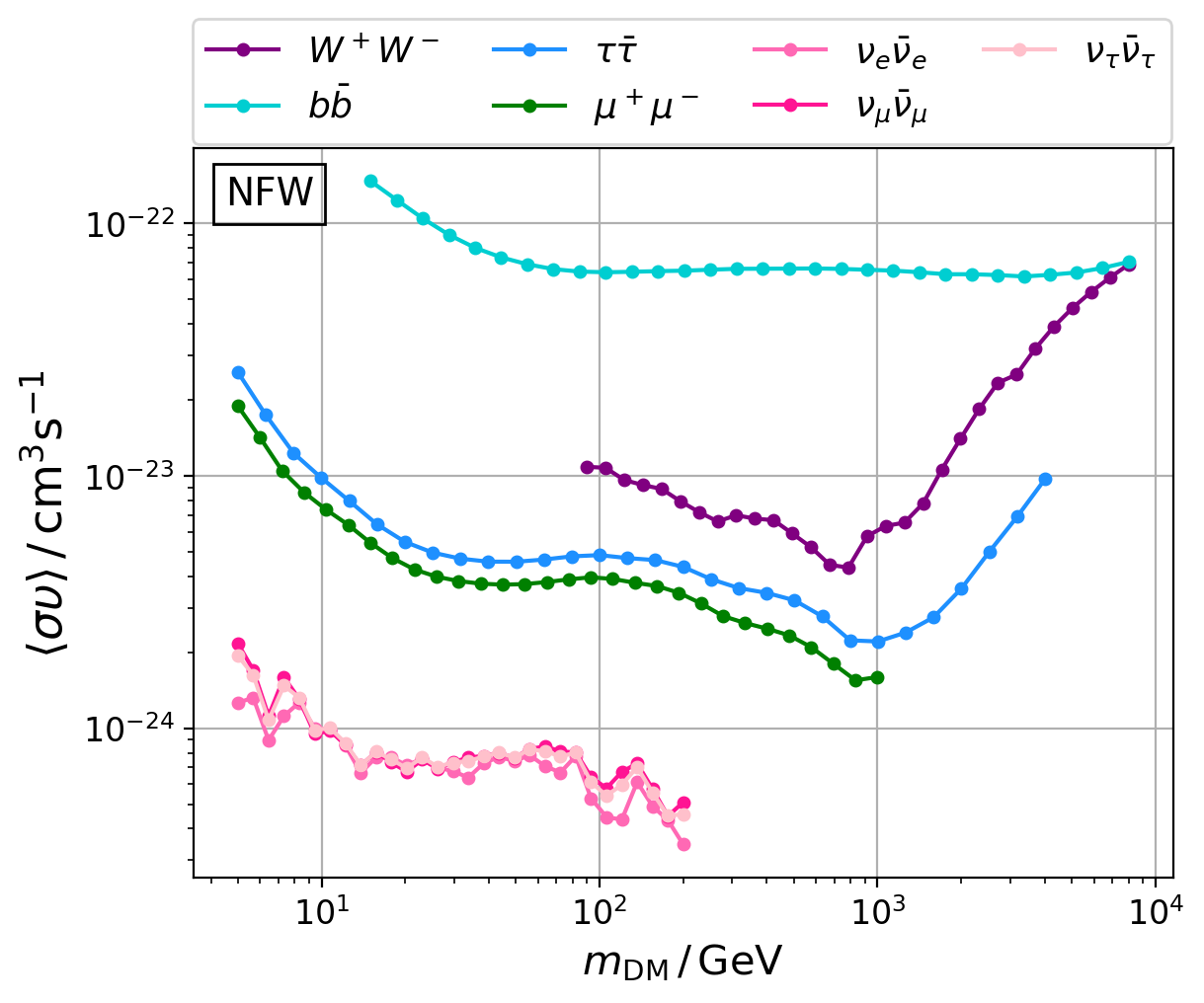}
    \end{minipage}
    \hfill
    \begin{minipage}{0.48\textwidth}
        \includegraphics[width=\linewidth]{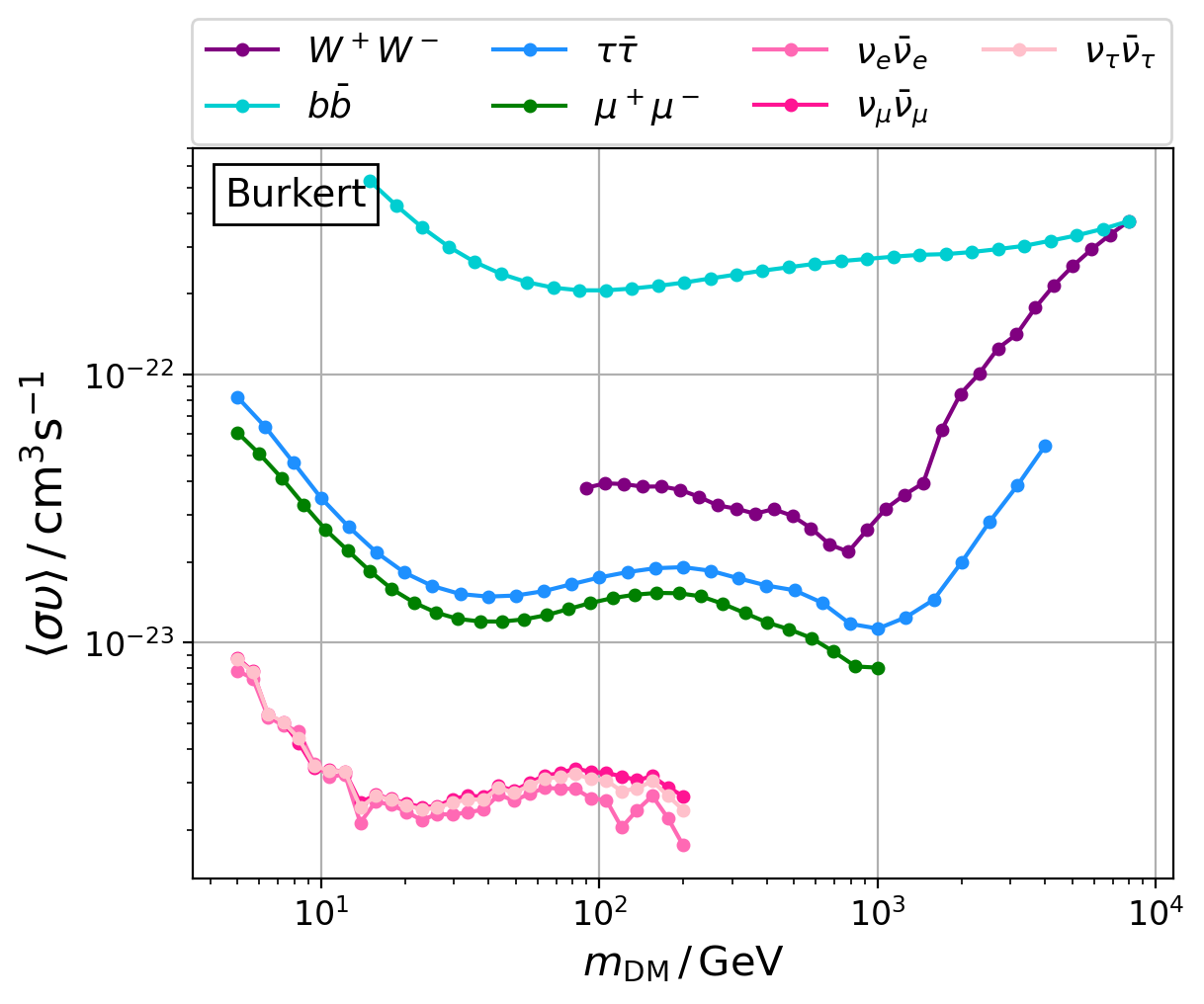}
    \end{minipage}

    \begin{minipage}{0.48\textwidth}
        \includegraphics[width=\linewidth]{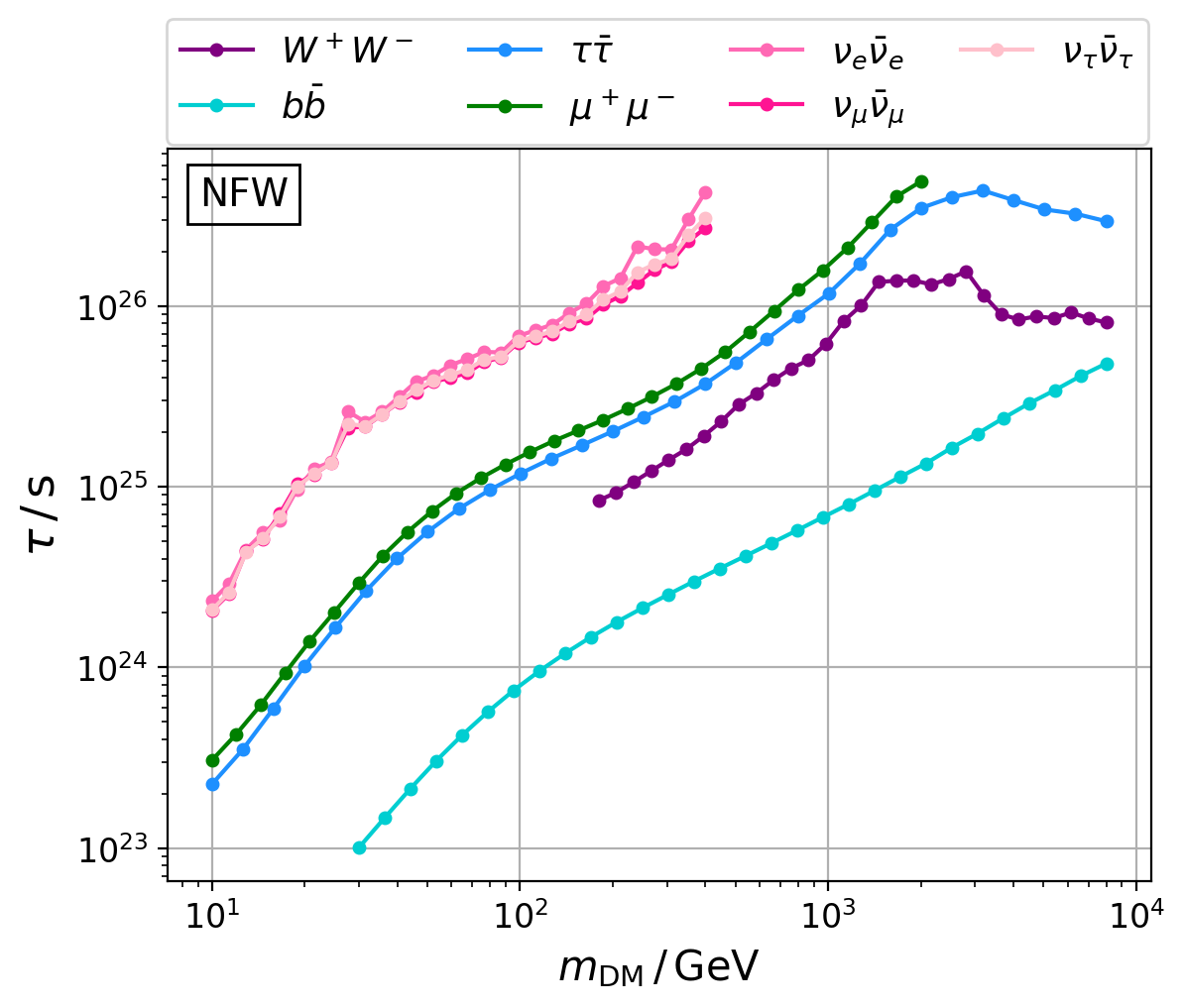}
    \end{minipage}
    \hfill
    \begin{minipage}{0.48\textwidth}
        \includegraphics[width=\linewidth]{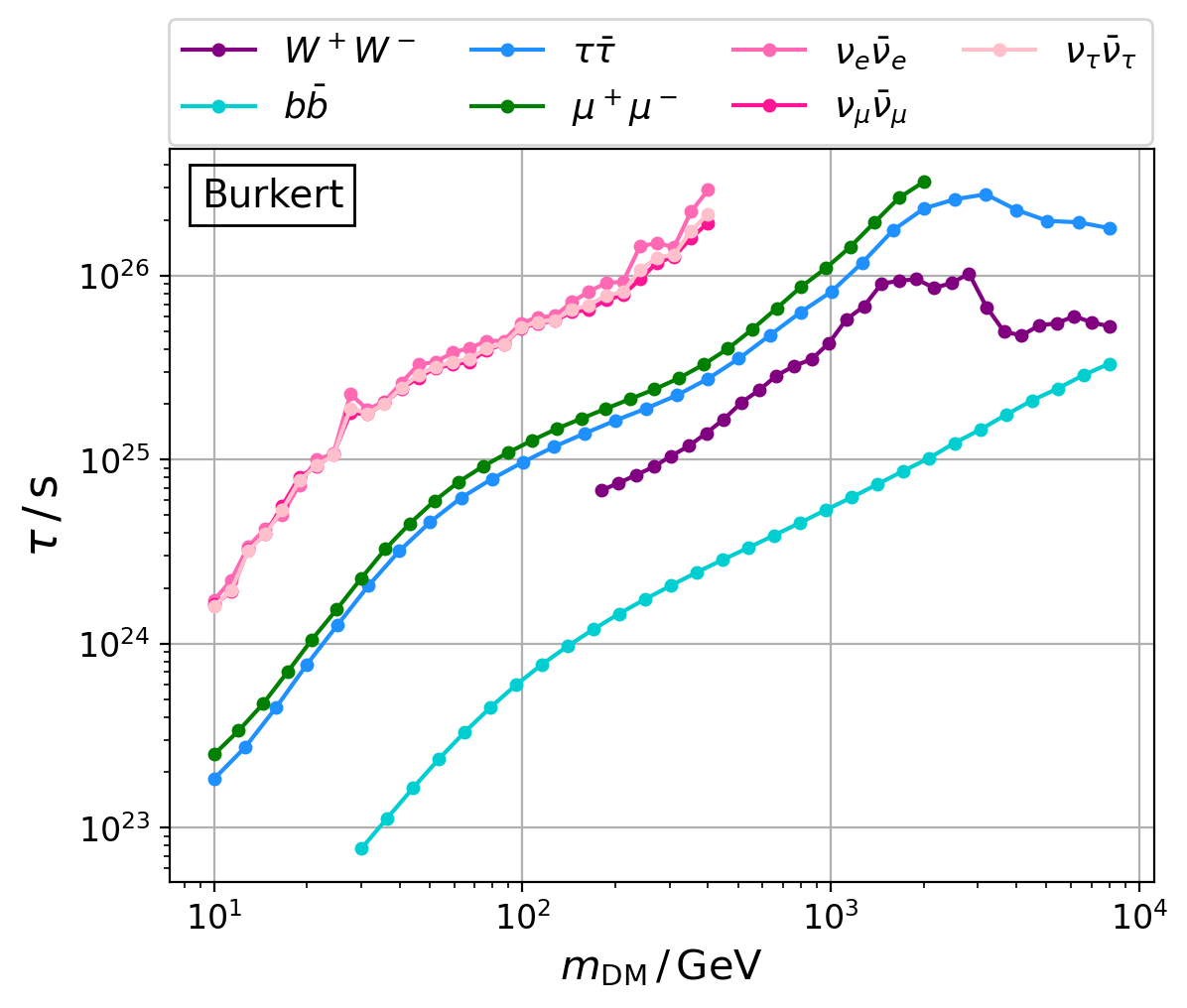}
    \end{minipage}

\caption{$90\%$ C.L. upper limits on the thermally-averaged self-annihilation cross-section $\langle \sigma \upsilon \rangle$ \textit{(upper)} and lower limit on lifetime $\tau$ \textit{(lower)} of DM as a function of its mass for both the NFW \textit{(left)} and Burkert \textit{(right)} halo profiles. The dots represent the mass values at which the limits are evaluated.}
\label{fig:90CL_limits}
\end{figure*}

Fig.~\ref{fig:90CL_limits} presents the 90\% C.L. upper limits on the thermally-averaged self-annihilation cross-section and the lower limits on the DM lifetime as functions of DM mass. The results are shown for all considered annihilation and decay channels, assuming both the NFW and Burkert halo profiles. The derived limits reach the level of $10^{-24} ~\mathrm{cm}^3 \mathrm{s}^{-1}$ for the DM self-annihilation cross-section while setting limits on the decay lifetime as high as $10^{26}~\mathrm{s}$. Among the tested channels, the neutrino lines provide the strongest constraints. As previously mentioned, these channels produce monochromatic peaks in the energy spectrum, making them particularly distinguishable from astrophysical backgrounds and thus enhance the sensitivity of searches with neutrino telescopes like IceCube.

These limits, alongside with the best-fit number of signal events, are reported in tables in the appendix. Additionally, the significance of each tested scenario, expressed as a z-score, is also included in these tables.

\begin{figure*}[tb!]
    % \centering
    \begin{minipage}{0.48\textwidth}
        \includegraphics[width=\linewidth]{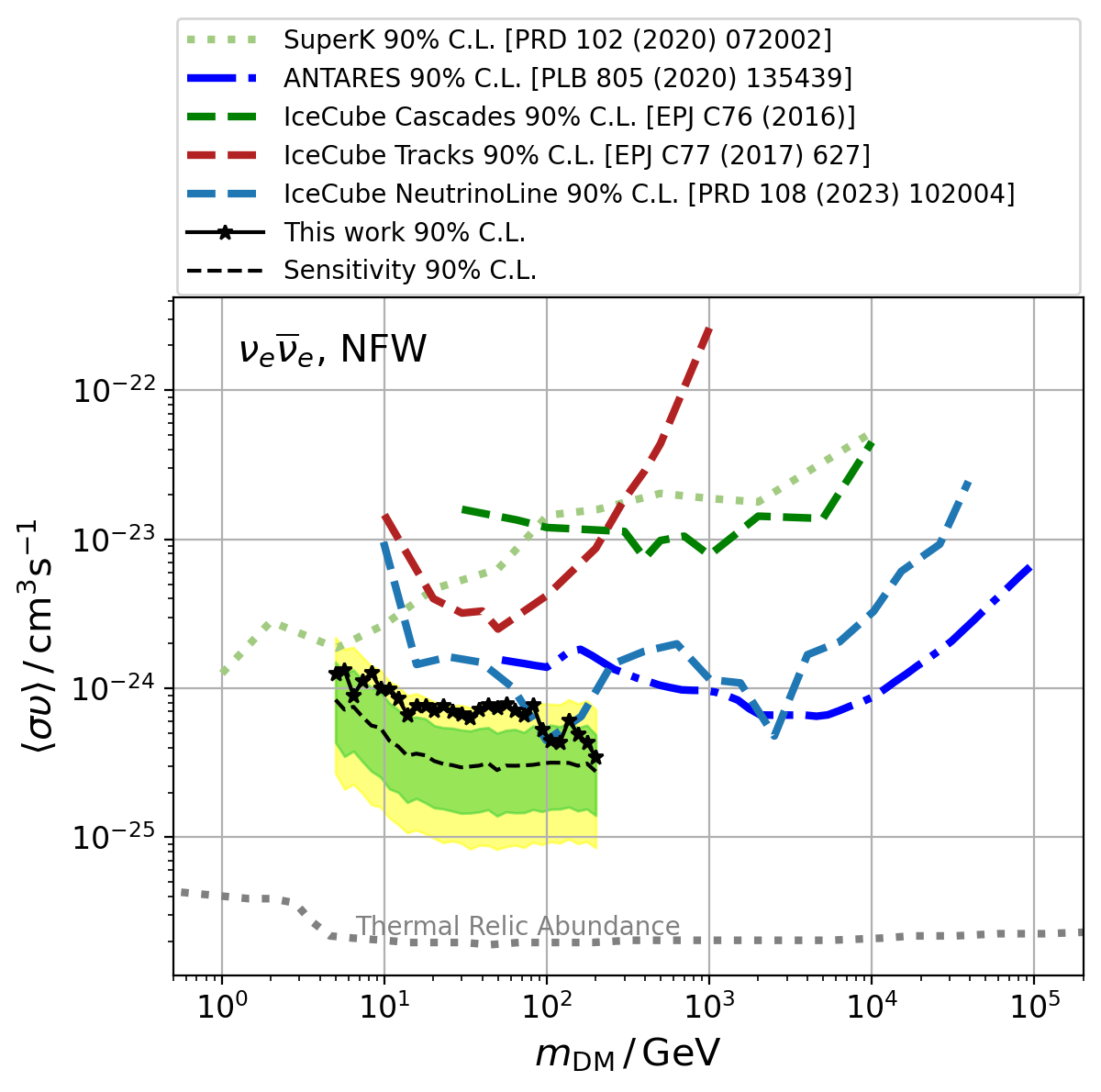}
    \end{minipage}
    \hfill
    \begin{minipage}{0.48\textwidth}
        \includegraphics[width=\linewidth]{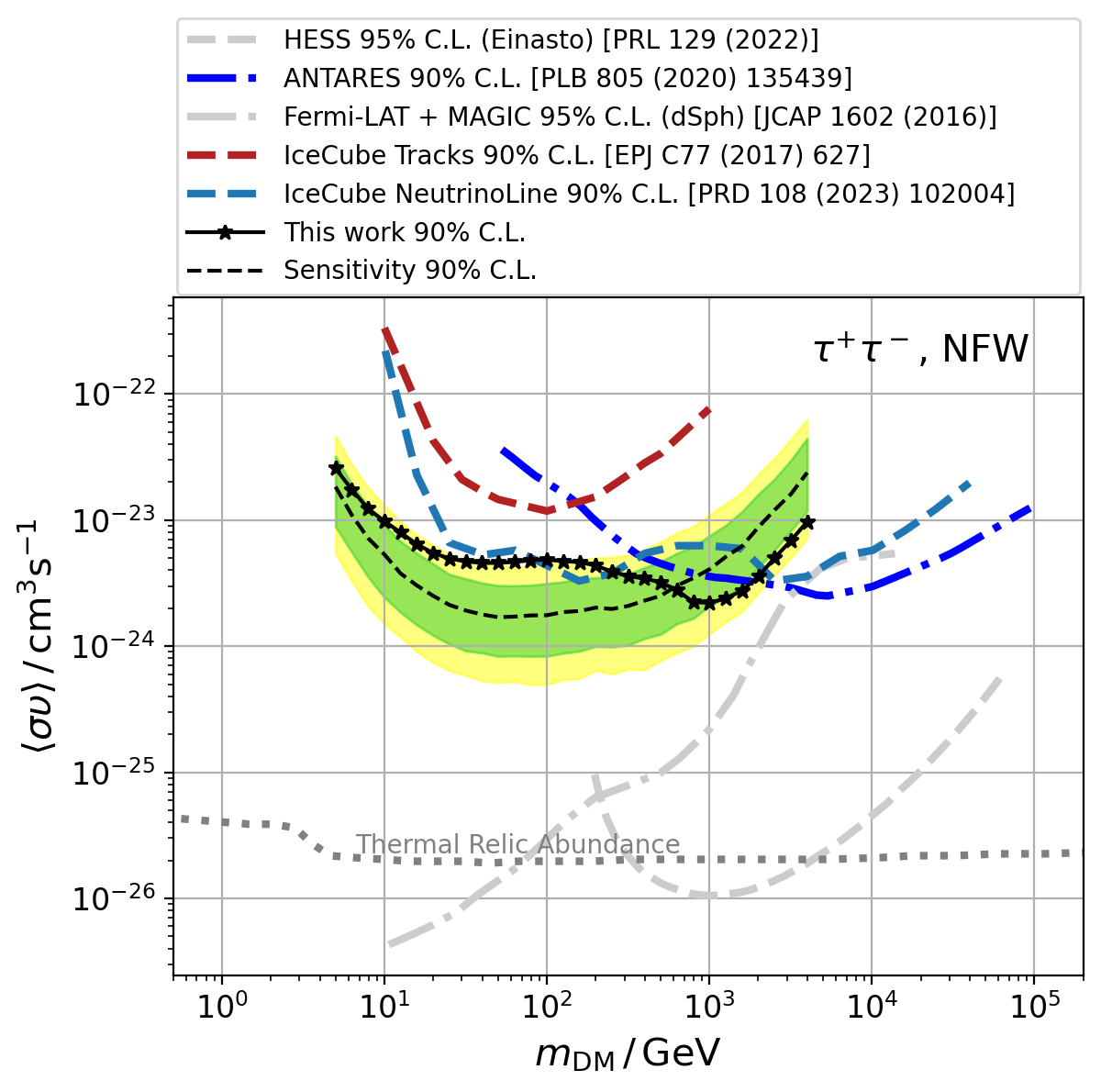}
    \end{minipage}
\caption{Sensitivity and upper limits on the thermally-averaged cross-section for the $\nu_e \overline{\nu}_e$ channel \textit{(left)} and $\tau \bar{\tau}$ channel \textit{(right)} compared to previous IceCube results \cite{IceCube:2016oqp,IceCube:2017rdn,IceCube:2023ies}, as well as Super-Kamiokande \cite{Super-Kamiokande:2020sgt}, ANTARES \cite{ANTARES:2019svn}, and the gamma-ray telescopes HESS \cite{HESS:2022ygk}, Fermi-LAT and MAGIC \cite{MAGIC:2016xys}. The dashed-black line shows the median sensitivity while the green and yellow represent 1 and 2$~\sigma$ containment bands of the statistical fluctuation assuming background-only hypothesis. The dotted grey line is the cross-section required to produce the relic abundance from thermal freeze-out computed in \cite{Steigman:2012nb}.}
    \label{fig:Limits_Annihilation_Comparison}
\end{figure*}

In Fig.~\ref{fig:Limits_Annihilation_Comparison}, we compare the limits on the DM annihilation cross-section obtained in this work with previous IceCube analyses, results from other experiments, and the theoretical prediction from a DM freeze-out model presented in Ref.~\cite{Steigman:2012nb}. The median sensitivity under the background-only hypothesis is also shown, along with the corresponding 1 and 2~$\sigma$ statistical fluctuations. For the $\nu_e \bar{\nu}_e$ channel, a mild excess in the data leads to a positive fluctuation above $1~\sigma$ across DM masses of 10–100 GeV. This same excess also causes a positive fluctuation observed in the $\tau \bar{\tau}$ channel, at slightly higher masses of 20–200 GeV due to the broader power-law energy spectrum, in contrast to the sharp monochromatic peaks of the neutrino line cases. 

\begin{figure*}[tb!]
    \begin{minipage}{0.48\textwidth}
        \includegraphics[width=\linewidth]{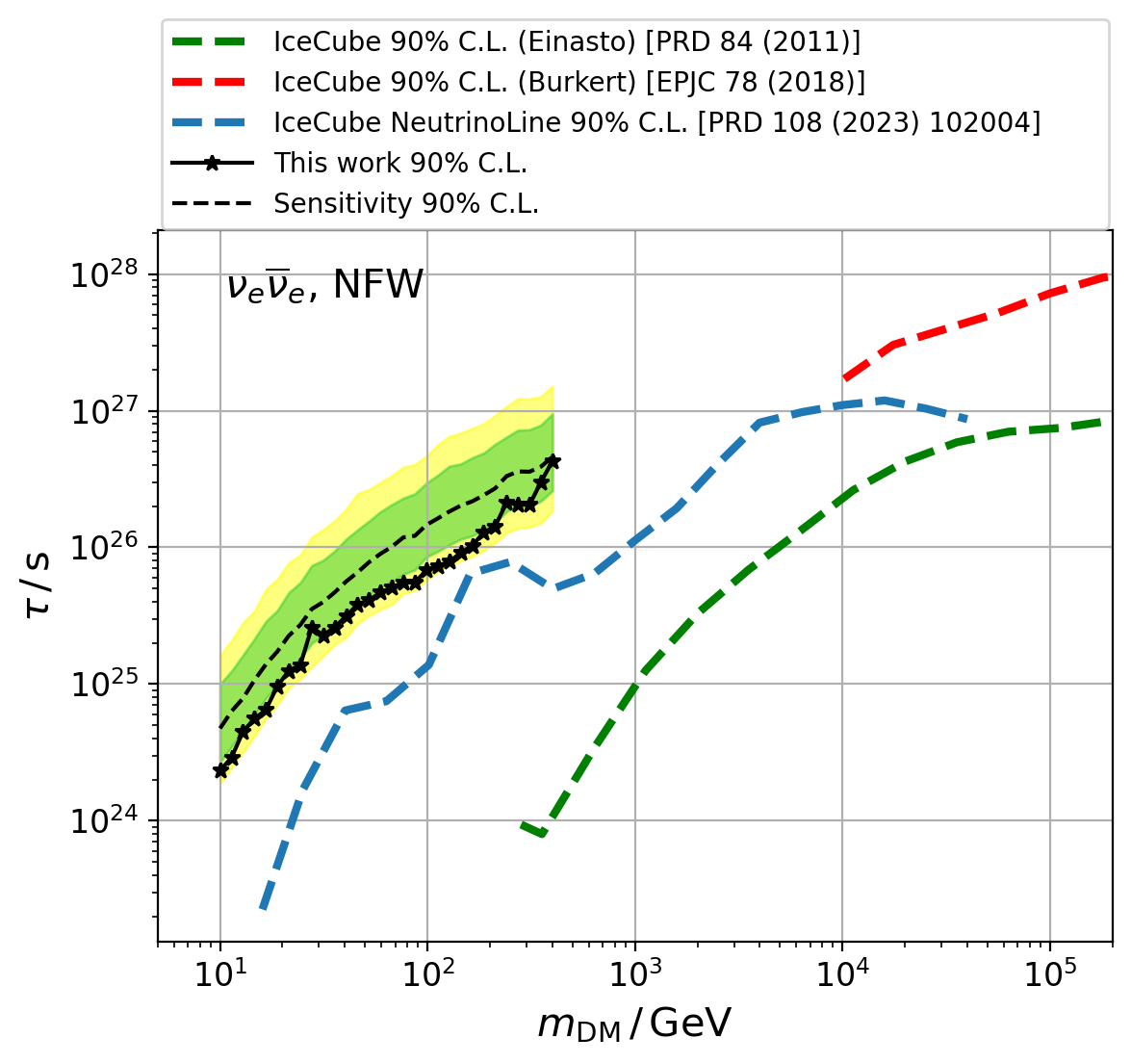}
    \end{minipage}
    \hfill
    \begin{minipage}{0.48\textwidth}
        \includegraphics[width=\linewidth]{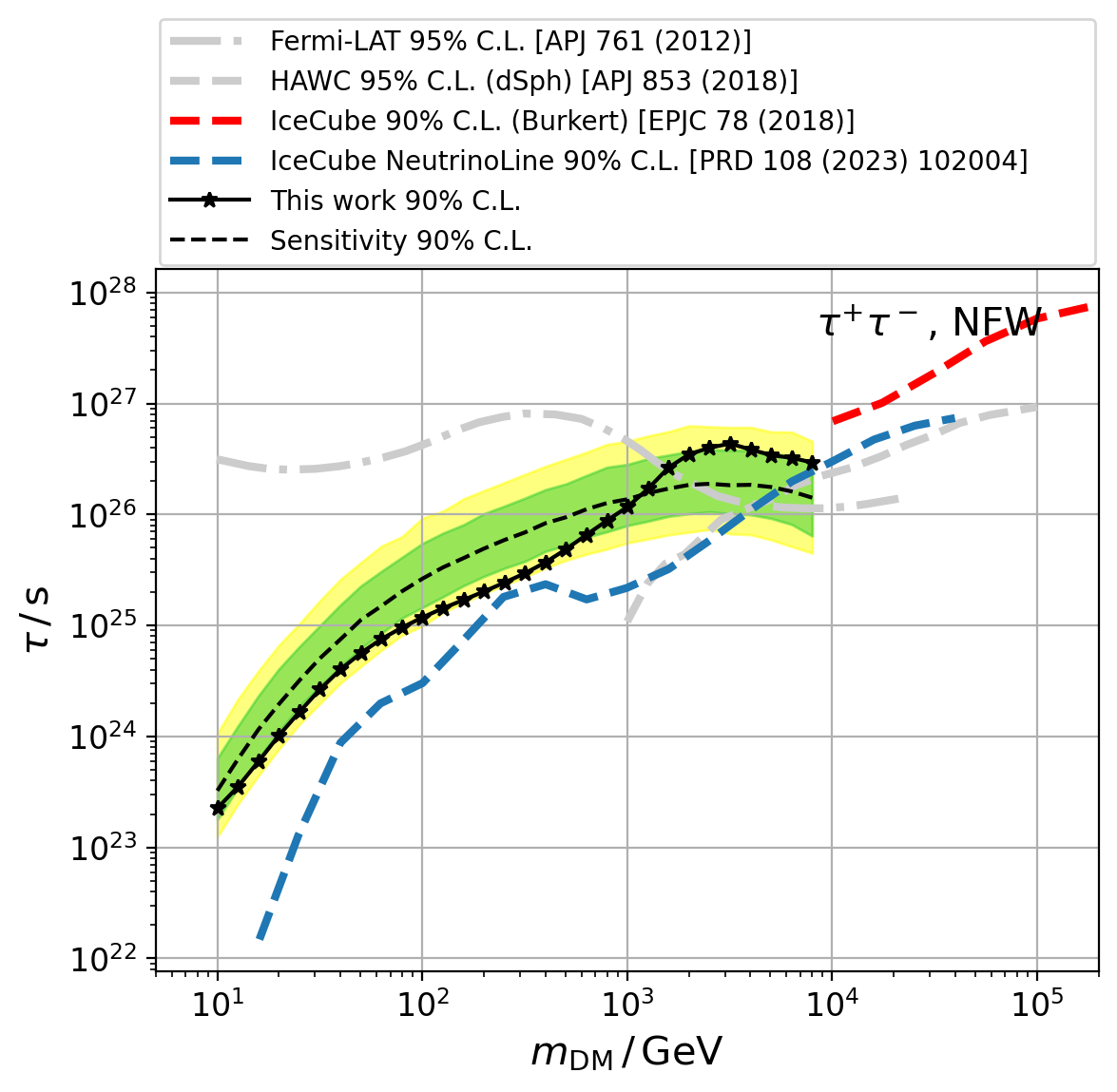}
    \end{minipage}

\caption{Sensitivity and lower limits on the DM lifetime for the $\nu_e \overline{\nu}_e$ channel (left) and $\tau \bar{\tau}$ channel (right) compared to previous IceCube results \cite{IceCube:2011kcp, IceCube:2018tkk, IceCube:2023ies}, and the gamma-ray telescopes HAWC \cite{HAWC:2017mfa} and Fermi-LAT \cite{Ackermann_2012}.   The dashed black line shows the median sensitivity while the green and yellow represent 1 and 2$~\sigma$ containment bands of the statistical fluctuation assuming background-only hypothesis.}
\label{fig:Limits_Decay_Comparison}
\label{fig:limits}
\end{figure*}

As shown in the figure, this analysis demonstrates an improvement over previous IceCube results. At a DM mass of around 10 GeV, the limits improve by an order of magnitude compared to the most recent IceCube study in Ref.~\cite{IceCube:2023ies}, which focused specifically on neutrino line signals from the Galactic Center. A similar comparison for the DM decay scenario is also shown in Fig.~\ref{fig:Limits_Decay_Comparison}. Overall, this work provides the strongest constraints among neutrino telescopes and establishes current world-leading limits on the neutrino line signal for DM masses ranging from several tens of GeV up to 100 GeV.

It is important to recall that the limits presented here adopt the commonly used model-independent framework in indirect DM searches with certain underlying assumptions. In particular, the limits are derived for specific halo profiles (NFW and Burkert), assuming $100\%$ branching into a single two-body annihilation or decay channel. More complex scenarios, such as three-body or $n$-body final states, or models with mixed branching ratios, are not considered in this work, but in some cases, constraints on these models could be obtained by properly recasting the model-independent building blocks provided here.

The studied benchmark cases also span a wide range of neutrino spectra and spatial features. This ensures that a large class of possible DM signals, even beyond the scenarios considered, would leave an imprint within the sensitivity of the search. Examples include mechanisms in which the signal originates not from halo annihilation or decay, but from alternative processes such as the annihilation of captured DM in the compact objects around the Galactic Center like neutron stars \cite{Bose:2021yhz, Nguyen:2022zwb, Acevedo:2024ttq}.

\subsection{Best-fit result}
The best-fit scenario corresponds to the annihilation of DM with mass of $201.6$ GeV into a pair of $b \bar{b}$ assuming NFW density profile of the Galactic halo. The local significance is found to be $2.47~\sigma$. To compute the post-trial correction that accounts for the look-elsewhere effect~\cite{Gross:2010qma} of testing different DM hypotheses (channel, mass, and halo profile combinations) defined in Section \ref{sec:analysis_method}, the analysis is repeated on $10\,000$ trials, achieved by scrambling the RA coordinates of the data. From these trials, the distribution of the most significant test statistics $q$ can be achieved. This allows the determination of the post-trial significance to be $1.08~\sigma$.

\begin{figure*}
    % \centering
    \begin{minipage}{0.48\textwidth}
        \includegraphics[width=\linewidth]{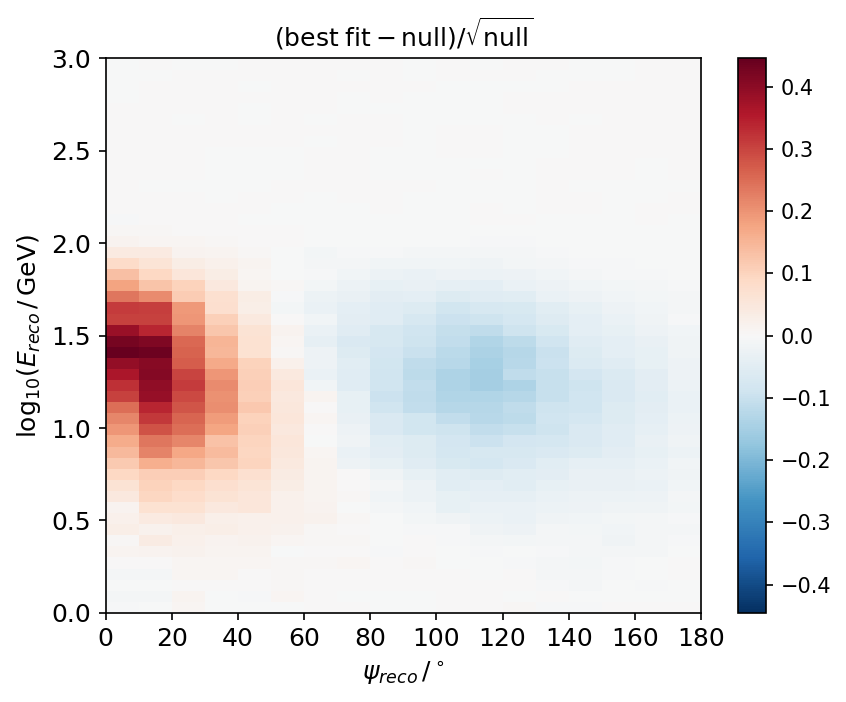}
    \end{minipage}
    \hfill
    \begin{minipage}{0.48\textwidth}
        \includegraphics[width=\linewidth]{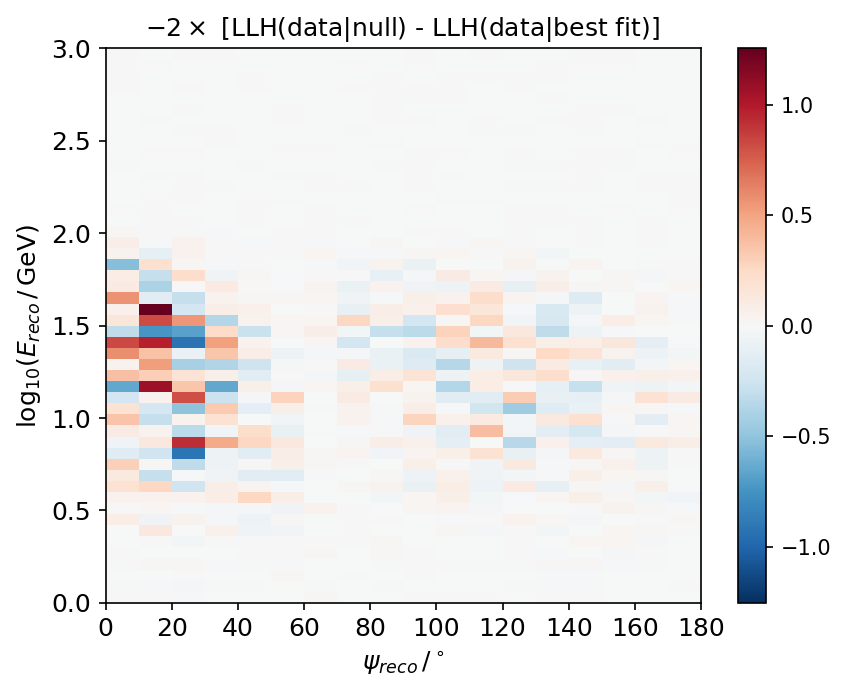}
    \end{minipage}
\caption{Pull distribution between null hypothesis and best-fit signal \textit{(left)} and the bin-wise test statistic distribution at best-fit signal \textit{(right)}, both shown as a function of reconstructed opening angle and reconstructed energy.}
    \label{fig:TSmap}
\end{figure*}

Fig.~\ref{fig:TSmap} illustrates the compatibility of the data with both the null and the best-fit signal hypotheses through a two-dimensional binned distribution of the observables: the reconstructed energy and opening angle from the Galactic Center. The left panel shows the pull distribution between the expectation of the null hypothesis $\mu^0$ and the best-fit signal $\hat{\mu}$, defined as: $(\hat{\mu}_i - \mu^0_i)/\sqrt{\mu^0_i}$ in each bin $i$. A pronounced discrepancy near the Galactic Center is attributed to the signal PDF, with a mild contribution at larger angular distances resulting from the RA-scrambled signal PDF. The right panel shows the distribution of test statistics, as defined in equation \eqref{eq:discovery_ts}, indicating the contribution of each bin in the observable space. The features observed in the test statistic map align with the expected pull distribution.

\begin{figure*}
    % \centering
    \begin{minipage}{0.48\textwidth}
        \includegraphics[width=\linewidth]{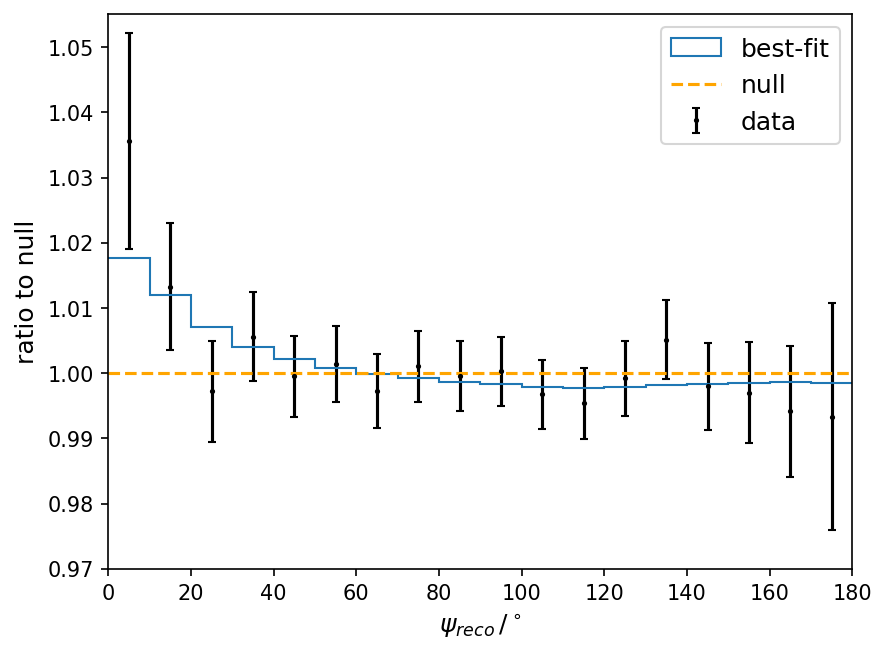}
    \end{minipage}
    \hfill
    \begin{minipage}{0.48\textwidth}
        \includegraphics[width=\linewidth]{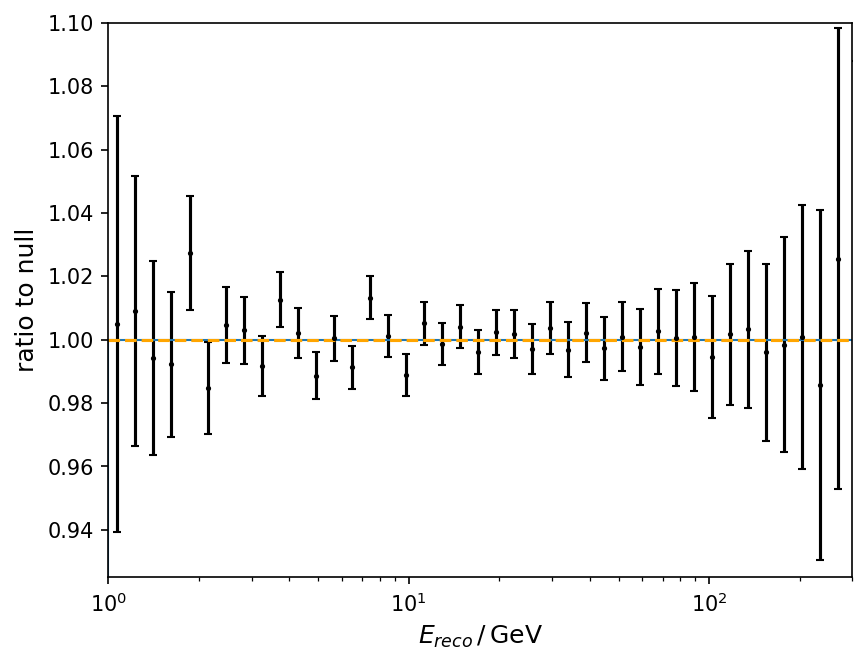}
    \end{minipage}
\caption{The ratio of the observed event distribution and best-fit hypothesis to the null hypothesis as a function of reconstructed opening angle \textit{(left)} and reconstructed energy \textit{(right)}. Error bars represent the statistical uncertainties of the data points.}
    \label{fig:MCdata_ratio}
\end{figure*}

The ratio of data and the best-fit hypothesis to the null hypothesis is shown in Fig.~\ref{fig:MCdata_ratio}, projected into angular distance (left panel) and energy (right panel). The figure demonstrates that the significance arises primarily from the angular distribution of the signal hypothesis rather than its energy distribution. This is because the background is constructed using RA-scrambled data, which dilutes potential energy-based signal features. However, incorporating energy remains crucial, as it plays an important role in improving the sensitivities especially in the case of neutrino lines, as indicated in Ref.~\cite{ElAisati:2015ugc, IceCube:2023ies}.

\begin{figure}[tb!]
 \includegraphics[width=\linewidth]{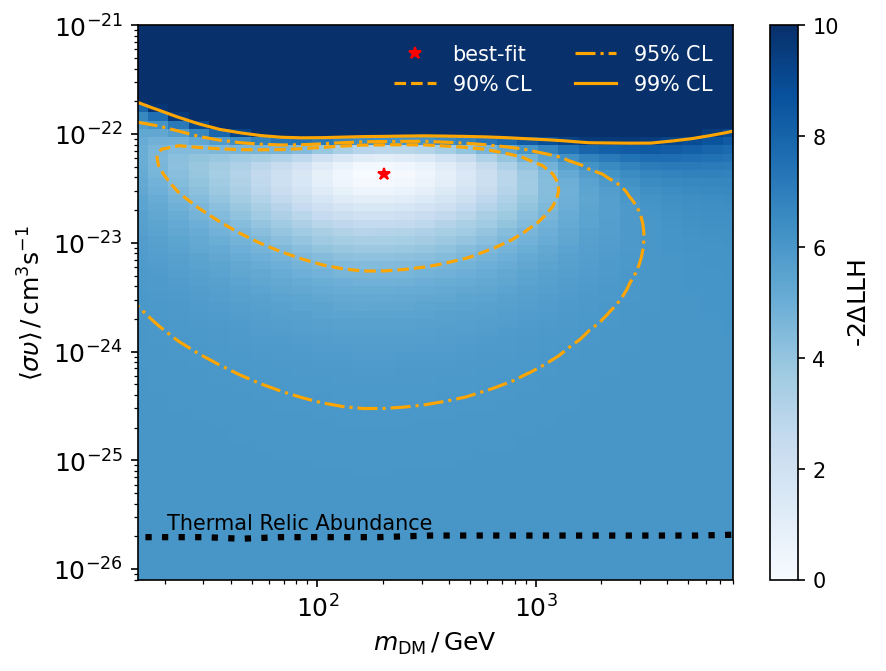}
\caption{$-2 \Delta \rm{LLH}$ scan for the best-fit DM scenario: annihilation of DM into $b \bar{b}$ with NFW profile. Contours are derived assuming Wilk's theorem~\cite{Wilks:1938dza}.}
    \label{fig:Contour}
\end{figure}

In Fig.~\ref{fig:Contour}, we present the log-likelihood difference ($\Delta \mathrm{LLH}$) relative to the best-fit point, scanned over two physical parameters of the DM: the DM mass $m_\mathrm{DM}$ and the thermally-averaged annihilation cross-section $\langle \sigma \upsilon \rangle$. The C.L. contours are derived assuming Wilk's theorem with two degrees of freedom \cite{Wilks:1938dza, Cowan:2010js}. It is important to emphasize that the $90\%$ C.L. upper limit obtained in the last section differs from and should not be confused with the C.L. contour depicted in Fig.~\ref{fig:Contour}. The upper limits establish the maximum allowed values of the signal fraction that remain consistent with the null hypothesis (background-only scenario) at 90\% C.L. On the other hand, the obtained contours represent regions of parameter space that are consistent with the observed data at a given confidence level and pertain to monitoring parameter space in the context of parameter estimation \cite{Cowan:1998ji, Cowan:2010js}.

\section{Conclusion}\label{sec:conclusion}
In this article, we present a search for neutrino signals from DM annihilation or decay in the Galactic Center. The analysis is performed using 9.28 years of IceCube-DeepCore data which targets GeV-scale DM masses. The analyzed data are consistent with the background-only hypothesis as no significant excess was found. Given the null result, we derive upper limits on the thermally-averaged cross-section for the DM self-annihilation as well as lower limits on the lifetime for DM decay. The result yields an order of magnitude improvement for the current IceCube limit at DM mass of $\sim$10 GeV. We put the most stringent limit among neutrino telescopes for DM masses from 10 GeV up to 100 GeV. In the same mass range, this work sets the current world-leading limits for the neutrino line channels.

The improvement in this analysis is due to not only more data taken but also the better understanding of the detector that resulted in the advancement in MC simulation, reconstruction, and event selection optimized for the detection of GeV neutrinos. Furthermore, the upcoming IceCube-Upgrade \cite{Ishihara:2019aao}, with new optical modules and improved calibration techniques, will enhance the detector’s sensitivity in the GeV regime. This will allow for even better constraints on DM properties in the GeV mass range.

%\section*{Acknowledgements}
\begin{acknowledgements}
The IceCube collaboration acknowledges the significant contributions to this manuscript from N. Chau. The authors gratefully acknowledge the support from the following agencies and institutions:
USA {\textendash} U.S. National Science Foundation-Office of Polar Programs,
U.S. National Science Foundation-Physics Division,
U.S. National Science Foundation-EPSCoR,
U.S. National Science Foundation-Office of Advanced Cyberinfrastructure,
Wisconsin Alumni Research Foundation,
Center for High Throughput Computing (CHTC) at the University of Wisconsin{\textendash}Madison,
Open Science Grid (OSG),
Partnership to Advance Throughput Computing (PATh),
Advanced Cyberinfrastructure Coordination Ecosystem: Services {\&} Support (ACCESS),
Frontera and Ranch computing project at the Texas Advanced Computing Center,
U.S. Department of Energy-National Energy Research Scientific Computing Center,
Particle astrophysics research computing center at the University of Maryland,
Institute for Cyber-Enabled Research at Michigan State University,
Astroparticle physics computational facility at Marquette University,
NVIDIA Corporation,
and Google Cloud Platform;
Belgium {\textendash} Funds for Scientific Research (FRS-FNRS and FWO),
FWO Odysseus and Big Science programmes,
and Belgian Federal Science Policy Office (Belspo);
Germany {\textendash} Bundesministerium f{\"u}r Bildung und Forschung (BMBF),
Deutsche Forschungsgemeinschaft (DFG),
Helmholtz Alliance for Astroparticle Physics (HAP),
Initiative and Networking Fund of the Helmholtz Association,
Deutsches Elektronen Synchrotron (DESY),
and High Performance Computing cluster of the RWTH Aachen;
Sweden {\textendash} Swedish Research Council,
Swedish Polar Research Secretariat,
Swedish National Infrastructure for Computing (SNIC),
and Knut and Alice Wallenberg Foundation;
European Union {\textendash} EGI Advanced Computing for research;
Australia {\textendash} Australian Research Council;
Canada {\textendash} Natural Sciences and Engineering Research Council of Canada,
Calcul Qu{\'e}bec, Compute Ontario, Canada Foundation for Innovation, WestGrid, and Digital Research Alliance of Canada;
Denmark {\textendash} Villum Fonden, Carlsberg Foundation, and European Commission;
New Zealand {\textendash} Marsden Fund;
Japan {\textendash} Japan Society for Promotion of Science (JSPS)
and Institute for Global Prominent Research (IGPR) of Chiba University;
Korea {\textendash} National Research Foundation of Korea (NRF);
Switzerland {\textendash} Swiss National Science Foundation (SNSF).
\end{acknowledgements}
% \clearpage
\bibliography{sample}

@article{Aartsen:2016nxy,
      author         = "Aartsen, M. G. and others",
      title          = "{The IceCube Neutrino Observatory: Instrumentation and
                        Online Systems}",
      collaboration  = "IceCube",
      journal        = "JINST",
      volume         = "12",
      year           = "2017",
      number         = "03",
      pages          = "P03012",
      doi            = "10.1088/1748-0221/12/03/P03012",
      eprint         = "1612.05093",
      archivePrefix  = "arXiv",
      primaryClass   = "astro-ph.IM",
      SLACcitation   = "%%CITATION = ARXIV:1612.05093;%%"
}

@article{Ishihara:2019aao,
    author = "Ishihara, Aya",
    collaboration = "IceCube",
    title = "{The IceCube Upgrade - Design and Science Goals}",
    eprint = "1908.09441",
    archivePrefix = "arXiv",
    primaryClass = "astro-ph.HE",
    reportNumber = "PoS-ICRC2019-1031",
    doi = "10.22323/1.358.1031",
    journal = "PoS",
    volume = "ICRC2019",
    pages = "1031",
    year = "2021"
}

@article{IceCube:2023ame,
    author = "Abbasi, R. and others",
    collaboration = "IceCube",
    title = "{Observation of high-energy neutrinos from the Galactic plane}",
    eprint = "2307.04427",
    archivePrefix = "arXiv",
    primaryClass = "astro-ph.HE",
    doi = "10.1126/science.adc9818",
    journal = "Science",
    volume = "380",
    number = "6652",
    pages = "adc9818",
    year = "2023"
}

@article{Fermi-LAT:2012edv,
    author = "Ackermann, M. and others",
    collaboration = "Fermi-LAT",
    title = "{Fermi-LAT Observations of the Diffuse Gamma-Ray Emission: Implications for Cosmic Rays and the Interstellar Medium}",
    eprint = "1202.4039",
    archivePrefix = "arXiv",
    primaryClass = "astro-ph.HE",
    doi = "10.1088/0004-637X/750/1/3",
    journal = "Astrophys. J.",
    volume = "750",
    pages = "3",
    year = "2012"
}

@article{Gaggero:2015xza,
    author = "Gaggero, Daniele and Grasso, Dario and Marinelli, Antonio and Urbano, Alfredo and Valli, Mauro",
    title = "{The gamma-ray and neutrino sky: A consistent picture of Fermi-LAT, Milagro, and IceCube results}",
    eprint = "1504.00227",
    archivePrefix = "arXiv",
    primaryClass = "astro-ph.HE",
    doi = "10.1088/2041-8205/815/2/L25",
    journal = "Astrophys. J. Lett.",
    volume = "815",
    number = "2",
    pages = "L25",
    year = "2015"
}

@article{Gaggero:2014xla,
    author = "Gaggero, Daniele and Urbano, Alfredo and Valli, Mauro and Ullio, Piero",
    title = "{Gamma-ray sky points to radial gradients in cosmic-ray transport}",
    eprint = "1411.7623",
    archivePrefix = "arXiv",
    primaryClass = "astro-ph.HE",
    doi = "10.1103/PhysRevD.91.083012",
    journal = "Phys. Rev. D",
    volume = "91",
    number = "8",
    pages = "083012",
    year = "2015"
}

@article{Abbasi:2024wpf,
    author = "Abbasi, R. and others",
    collaboration = "IceCube",
    title = "{Improved modeling of in-ice particle showers for IceCube event reconstruction}",
    eprint = "2403.02470",
    archivePrefix = "arXiv",
    primaryClass = "astro-ph.HE",
    doi = "10.1088/1748-0221/19/06/P06026",
    journal = "JINST",
    volume = "19",
    number = "06",
    pages = "P06026",
    year = "2024"
}

@article{IceCube:2020nwx,
    author = "Aartsen, M. G. and others",
    collaboration = "IceCube",
    title = "{In-situ calibration of the single-photoelectron charge response of the IceCube photomultiplier tubes}",
    eprint = "2002.00997",
    archivePrefix = "arXiv",
    primaryClass = "physics.ins-det",
    doi = "10.1088/1748-0221/15/06/P06032",
    journal = "JINST",
    volume = "15",
    number = "06",
    pages = "P06032",
    year = "2020"
}

@article{IceCube:2022kff,
    author = "Abbasi, R. and others",
    collaboration = "IceCube",
    title = "{Low energy event reconstruction in IceCube DeepCore}",
    eprint = "2203.02303",
    archivePrefix = "arXiv",
    primaryClass = "hep-ex",
    doi = "10.1140/epjc/s10052-022-10721-2",
    journal = "Eur. Phys. J. C",
    volume = "82",
    number = "9",
    pages = "807",
    year = "2022"
}

@article{IceCube:2008qbc,
    author = "Abbasi, R. and others",
    collaboration = "IceCube",
    title = "{The IceCube Data Acquisition System: Signal Capture, Digitization, and Timestamping}",
    eprint = "0810.4930",
    archivePrefix = "arXiv",
    primaryClass = "physics.ins-det",
    doi = "10.1016/j.nima.2009.01.001",
    journal = "Nucl. Instrum. Meth. A",
    volume = "601",
    pages = "294--316",
    year = "2009"
}

@article{Abbasi_2010,
   title="{Calibration and characterization of the IceCube photomultiplier tube}",
   author = "Abbasi, R. and others",
   volume={618},
   ISSN={0168-9002},
   url={http://dx.doi.org/10.1016/j.nima.2010.03.102},
    archivePrefix="arXiv",
    eprint="1002.2442v1",
   DOI={10.1016/j.nima.2010.03.102},
   number={1–3},
   journal={Nucl. Instrum. Meth. A},
   publisher={Elsevier BV},
   collaboration  = "IceCube",
   year={2010},
   month=jun, pages={139–152} }

@article{IceCube:2011ucd,
    author = "Abbasi, R. and others",
    collaboration = "IceCube",
    title = "{The Design and Performance of IceCube DeepCore}",
    eprint = "1109.6096",
    archivePrefix = "arXiv",
    primaryClass = "astro-ph.IM",
    doi = "10.1016/j.astropartphys.2012.01.004",
    journal = "Astropart. Phys.",
    volume = "35",
    pages = "615--624",
    year = "2012"
}

@article{IceCube:2023ewe,
    author = "Abbasi, R. and others",
    collaboration = "IceCube",
    title = "{Measurement of atmospheric neutrino mixing with improved IceCube DeepCore calibration and data processing}",
    eprint = "2304.12236",
    archivePrefix = "arXiv",
    primaryClass = "hep-ex",
    doi = "10.1103/PhysRevD.108.012014",
    journal = "Phys. Rev. D",
    volume = "108",
    number = "1",
    pages = "012014",
    year = "2023"
}

@article{IceCube:2024xjj,
    author = "Abbasi, R. and others",
    collaboration = "IceCube",
    title = "{Measurement of Atmospheric Neutrino Oscillation Parameters Using Convolutional Neural Networks with 9.3 Years of Data in IceCube DeepCore}",
    eprint = "2405.02163",
    archivePrefix = "arXiv",
    primaryClass = "hep-ex",
    doi = "10.1103/PhysRevLett.134.091801",
    journal = "Phys. Rev. Lett.",
    volume = "134",
    number = "9",
    pages = "091801",
    year = "2025"
}

@article{RevModPhys.90.045002,
    author = "Bertone, Gianfranco and Hooper, Dan",
    title = "{History of dark matter}",
    eprint = "1605.04909",
    archivePrefix = "arXiv",
    primaryClass = "astro-ph.CO",
    reportNumber = "FERMILAB-PUB-16-157-A",
    doi = "10.1103/RevModPhys.90.045002",
    journal = "Rev. Mod. Phys.",
    volume = "90",
    number = "4",
    pages = "045002",
    year = "2018"
}

@article{Cirelli:2024ssz,
    author = "Cirelli, Marco and Strumia, Alessandro and Zupan, Jure",
    title = "{Dark Matter}",
    eprint = "2406.01705",
    archivePrefix = "arXiv",
    primaryClass = "hep-ph",
    month = "6",
    year = "2024"
}

@article{Persic:1995ru,
    author = "Persic, Massimo and Salucci, Paolo and Stel, Fulvio",
    title = "{The Universal rotation curve of spiral galaxies: 1. The Dark matter connection}",
    eprint = "astro-ph/9506004",
    archivePrefix = "arXiv",
    reportNumber = "SISSA-60-95-A",
    doi = "10.1093/mnras/278.1.27",
    journal = "Mon. Not. Roy. Astron. Soc.",
    volume = "281",
    pages = "27",
    year = "1996"
}

@article{Rubin:1970zza,
    author = "Rubin, Vera C. and Ford, Jr., W. Kent",
    title = "{Rotation of the Andromeda Nebula from a Spectroscopic Survey of Emission Regions}",
    doi = "10.1086/150317",
    journal = "Astrophys. J.",
    volume = "159",
    pages = "379--403",
    year = "1970"
}

@article{Zwicky:1937zza,
    author = "Zwicky, F.",
    title = "{On the Masses of Nebulae and of Clusters of Nebulae}",
    doi = "10.1086/143864",
    journal = "Astrophys. J.",
    volume = "86",
    pages = "217--246",
    year = "1937"
}

@article{Jee:2007nx,
    author = "Jee, Myungkook James and others",
    title = "{Discovery of a Ringlike Dark Matter Structure in the Core of the Galaxy Cluster Cl 0024+17}",
    eprint = "0705.2171",
    archivePrefix = "arXiv",
    primaryClass = "astro-ph",
    doi = "10.1086/517498",
    journal = "Astrophys. J.",
    volume = "661",
    pages = "728--749",
    year = "2007"
}

@article{Jee:2008qj,
    author = "Jee, M. J. and Tyson, J. A.",
    title = "{Dark Matter in the Galaxy Cluster CL J1226+3332 at Z=0.89}",
    eprint = "0810.0709",
    archivePrefix = "arXiv",
    primaryClass = "astro-ph",
    doi = "10.1088/0004-637X/691/2/1337",
    journal = "Astrophys. J.",
    volume = "691",
    pages = "1337--1347",
    year = "2009"
}

@article{Planck:2013pxb,
    author = "Ade, P. A. R. and others",
    collaboration = "Planck",
    title = "{Planck 2013 results. XVI. Cosmological parameters}",
    eprint = "1303.5076",
    archivePrefix = "arXiv",
    primaryClass = "astro-ph.CO",
    reportNumber = "CERN-PH-TH-2013-129",
    doi = "10.1051/0004-6361/201321591",
    journal = "Astron. Astrophys.",
    volume = "571",
    pages = "A16",
    year = "2014"
}

@article{Bertone:2004pz,
    author = "Bertone, Gianfranco and Hooper, Dan and Silk, Joseph",
    title = "{Particle dark matter: Evidence, candidates and constraints}",
    eprint = "hep-ph/0404175",
    archivePrefix = "arXiv",
    reportNumber = "FERMILAB-PUB-04-047-A",
    doi = "10.1016/j.physrep.2004.08.031",
    journal = "Phys. Rept.",
    volume = "405",
    pages = "279--390",
    year = "2005"
}

@article{PerezdelosHeros:2020qyt,
    author = "P\'erez de los Heros, Carlos",
    title = "{Status, Challenges and Directions in Indirect Dark Matter Searches}",
    eprint = "2008.11561",
    archivePrefix = "arXiv",
    primaryClass = "astro-ph.HE",
    doi = "10.3390/sym12101648",
    journal = "Symmetry",
    volume = "12",
    number = "10",
    pages = "1648",
    year = "2020"
}

@article{Gaskins:2016cha,
    author = "Gaskins, Jennifer M.",
    title = "{A review of indirect searches for particle dark matter}",
    eprint = "1604.00014",
    archivePrefix = "arXiv",
    primaryClass = "astro-ph.HE",
    doi = "10.1080/00107514.2016.1175160",
    journal = "Contemp. Phys.",
    volume = "57",
    number = "4",
    pages = "496--525",
    year = "2016"
}

@article{Arcadi:2017kky,
    author = "Arcadi, Giorgio and Dutra, Ma\'\i{}ra and Ghosh, Pradipta and Lindner, Manfred and Mambrini, Yann and Pierre, Mathias and Profumo, Stefano and Queiroz, Farinaldo S.",
    title = "{The waning of the WIMP? A review of models, searches, and constraints}",
    eprint = "1703.07364",
    archivePrefix = "arXiv",
    primaryClass = "hep-ph",
    doi = "10.1140/epjc/s10052-018-5662-y",
    journal = "Eur. Phys. J. C",
    volume = "78",
    number = "3",
    pages = "203",
    year = "2018"
}

@article{ElAisati:2015ugc,
    author = {El Aisati, Cha\"\i{}mae and Gustafsson, Michael and Hambye, Thomas},
    title = "{New Search for Monochromatic Neutrinos from Dark Matter Decay}",
    eprint = "1506.02657",
    archivePrefix = "arXiv",
    primaryClass = "hep-ph",
    reportNumber = "ULB-TH-15-06",
    doi = "10.1103/PhysRevD.92.123515",
    journal = "Phys. Rev. D",
    volume = "92",
    number = "12",
    pages = "123515",
    year = "2015"
}

@article{Queiroz:2016sxf,
    author = "Queiroz, Farinaldo S. and Rodejohann, Werner and Yaguna, Carlos E.",
    title = "{Is the dark matter particle its own antiparticle?}",
    eprint = "1610.06581",
    archivePrefix = "arXiv",
    primaryClass = "hep-ph",
    doi = "10.1103/PhysRevD.95.095010",
    journal = "Phys. Rev. D",
    volume = "95",
    number = "9",
    pages = "095010",
    year = "2017"
}

@article{Fermi-LAT:2015sau,
    author = "Ajello, M. and others",
    collaboration = "Fermi-LAT",
    title = "{Fermi-LAT Observations of High-Energy $\gamma$-Ray Emission Toward the Galactic Center}",
    eprint = "1511.02938",
    archivePrefix = "arXiv",
    primaryClass = "astro-ph.HE",
    doi = "10.3847/0004-637X/819/1/44",
    journal = "Astrophys. J.",
    volume = "819",
    number = "1",
    pages = "44",
    year = "2016"
}

@article{Fermi-LAT:2017opo,
    author = "Ackermann, M. and others",
    collaboration = "Fermi-LAT",
    title = "{The Fermi Galactic Center GeV Excess and Implications for Dark Matter}",
    eprint = "1704.03910",
    archivePrefix = "arXiv",
    primaryClass = "astro-ph.HE",
    doi = "10.3847/1538-4357/aa6cab",
    journal = "Astrophys. J.",
    volume = "840",
    number = "1",
    pages = "43",
    year = "2017"
}

@article{Daylan:2014rsa,
    author = "Daylan, Tansu and Finkbeiner, Douglas P. and Hooper, Dan and Linden, Tim and Portillo, Stephen K. N. and Rodd, Nicholas L. and Slatyer, Tracy R.",
    title = "{The characterization of the gamma-ray signal from the central Milky Way: A case for annihilating dark matter}",
    eprint = "1402.6703",
    archivePrefix = "arXiv",
    primaryClass = "astro-ph.HE",
    reportNumber = "FERMILAB-PUB-14-032-A, MIT-CTP-4533",
    doi = "10.1016/j.dark.2015.12.005",
    journal = "Phys. Dark Univ.",
    volume = "12",
    pages = "1--23",
    year = "2016"
}

@article{Murgia:2020dzu,
    author = "Murgia, Simona",
    title = "{The Fermi\textendash{}LAT Galactic Center Excess: Evidence of Annihilating Dark Matter?}",
    doi = "10.1146/annurev-nucl-101916-123029",
    journal = "Ann. Rev. Nucl. Part. Sci.",
    volume = "70",
    pages = "455--483",
    year = "2020"
}

@article{Cholis:2021rpp,
    author = "Cholis, Ilias and Zhong, Yi-Ming and McDermott, Samuel D. and Surdutovich, Joseph P.",
    title = "{Return of the templates: Revisiting the Galactic Center excess with multimessenger observations}",
    eprint = "2112.09706",
    archivePrefix = "arXiv",
    primaryClass = "astro-ph.HE",
    reportNumber = "FERMILAB-PUB-21-709-T",
    doi = "10.1103/PhysRevD.105.103023",
    journal = "Phys. Rev. D",
    volume = "105",
    number = "10",
    pages = "103023",
    year = "2022"
}

@article{Caron:2022akb,
    author = "Caron, Sascha and Eckner, Christopher and Hendriks, Luc and J{\'o}hannesson, Gu{\dh}laugur and Ruiz de Austri, Roberto and Zaharijas, Gabrijela",
    title = "{Mind the gap: the discrepancy between simulation and reality drives interpretations of the Galactic Center Excess}",
    eprint = "2211.09796",
    archivePrefix = "arXiv",
    primaryClass = "astro-ph.HE",
    reportNumber = "LAPTH-068/22",
    doi = "10.1088/1475-7516/2023/06/013",
    journal = "JCAP",
    volume = "06",
    pages = "013",
    year = "2023"
}

@article{Holst:2024fvb,
    author = "Holst, Ian and Hooper, Dan",
    title = "{New determination of the millisecond pulsar gamma-ray luminosity function and implications for the Galactic Center gamma-ray excess}",
    eprint = "2403.00978",
    archivePrefix = "arXiv",
    primaryClass = "astro-ph.HE",
    reportNumber = "FERMILAB-PUB-23-583-T",
    doi = "10.1103/PhysRevD.111.023048",
    journal = "Phys. Rev. D",
    volume = "111",
    number = "2",
    pages = "023048",
    year = "2025"
}

@article{Hooper:2022bec,
    author = "Hooper, Dan",
    title = "{The status of the galactic center gamma-ray excess}",
    eprint = "2209.14370",
    archivePrefix = "arXiv",
    primaryClass = "astro-ph.HE",
    reportNumber = "FERMILAB-CONF-22-732-T",
    doi = "10.21468/SciPostPhysProc.12.006",
    journal = "SciPost Phys. Proc.",
    volume = "12",
    pages = "006",
    year = "2023"
}

@article{Beacom:2006tt,
    author = "Beacom, John F. and Bell, Nicole F. and Mack, Gregory D.",
    title = "{General Upper Bound on the Dark Matter Total Annihilation Cross Section}",
    eprint = "astro-ph/0608090",
    archivePrefix = "arXiv",
    reportNumber = "KRL-MAP-322",
    doi = "10.1103/PhysRevLett.99.231301",
    journal = "Phys. Rev. Lett.",
    volume = "99",
    pages = "231301",
    year = "2007"
}

@article{Yuksel:2007ac,
    author = "Yuksel, Hasan and Horiuchi, Shunsaku and Beacom, John F. and Ando, Shin'ichiro",
    title = "{Neutrino Constraints on the Dark Matter Total Annihilation Cross Section}",
    eprint = "0707.0196",
    archivePrefix = "arXiv",
    primaryClass = "astro-ph",
    doi = "10.1103/PhysRevD.76.123506",
    journal = "Phys. Rev. D",
    volume = "76",
    pages = "123506",
    year = "2007"
}

@article{Arguelles:2022nbl,
    author = {Arg{\"u}elles, Carlos A. and Delgado, Diyaselis and Friedlander, Avi and Kheirandish, Ali and Safa, Ibrahim and Vincent, Aaron C. and White, Henry},
    title = "{Dark matter decay to neutrinos}",
    eprint = "2210.01303",
    archivePrefix = "arXiv",
    primaryClass = "hep-ph",
    doi = "10.1103/PhysRevD.108.123021",
    journal = "Phys. Rev. D",
    volume = "108",
    number = "12",
    pages = "123021",
    year = "2023"
}

@article{Liu_2020,
    author = {Liu, Qinrui and Lazar, Jeffrey and Arg{\"u}elles, Carlos A. and Kheirandish, Ali},
    title = "{$\chi$aro$\nu$: a tool for neutrino flux generation from WIMPs}",
    eprint = "2007.15010",
    archivePrefix = "arXiv",
    primaryClass = "hep-ph",
    doi = "10.1088/1475-7516/2020/10/043",
    journal = "JCAP",
    volume = "10",
    pages = "043",
    year = "2020"
}

@article{Bauer:2020jay,
    author = "Bauer, Christian W. and Rodd, Nicholas L. and Webber, Bryan R.",
    title = "{Dark matter spectra from the electroweak to the Planck scale}",
    eprint = "2007.15001",
    archivePrefix = "arXiv",
    primaryClass = "hep-ph",
    doi = "10.1007/JHEP06(2021)121",
    journal = "JHEP",
    volume = "06",
    pages = "121",
    year = "2021"
}

@article{Esteban:2020cvm,
    author = "Esteban, Ivan and Gonzalez-Garcia, M. C. and Maltoni, Michele and Schwetz, Thomas and Zhou, Albert",
    title = "{The fate of hints: updated global analysis of three-flavor neutrino oscillations}",
    eprint = "2007.14792",
    archivePrefix = "arXiv",
    primaryClass = "hep-ph",
    reportNumber = "IFT-UAM/CSIC-112, YITP-SB-2020-21",
    doi = "10.1007/JHEP09(2020)178",
    journal = "JHEP",
    volume = "09",
    pages = "178",
    year = "2020"
}

@ARTICLE{NFW_1996,
       author = {{Navarro}, Julio F. and {Frenk}, Carlos S. and {White}, Simon D.~M.},
        title = "{The Structure of Cold Dark Matter Halos}",
      journal = {ApJ},
         year = {1996},
        month = {may},
       volume = {462},
        pages = {563},
          doi = {10.1086/177173},
archivePrefix = {arXiv},
       eprint = {astro-ph/9508025},
 primaryClass = {astro-ph},
       adsurl = {https://ui.adsabs.harvard.edu/abs/1996ApJ...462..563N}
}

@article{Burkert_1995,
    author = "Burkert, A.",
    title = "{The Structure of dark matter halos in dwarf galaxies}",
    eprint = "astro-ph/9504041",
    archivePrefix = "arXiv",
    doi = "10.1086/309560",
    journal = "Astrophys. J. Lett.",
    volume = "447",
    pages = "L25",
    year = "1995"
}

@article{Nesti:2013uwa,
    author = "Nesti, Fabrizio and Salucci, Paolo",
    title = "{The Dark Matter halo of the  Milky Way, AD 2013}",
    eprint = "1304.5127",
    archivePrefix = "arXiv",
    primaryClass = "astro-ph.GA",
    doi = "10.1088/1475-7516/2013/07/016",
    journal = "JCAP",
    volume = "07",
    pages = "016",
    year = "2013"
}

@article{Benito_2019,
    author = "Benito, Maria and Cuoco, Alessandro and Iocco, Fabio",
    title = "{Handling the Uncertainties in the Galactic Dark Matter Distribution for Particle Dark Matter Searches}",
    eprint = "1901.02460",
    archivePrefix = "arXiv",
    primaryClass = "astro-ph.GA",
    doi = "10.1088/1475-7516/2019/03/033",
    journal = "JCAP",
    volume = "03",
    pages = "033",
    year = "2019"
}

@article{S_ding_2025,
   title={Local dark matter density from <i>Gaia</i> DR3 K-dwarfs using Gaussian processes},
   volume={542},
   ISSN={1365-2966},
   url={http://dx.doi.org/10.1093/mnras/staf1391},
   DOI={10.1093/mnras/staf1391},
   number={4},
   journal={Monthly Notices of the Royal Astronomical Society},
   publisher={Oxford University Press (OUP)},
   author={Söding, Laurin and Bartel, Ruben L and Mertsch, Philipp},
   year={2025},
   month=Aug, pages={2987–2997} }

@article{deBlok:2009sp,
    author = "de Blok, W. J. G.",
    title = "{The Core-Cusp Problem}",
    eprint = "0910.3538",
    archivePrefix = "arXiv",
    primaryClass = "astro-ph.CO",
    doi = "10.1155/2010/789293",
    journal = "Adv. Astron.",
    volume = "2010",
    pages = "789293",
    year = "2010"
}

@article{An:2012pv,
    author = "An, J. and Zhao, H.",
    title = "{Fitting functions for dark matter density profiles}",
    eprint = "1209.6220",
    archivePrefix = "arXiv",
    primaryClass = "astro-ph.CO",
    doi = "10.1093/mnras/sts175",
    journal = "Mon. Not. Roy. Astron. Soc.",
    volume = "428",
    pages = "2805--2811",
    year = "2013"
}

@article{Cowan:2010js,
    author = "Cowan, Glen and Cranmer, Kyle and Gross, Eilam and Vitells, Ofer",
    title = "{Asymptotic formulae for likelihood-based tests of new physics}",
    eprint = "1007.1727",
    archivePrefix = "arXiv",
    primaryClass = "physics.data-an",
    doi = "10.1140/epjc/s10052-011-1554-0",
    journal = "Eur. Phys. J. C",
    volume = "71",
    pages = "1554",
    year = "2011",
    note = "[Erratum: Eur.Phys.J.C 73, 2501 (2013)]"
}

@article{10.1214/aoms/1177728725,
author = {Herman Chernoff},
title = {{On the Distribution of the Likelihood Ratio}},
volume = {25},
journal = {Annals Math. Statist.},
number = {3},
publisher = {Institute of Mathematical Statistics},
pages = {573 -- 578},
year = {1954},
doi = {10.1214/aoms/1177728725},
URL = {https://doi.org/10.1214/aoms/1177728725}
}

@article{Gross:2010qma,
    author = "Gross, Eilam and Vitells, Ofer",
    title = "{Trial factors for the look elsewhere effect in high energy physics}",
    eprint = "1005.1891",
    archivePrefix = "arXiv",
    primaryClass = "physics.data-an",
    doi = "10.1140/epjc/s10052-010-1470-8",
    journal = "Eur. Phys. J. C",
    volume = "70",
    pages = "525--530",
    year = "2010"
}

@article{Wilks:1938dza,
    author = "Wilks, S. S.",
    title = "{The Large-Sample Distribution of the Likelihood Ratio for Testing Composite Hypotheses}",
    doi = "10.1214/aoms/1177732360",
    journal = "Annals Math. Statist.",
    volume = "9",
    number = "1",
    pages = "60--62",
    year = "1938"
}

@book{Cowan:1998ji,
    author = "Cowan, G.",
    title = "{Statistical data analysis}",
  publisher = {Oxford University Press, USA},
    isbn = "978-0-19-850156-5",
    year = "1998"
}

@article{IceCube:2023ies,
    author = "Abbasi, R. and others",
    collaboration = "IceCube",
    title = "{Search for neutrino lines from dark matter annihilation and decay with IceCube}",
    eprint = "2303.13663",
    archivePrefix = "arXiv",
    primaryClass = "astro-ph.HE",
    doi = "10.1103/PhysRevD.108.102004",
    journal = "Phys. Rev. D",
    volume = "108",
    number = "10",
    pages = "102004",
    year = "2023"
}

@article{IceCube:2017rdn,
    author = "Aartsen, M. G. and others",
    collaboration = "IceCube",
    title = "{Search for Neutrinos from Dark Matter Self-Annihilations in the center of the Milky Way with 3 years of IceCube/DeepCore}",
    eprint = "1705.08103",
    archivePrefix = "arXiv",
    primaryClass = "hep-ex",
    doi = "10.1140/epjc/s10052-017-5213-y",
    journal = "Eur. Phys. J. C",
    volume = "77",
    number = "9",
    pages = "627",
    year = "2017"
}

@article{IceCube:2016oqp,
    author = "Aartsen, M. G. and others",
    collaboration = "IceCube",
    title = "{All-flavour Search for Neutrinos from Dark Matter Annihilations in the Milky Way with IceCube/DeepCore}",
    eprint = "1606.00209",
    archivePrefix = "arXiv",
    primaryClass = "astro-ph.HE",
    doi = "10.1140/epjc/s10052-016-4375-3",
    journal = "Eur. Phys. J. C",
    volume = "76",
    number = "10",
    pages = "531",
    year = "2016"
}

@article{IceCube:2011kcp,
    author = "Abbasi, R. and others",
    collaboration = "IceCube",
    title = "{Search for dark matter from the Galactic halo with the IceCube Neutrino Telescope}",
    eprint = "1101.3349",
    archivePrefix = "arXiv",
    primaryClass = "astro-ph.HE",
    doi = "10.1103/PhysRevD.84.022004",
    journal = "Phys. Rev. D",
    volume = "84",
    pages = "022004",
    year = "2011"
}

@article{IceCube:2018tkk,
    author = "Aartsen, M. G. and others",
    collaboration = "IceCube",
    title = "{Search for neutrinos from decaying dark matter with IceCube}",
    eprint = "1804.03848",
    archivePrefix = "arXiv",
    primaryClass = "astro-ph.HE",
    doi = "10.1140/epjc/s10052-018-6273-3",
    journal = "Eur. Phys. J. C",
    volume = "78",
    number = "10",
    pages = "831",
    year = "2018"
}

@article{HESS:2022ygk,
    author = "Abdalla, H. and others",
    collaboration = "H.E.S.S.",
    title = "{Search for Dark Matter Annihilation Signals in the H.E.S.S. Inner Galaxy Survey}",
    eprint = "2207.10471",
    archivePrefix = "arXiv",
    primaryClass = "astro-ph.HE",
    doi = "10.1103/PhysRevLett.129.111101",
    journal = "Phys. Rev. Lett.",
    volume = "129",
    number = "11",
    pages = "111101",
    year = "2022"
}

@article{MAGIC:2016xys,
    author = "Ahnen, M. L. and others",
    collaboration = "MAGIC, Fermi-LAT",
    title = "{Limits to Dark Matter Annihilation Cross-Section from a Combined Analysis of MAGIC and Fermi-LAT Observations of Dwarf Satellite Galaxies}",
    eprint = "1601.06590",
    archivePrefix = "arXiv",
    primaryClass = "astro-ph.HE",
    reportNumber = "FERMILAB-PUB-16-283-AE",
    doi = "10.1088/1475-7516/2016/02/039",
    journal = "JCAP",
    volume = "02",
    pages = "039",
    year = "2016"
}

@article{Ackermann_2012,
   title={CONSTRAINTS ON THE GALACTIC HALO DARK MATTER FROMFERMI-LAT DIFFUSE MEASUREMENTS},
    collaboration = "Fermi-LAT",
   volume={761},
   ISSN={1538-4357},
   url={http://dx.doi.org/10.1088/0004-637X/761/2/91},
   DOI={10.1088/0004-637x/761/2/91},
   number={2},
   journal={The Astrophysical Journal},
   publisher={American Astronomical Society},
   author={Ackermann, M. and others},
   year={2012},
   month=nov, pages={91},
    eprint = "1205.6474",
    archivePrefix="arXiv",
    primaryClass="astro-ph.CO"}

@article{HAWC:2017mfa,
    author = "Albert, A. and others",
    collaboration = "HAWC",
    title = "{Dark Matter Limits From Dwarf Spheroidal Galaxies with The HAWC Gamma-Ray Observatory}",
    eprint = "1706.01277",
    archivePrefix = "arXiv",
    primaryClass = "astro-ph.HE",
    doi = "10.3847/1538-4357/aaa6d8",
    journal = "Astrophys. J.",
    volume = "853",
    number = "2",
    pages = "154",
    year = "2018"
}

@article{Super-Kamiokande:2020sgt,
    author = "Abe, K. and others",
    collaboration = "Super-Kamiokande",
    title = "{Indirect search for dark matter from the Galactic Center and halo with the Super-Kamiokande detector}",
    eprint = "2005.05109",
    archivePrefix = "arXiv",
    primaryClass = "hep-ex",
    doi = "10.1103/PhysRevD.102.072002",
    journal = "Phys. Rev. D",
    volume = "102",
    number = "7",
    pages = "072002",
    year = "2020"
}

@article{ANTARES:2019svn,
    author = "Albert, A. and others",
    collaboration = "ANTARES",
    title = "{Search for dark matter towards the Galactic Centre with 11 years of ANTARES data}",
    eprint = "1912.05296",
    archivePrefix = "arXiv",
    primaryClass = "astro-ph.HE",
    doi = "10.1016/j.physletb.2020.135439",
    journal = "Phys. Lett. B",
    volume = "805",
    pages = "135439",
    year = "2020"
}

@article{KM3NeT:2024xca,
    author = "Aiello, S. and others",
    collaboration = "KM3NeT",
    title = "{First searches for dark matter with the KM3NeT neutrino telescopes}",
    eprint = "2411.10092",
    archivePrefix = "arXiv",
    primaryClass = "astro-ph.HE",
    doi = "10.1088/1475-7516/2025/03/058",
    journal = "JCAP",
    volume = "03",
    pages = "058",
    year = "2025"
}

@article{Hutten:2018aix,
    author = {H\"utten, Moritz and Combet, C\'eline and Maurin, David},
    title = "{CLUMPY v3: $\gamma$-ray and $\nu$ signals from dark matter at all scales}",
    eprint = "1806.08639",
    archivePrefix = "arXiv",
    primaryClass = "astro-ph.CO",
    doi = "10.1016/j.cpc.2018.10.001",
    journal = "Comput. Phys. Commun.",
    volume = "235",
    pages = "336--345",
    year = "2019"
}

@article{Steigman:2012nb,
    author = "Steigman, Gary and Dasgupta, Basudeb and Beacom, John F.",
    title = "{Precise Relic WIMP Abundance and its Impact on Searches for Dark Matter Annihilation}",
    eprint = "1204.3622",
    archivePrefix = "arXiv",
    primaryClass = "hep-ph",
    doi = "10.1103/PhysRevD.86.023506",
    journal = "Phys. Rev. D",
    volume = "86",
    pages = "023506",
    year = "2012"
}

@software{KDEpy,
  author       = {Tommy Odland},
  title        = {tommyod/KDEpy: Kernel Density Estimation in Python},
  month        = dec,
  year         = 2018,
  publisher    = {Zenodo},
  version      = {v0.9.10},
  doi          = {10.5281/zenodo.2392268},
  url          = {https://doi.org/10.5281/zenodo.2392268},
}

@article{Bose:2021yhz,
    author = "Bose, Debajit and Maity, Tarak Nath and Ray, Tirtha Sankar",
    title = "{Neutrinos from captured dark matter annihilation in a galactic population of neutron stars}",
    eprint = "2108.12420",
    archivePrefix = "arXiv",
    primaryClass = "hep-ph",
    doi = "10.1088/1475-7516/2022/05/001",
    journal = "JCAP",
    volume = "05",
    number = "05",
    pages = "001",
    year = "2022"
}

@article{Nguyen:2022zwb,
    author = "Nguyen, Thong T. Q. and Tait, Tim M. P.",
    title = "{Bounds on long-lived dark matter mediators from neutron stars}",
    eprint = "2212.12547",
    archivePrefix = "arXiv",
    primaryClass = "hep-ph",
    doi = "10.1103/PhysRevD.107.115016",
    journal = "Phys. Rev. D",
    volume = "107",
    number = "11",
    pages = "115016",
    year = "2023"
}

@article{Acevedo:2024ttq,
    author = "Acevedo, Javier F. and Bramante, Joseph and Liu, Qinrui and Tyagi, Narayani",
    title = "{Neutrino and gamma-ray signatures of inelastic dark matter annihilating outside neutron stars}",
    eprint = "2404.10039",
    archivePrefix = "arXiv",
    primaryClass = "hep-ph",
    reportNumber = "SLAC-PUB-17767",
    doi = "10.1088/1475-7516/2025/03/028",
    journal = "JCAP",
    volume = "03",
    pages = "028",
    year = "2025"
}

@article{Ou:2023adg,
    author = "Ou, Xiaowei and Eilers, Anna-Christina and Necib, Lina and Frebel, Anna",
    title = "{The dark matter profile of the Milky Way inferred from its circular velocity curve}",
    eprint = "2303.12838",
    archivePrefix = "arXiv",
    primaryClass = "astro-ph.GA",
    doi = "10.1093/mnras/stae034",
    journal = "Mon. Not. Roy. Astron. Soc.",
    volume = "528",
    number = "1",
    pages = "693--710",
    year = "2024"
}

@article{de_Salas_2019,
   title={On the estimation of the local dark matter density using the rotation curve of the Milky Way},
   volume={2019},
   ISSN={1475-7516},
   url={http://dx.doi.org/10.1088/1475-7516/2019/10/037},
   DOI={10.1088/1475-7516/2019/10/037},
   number={10},
   journal={Journal of Cosmology and Astroparticle Physics},
   publisher={IOP Publishing},
   author={de Salas, P.F. and Malhan, K. and Freese, K. and Hattori, K. and Valluri, M.},
   year={2019},
   month=oct, pages={037–037} }

@article{Pontzen:2014lma,
    author = "Pontzen, Andrew and Governato, Fabio",
    title = "{Cold dark matter heats up}",
    eprint = "1402.1764",
    archivePrefix = "arXiv",
    primaryClass = "astro-ph.CO",
    doi = "10.1038/nature12953",
    journal = "Nature",
    volume = "506",
    pages = "171--178",
    year = "2014"
}

@article{Chan:2015tna,
    author = "Chan, T. K. and Kere{\v{s}}, D. and O{\~n}orbe, J. and Hopkins, P. F. and Muratov, A. L. and Faucher-Gigu{\`e}re, C. -A. and Quataert, E.",
    title = "{The impact of baryonic physics on the structure of dark matter haloes: the view from the FIRE cosmological simulations}",
    eprint = "1507.02282",
    archivePrefix = "arXiv",
    primaryClass = "astro-ph.GA",
    doi = "10.1093/mnras/stv2165",
    journal = "Mon. Not. Roy. Astron. Soc.",
    volume = "454",
    number = "3",
    pages = "2981--3001",
    year = "2015"
}

@article{DiCintio:2013qxa,
    author = "Di Cintio, Arianna and Brook, Chris B. and Macci{\`o}, Andrea V. and Stinson, Greg S. and Knebe, Alexander and Dutton, Aaron A. and Wadsley, James",
    title = "{The dependence of dark matter profiles on the stellar-to-halo mass ratio: a prediction for cusps versus cores}",
    eprint = "1306.0898",
    archivePrefix = "arXiv",
    primaryClass = "astro-ph.CO",
    doi = "10.1093/mnras/stt1891",
    journal = "Mon. Not. Roy. Astron. Soc.",
    volume = "437",
    number = "1",
    pages = "415--423",
    year = "2014"
}

\appendix
\section{}
\label{Appendix A}
In this appendix, we present the expected diffuse GP PDFs as functions of the two observables used in the analysis: the reconstructed energy and the reconstructed opening angle with respect to the Galactic Center. These distributions are compared to the expected DM signal evaluated at the 90$\%$ C.L. sensitivity.

\begin{figure}[tb!]
    % \centering
    % \begin{minipage}{0.48\textwidth}
        \includegraphics[width=1.\linewidth]{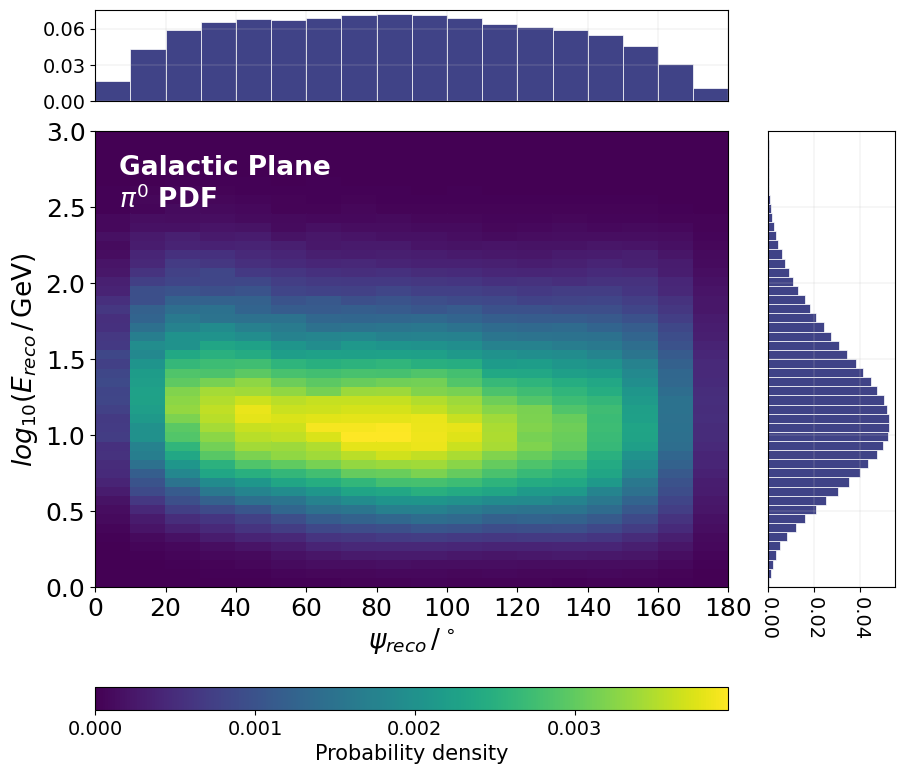}
    % \end{minipage}
    % \hfill
    % \begin{minipage}{0.48\textwidth}
        \includegraphics[width=1.\linewidth]{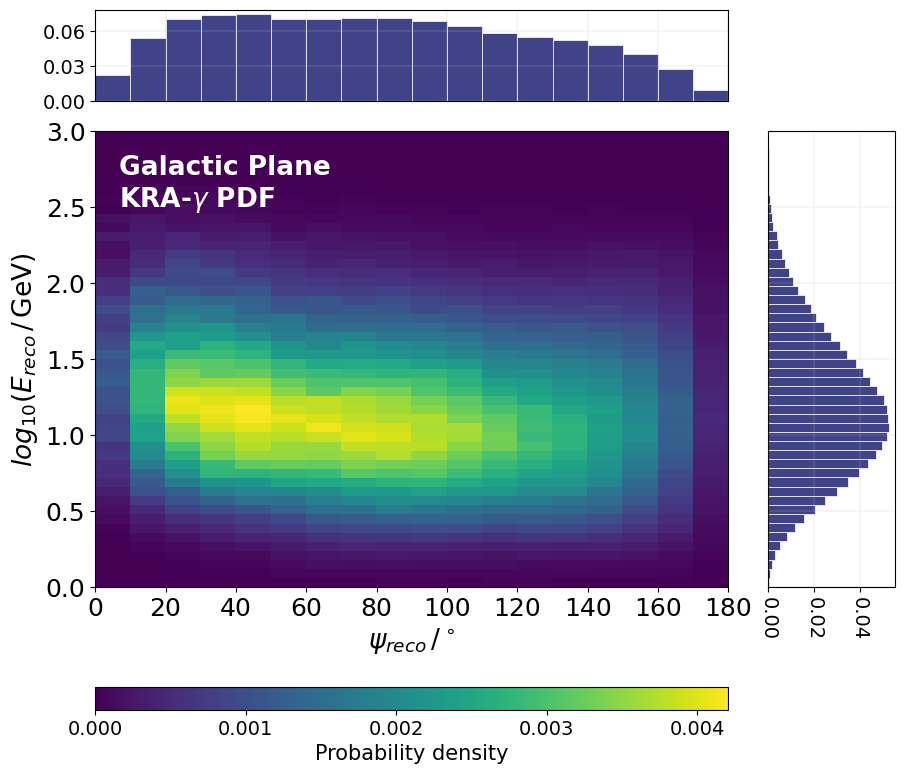}
    % \end{minipage}
\caption{Two-dimensional PDFs as a function of reconstructed energy ($E_{reco}$) and reconstructed opening angle relative to the Galactic Center ($\psi_{reco}$) for the diffuse astrophysical neutrino emission from GP assuming $\pi^0$ (\textit{upper}) and KRA-$\gamma$ models (\textit{lower}).}
\label{fig:GP_pdfs}
\end{figure}

Fig.~\ref{fig:GP_pdfs} shows the GP PDFs for the two templates introduced in Section~\ref{sec:systematics}. In terms of the opening-angle distribution, the GP emission exhibits a pattern broadly similar to that of the atmospheric background shown in Fig.~\ref{fig:background_pdf}. As expected, the KRA-$\gamma$ template predicts slightly stronger emission toward the Galactic Center compared to the $\pi^0$ template~\cite{Gaggero:2015xza}. Nevertheless, both GP models resemble the background much more closely than the DM signals, which display a pronounced excess toward the Galactic Center. This behavior arises from the limited angular resolution and the choice of observables: when projected onto the opening angle relative to the Galactic Center, the long extended Galactic disk emission effectively spreads over a wide angular range, thereby reducing its distinguishing with respect to the isotropic background.

\begin{figure*}[!tb]
    \begin{minipage}{1.\textwidth}
\includegraphics[width=\linewidth]{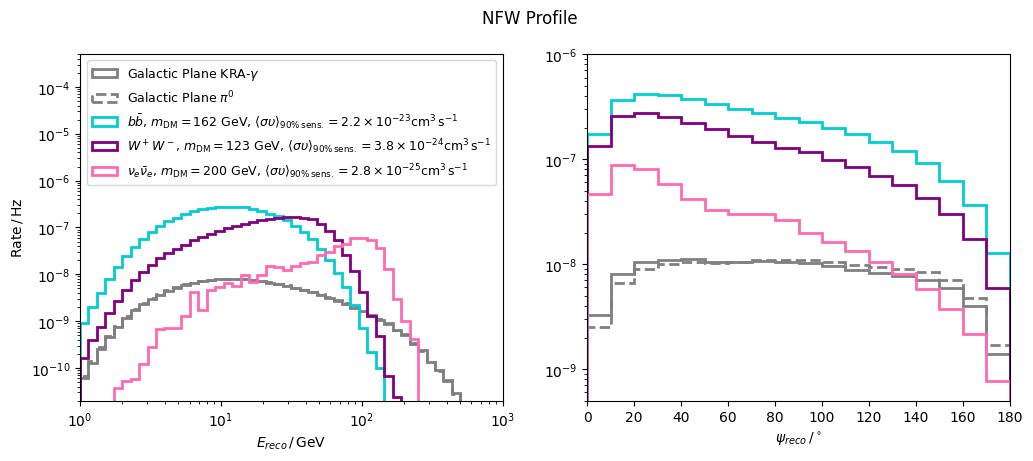}
    \end{minipage}
    \hfill
    \begin{minipage}{1.\textwidth}
\includegraphics[width=\linewidth]{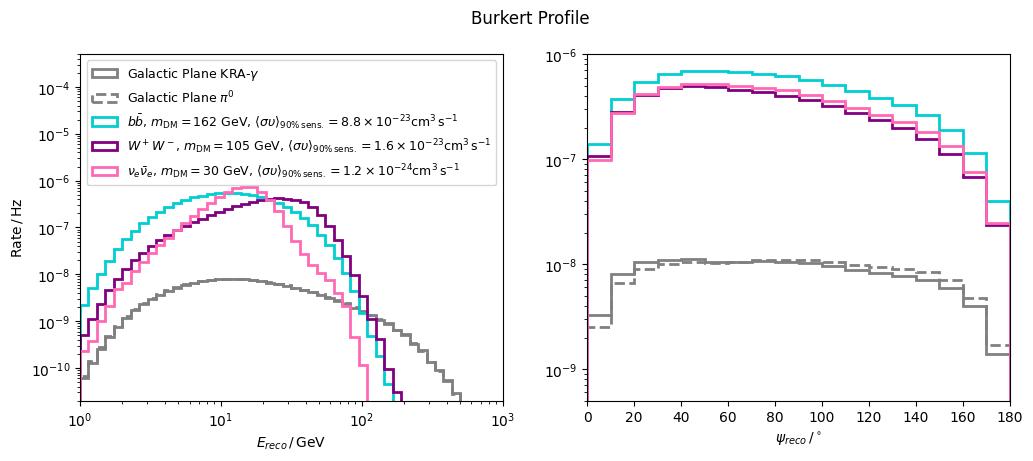}
    \end{minipage}
\caption{Expected event distribution as a function of reconstructed energy \textit{(left)} and reconstructed opening angle \textit{(right)} for diffuse astrophysical neutrino emission in GP in comparison with DM signals at $90 \%$ C.L. sensitivity assuming NFW (\textit{upper}) and Burkert (\textit{lower}) profile.}
\label{fig:GPvsDM}
\end{figure*}

Fig.~\ref{fig:GPvsDM} compares the expected event rates from the two GP templates with those from DM annihilation in the Galactic Center. Three representative annihilation channels are shown, corresponding to soft ($b\bar{b}$), hard ($W^+W^-$), and line-like ($\nu_e\bar{\nu}_e$) neutrino spectra. For each channel, the DM mass is chosen to provide the strongest sensitivity (i.e., the lowest upper limit on the annihilation cross section), and the signal normalization corresponds to the 90$\%$ C.L. sensitivity. In the regions of phase space where the DM contribution is most significant, the expected DM event rate at the sensitivity level exceeds the GP emission by orders of magnitude. In particular, the DM-induced excess is more concentrated toward small opening angles, in contrast to the more extended GP component. These differences explain why the inclusion of GP emission has a subdominant impact on the derived DM sensitivity compared to statistical fluctuations, as discussed in Section~\ref{sec:systematics}.

\section{}
\label{sec:appendixB}
The following tables report, for each DM channel, the 90\% C.L. limits, the best-fit number of signal events alongside with the significance of each tested scenario, expressed as a z-score.

% #############################################
% Table bb
\begin{center}
\begin{table*}
\scriptsize
\centering
\begin{tabular}{|r| c c c | c c c ||r| c c c |  c c c | }
\toprule
%1st
 \multicolumn{7}{|c||}{{\bf Annihilation}} & \multicolumn{7}{c|}{{\bf Decay}} \\
%2nd
  & \multicolumn{3}{c|}{{\bf NFW }} &  \multicolumn{3}{c||}{{\bf Burkert}} & &\multicolumn{3}{c|}{{\bf NFW}} &  \multicolumn{3}{c|}{\bf{Burkert}} \\
%3rd
{$m_\chi$} &  \multirow{2}{*}{$\hat{n}_s$} & {$\langle \sigma v \rangle^{90\%}_{u.l.} $} & \multirow{2}{*}{$z$-score} &  \multirow{2}{*}{$\hat{n}_s$} & {$\langle \sigma v \rangle^{90\%}_{u.l.}$}   & \multirow{2}{*}{$z$-score} & {$m_\chi$} &  \multirow{2}{*}{$\hat{n}_s$} & {$\tau^{90\%}_{l.l.}$} & \multirow{2}{*}{$z$-score} &  \multirow{2}{*}{$\hat{n}_s$} & {$\tau^{90\%}_{l.l.}$} & \multirow{2}{*}{$z$-score} \\ 
%4th
${\rm [GeV]}$ &   & {$10^{-24} {\rm \;[cm^3 s^{-1}]}$} &   &   & {$10^{-24} {\rm \;[cm^3 s^{-1}]}$} &  & ${\rm [GeV]}$ & &  {$10^{24} {\rm \;[s]}$} &   &    & {$10^{24} {\rm \;[s]}$} & \\
\midrule

15.0 & 1171 & 147.32 & 1.04 & 2299 & 526.48 & 1.08 & 30.0 & 3830 & 0.10 & 1.16 & 4852 & 0.08 & 1.18 \\
18.6 & 1403 & 124.07 & 1.22 & 2476 & 427.46 & 1.15 & 36.4 & 4042 & 0.15 & 1.21 & 5048 & 0.11 & 1.21 \\
23.1 & 1575 & 104.78 & 1.37 & 2573 & 353.28 & 1.19 & 44.1 & 4117 & 0.21 & 1.23 & 5022 & 0.16 & 1.20 \\
28.7 & 1686 & 90.25 & 1.48 & 2625 & 300.09 & 1.22 & 53.5 & 4109 & 0.30 & 1.24 & 4861 & 0.24 & 1.16 \\
35.7 & 1791 & 80.16 & 1.59 & 2678 & 263.01 & 1.25 & 64.8 & 4099 & 0.42 & 1.25 & 4687 & 0.33 & 1.12 \\
44.3 & 1902 & 73.46 & 1.72 & 2743 & 237.60 & 1.30 & 78.6 & 4142 & 0.57 & 1.27 & 4588 & 0.45 & 1.11 \\
55.0 & 2004 & 68.97 & 1.86 & 2826 & 221.12 & 1.37 & 95.3 & 4237 & 0.75 & 1.32 & 4573 & 0.60 & 1.12 \\
68.3 & 2088 & 65.98 & 1.99 & 2916 & 211.00 & 1.44 & 115.5 & 4388 & 0.95 & 1.40 & 4669 & 0.77 & 1.16 \\
84.8 & 2166 & 64.51 & 2.14 & 3010 & 206.29 & 1.52 & 140.1 & 4529 & 1.19 & 1.47 & 4777 & 0.97 & 1.21 \\
105.3 & 2233 & 64.14 & 2.28 & 3119 & 206.16 & 1.61 & 169.8 & 4679 & 1.47 & 1.55 & 4900 & 1.20 & 1.26 \\
130.7 & 2260 & 64.40 & 2.39 & 3206 & 209.36 & 1.71 & 205.9 & 4832 & 1.78 & 1.64 & 5039 & 1.45 & 1.32 \\
162.4 & 2235 & 64.79 & 2.46 & 3247 & 214.23 & 1.78 & 249.6 & 5002 & 2.13 & 1.74 & 5241 & 1.74 & 1.41 \\
201.6 & 2147 & 65.07 & 2.47 & 3227 & 220.51 & 1.83 & 302.7 & 5062 & 2.53 & 1.81 & 5311 & 2.07 & 1.47 \\
250.3 & 2028 & 65.63 & 2.43 & 3159 & 228.36 & 1.86 & 367.0 & 5057 & 2.99 & 1.86 & 5359 & 2.44 & 1.52 \\
310.9 & 1880 & 66.17 & 2.37 & 3028 & 236.33 & 1.86 & 444.9 & 4983 & 3.52 & 1.89 & 5325 & 2.85 & 1.56 \\
386.0 & 1695 & 66.23 & 2.27 & 2838 & 244.24 & 1.82 & 539.4 & 4859 & 4.13 & 1.91 & 5286 & 3.32 & 1.60 \\
479.3 & 1500 & 66.29 & 2.15 & 2591 & 251.46 & 1.75 & 654.0 & 4667 & 4.86 & 1.91 & 5132 & 3.87 & 1.61 \\
595.2 & 1265 & 66.40 & 1.97 & 2245 & 259.05 & 1.63 & 792.9 & 4358 & 5.73 & 1.86 & 4826 & 4.53 & 1.57 \\
739.1 & 1061 & 66.11 & 1.79 & 1943 & 265.08 & 1.50 & 961.4 & 4000 & 6.77 & 1.79 & 4451 & 5.32 & 1.51 \\
917.8 & 859 & 65.59 & 1.56 & 1623 & 270.19 & 1.34 & 1165.6 & 3525 & 7.98 & 1.68 & 3939 & 6.22 & 1.42 \\
1139.7 & 678 & 65.10 & 1.34 & 1324 & 275.38 & 1.17 & 1413.2 & 3138 & 9.49 & 1.58 & 3535 & 7.33 & 1.34 \\
1415.2 & 502 & 64.13 & 1.08 & 1026 & 279.82 & 0.98 & 1713.4 & 2653 & 11.32 & 1.42 & 3007 & 8.66 & 1.21 \\
1757.3 & 355 & 62.98 & 0.83 & 753 & 281.54 & 0.78 & 2077.4 & 2262 & 13.43 & 1.29 & 2591 & 10.17 & 1.10 \\
2182.1 & 247 & 62.98 & 0.63 & 540 & 286.62 & 0.60 & 2518.6 & 1796 & 16.41 & 1.09 & 2074 & 12.31 & 0.94 \\
2709.7 & 153 & 62.52 & 0.42 & 362 & 293.88 & 0.43 & 3053.7 & 1379 & 19.62 & 0.90 & 1575 & 14.56 & 0.76 \\
3364.7 & 70 & 61.70 & 0.22 & 219 & 302.23 & 0.28 & 3702.3 & 1023 & 23.88 & 0.71 & 1166 & 17.55 & 0.60 \\
4178.2 & 15 & 62.66 & 0.05 & 97 & 315.84 & 0.14 & 4488.8 & 723 & 28.84 & 0.54 & 823 & 20.94 & 0.45 \\
5188.3 & 0 & 63.98 & 0.00 & 4 & 330.56 & 0.01 & 5442.3 & 480 & 34.12 & 0.38 & 549 & 24.36 & 0.32 \\
6442.5 & 0 & 66.66 & 0.00 & 0 & 349.47 & 0.00 & 6598.4 & 274 & 40.90 & 0.24 & 315 & 28.79 & 0.20 \\
8000.0 & 0 & 70.50 & 0.00 & 0 & 373.88 & 0.00 & 8000.0 & 116 & 48.03 & 0.11 & 135 & 33.24 & 0.09 \\

\bottomrule
\end{tabular}
\caption{Results for the final state channel $b\bar{b}$ for both the annihilation and decaying mode and for both the NFW and Burkert profile. The best fit value on the number of signal events $\hat{n}_s$ is shown together with the resulting upper limit in $\langle \sigma v \rangle^{90\%}_{u.l.}$ and lower limit on $\tau^{90\%}_{l.l.}$ along with the significance given in number of sigmas, $z$-score.}
\label{tb:bb}
\end{table*}
\end{center}

% #############################################
% Table WW
\begin{center}
\begin{table*}
\scriptsize
\centering
\begin{tabular}{|r| c c c | c c c ||r| c c c |  c c c | }
\toprule
%1st
 \multicolumn{7}{|c||}{{\bf Annihilation}} & \multicolumn{7}{c|}{{\bf Decay}} \\
%2nd
  & \multicolumn{3}{c|}{{\bf NFW }} &  \multicolumn{3}{c||}{{\bf Burkert}} & &\multicolumn{3}{c|}{{\bf NFW}} &  \multicolumn{3}{c|}{\bf{Burkert}} \\
%3rd
{$m_\chi$} &  \multirow{2}{*}{$\hat{n}_s$} & {$\langle \sigma v \rangle^{90\%}_{u.l.} $} & \multirow{2}{*}{$z$-score} &  \multirow{2}{*}{$\hat{n}_s$} & {$\langle \sigma v \rangle^{90\%}_{u.l.}$}   & \multirow{2}{*}{$z$-score} & {$m_\chi$} &  \multirow{2}{*}{$\hat{n}_s$} & {$\tau^{90\%}_{l.l.}$} & \multirow{2}{*}{$z$-score} &  \multirow{2}{*}{$\hat{n}_s$} & {$\tau^{90\%}_{l.l.}$} & \multirow{2}{*}{$z$-score} \\ 
%4th
${\rm [GeV]}$ &   & {$10^{-24} {\rm \;[cm^3 s^{-1}]}$} &   &   & {$10^{-24} {\rm \;[cm^3 s^{-1}]}$} &  & ${\rm [GeV]}$ & &  {$10^{24} {\rm \;[s]}$} &   &    & {$10^{24} {\rm \;[s]}$} & \\
\midrule

90.0 & 1450 & 10.87 & 2.33 & 2196 & 37.81 & 1.70 & 180.0 & 3458 & 8.38 & 1.78 & 3644 & 6.80 & 1.44 \\
105.1 & 1313 & 10.77 & 2.28 & 2149 & 39.26 & 1.76 & 205.2 & 3388 & 9.27 & 1.82 & 3620 & 7.47 & 1.50 \\
122.6 & 1001 & 9.64 & 1.88 & 1915 & 39.01 & 1.68 & 233.8 & 3096 & 10.61 & 1.76 & 3496 & 8.25 & 1.52 \\
143.2 & 815 & 9.20 & 1.65 & 1629 & 38.14 & 1.54 & 266.5 & 2753 & 12.20 & 1.66 & 3295 & 9.15 & 1.51 \\
167.1 & 680 & 8.88 & 1.52 & 1434 & 38.23 & 1.47 & 303.8 & 2429 & 14.05 & 1.57 & 2936 & 10.48 & 1.44 \\
195.1 & 495 & 7.94 & 1.26 & 1160 & 37.11 & 1.29 & 346.2 & 2149 & 16.07 & 1.49 & 2627 & 11.87 & 1.38 \\
227.7 & 360 & 7.19 & 1.04 & 864 & 34.99 & 1.06 & 394.6 & 1746 & 19.06 & 1.30 & 2143 & 13.94 & 1.19 \\
265.9 & 239 & 6.60 & 0.76 & 597 & 32.48 & 0.80 & 449.8 & 1355 & 22.93 & 1.09 & 1657 & 16.59 & 0.98 \\
310.4 & 210 & 6.98 & 0.70 & 415 & 31.58 & 0.59 & 512.7 & 951 & 28.46 & 0.83 & 1155 & 20.41 & 0.73 \\
362.3 & 146 & 6.77 & 0.50 & 253 & 30.25 & 0.38 & 584.3 & 734 & 32.85 & 0.68 & 835 & 23.82 & 0.56 \\
422.9 & 81 & 6.68 & 0.31 & 192 & 31.44 & 0.30 & 666.0 & 490 & 38.95 & 0.47 & 496 & 28.57 & 0.34 \\
493.7 & 9 & 5.91 & 0.04 & 70 & 29.70 & 0.12 & 759.1 & 332 & 45.14 & 0.34 & 334 & 32.29 & 0.24 \\
576.3 & 0 & 5.22 & 0.00 & 0 & 26.69 & 0.00 & 865.2 & 243 & 50.22 & 0.26 & 278 & 35.11 & 0.21 \\
672.8 & 0 & 4.44 & 0.00 & 0 & 23.21 & 0.00 & 986.2 & 67 & 61.97 & 0.08 & 60 & 42.94 & 0.05 \\
785.4 & 0 & 4.33 & 0.00 & 0 & 21.80 & 0.00 & 1124.0 & 0 & 82.48 & 0.00 & 0 & 57.91 & 0.00 \\
916.8 & 0 & 5.77 & 0.00 & 0 & 26.33 & 0.00 & 1281.1 & 0 & 100.52 & 0.00 & 0 & 68.14 & 0.00 \\
1070.2 & 0 & 6.34 & 0.00 & 0 & 31.56 & 0.00 & 1460.2 & 0 & 135.67 & 0.00 & 0 & 90.44 & 0.00 \\
1249.3 & 0 & 6.53 & 0.00 & 0 & 35.58 & 0.00 & 1664.3 & 0 & 137.82 & 0.00 & 0 & 93.76 & 0.00 \\
1458.4 & 0 & 7.78 & 0.00 & 0 & 39.31 & 0.00 & 1896.9 & 0 & 138.28 & 0.00 & 0 & 95.76 & 0.00 \\
1702.4 & 0 & 10.60 & 0.00 & 0 & 62.34 & 0.00 & 2162.1 & 0 & 131.38 & 0.00 & 0 & 85.72 & 0.00 \\
1987.4 & 0 & 14.04 & 0.00 & 0 & 84.74 & 0.00 & 2464.3 & 0 & 140.84 & 0.00 & 0 & 91.16 & 0.00 \\
2319.9 & 0 & 18.42 & 0.00 & 0 & 100.79 & 0.00 & 2808.8 & 0 & 155.52 & 0.00 & 0 & 102.62 & 0.00 \\
2708.2 & 0 & 23.40 & 0.00 & 0 & 125.25 & 0.00 & 3201.4 & 0 & 114.65 & 0.00 & 0 & 67.45 & 0.00 \\
3161.4 & 0 & 25.20 & 0.00 & 0 & 141.41 & 0.00 & 3648.9 & 0 & 89.84 & 0.00 & 0 & 49.65 & 0.00 \\
3690.5 & 0 & 31.93 & 0.00 & 0 & 177.74 & 0.00 & 4158.9 & 0 & 84.08 & 0.00 & 0 & 47.30 & 0.00 \\
4308.1 & 0 & 39.06 & 0.00 & 0 & 215.67 & 0.00 & 4740.3 & 0 & 87.45 & 0.00 & 0 & 53.67 & 0.00 \\
5029.0 & 0 & 46.32 & 0.00 & 0 & 254.17 & 0.00 & 5402.9 & 0 & 85.66 & 0.00 & 0 & 55.08 & 0.00 \\
5870.7 & 0 & 53.64 & 0.00 & 0 & 293.02 & 0.00 & 6158.1 & 0 & 91.95 & 0.00 & 0 & 60.23 & 0.00 \\
6853.1 & 0 & 61.08 & 0.00 & 0 & 332.65 & 0.00 & 7018.9 & 0 & 85.17 & 0.00 & 0 & 55.77 & 0.00 \\
8000.0 & 0 & 68.83 & 0.00 & 0 & 374.23 & 0.00 & 8000.0 & 0 & 80.98 & 0.00 & 0 & 53.06 & 0.00 \\

\bottomrule
\end{tabular}
\caption{Same as Table \ref{tb:bb}, but showing the results for the $W^+W^-$ channel.}
\label{tb:WW}
\end{table*}
\end{center}

% #############################################
% Table tautau

\begin{center}
\begin{table*}
\scriptsize
\begin{tabular}{|r| c c c | c c c ||r| c c c |  c c c | }
\toprule
%1st
 \multicolumn{7}{|c||}{{\bf Annihilation}} & \multicolumn{7}{c|}{{\bf Decay}} \\
%2nd
  & \multicolumn{3}{c|}{{\bf NFW }} &  \multicolumn{3}{c||}{{\bf Burkert}} & &\multicolumn{3}{c|}{{\bf NFW}} &  \multicolumn{3}{c|}{\bf{Burkert}} \\
%3rd
{$m_\chi$} &  \multirow{2}{*}{$\hat{n}_s$} & {$\langle \sigma v \rangle^{90\%}_{u.l.} $} & \multirow{2}{*}{$z$-score} &  \multirow{2}{*}{$\hat{n}_s$} & {$\langle \sigma v \rangle^{90\%}_{u.l.}$}   & \multirow{2}{*}{$z$-score} & {$m_\chi$} &  \multirow{2}{*}{$\hat{n}_s$} & {$\tau^{90\%}_{l.l.}$} & \multirow{2}{*}{$z$-score} &  \multirow{2}{*}{$\hat{n}_s$} & {$\tau^{90\%}_{l.l.}$} & \multirow{2}{*}{$z$-score} \\ 
%4th
${\rm [GeV]}$ &   & {$10^{-24} {\rm \;[cm^3 s^{-1}]}$} &   &   & {$10^{-24} {\rm \;[cm^3 s^{-1}]}$} &  & ${\rm [GeV]}$ & &  {$10^{24} {\rm \;[s]}$} &   &    & {$10^{24} {\rm \;[s]}$} & \\
\midrule

5.0 & 533 & 25.76 & 0.54 & 966 & 82.38 & 0.52 & 10.0 & 1506 & 0.23 & 0.52 & 1716 & 0.18 & 0.50 \\
6.3 & 816 & 17.46 & 0.78 & 1801 & 63.76 & 0.90 & 12.6 & 3017 & 0.35 & 0.96 & 3777 & 0.27 & 0.98 \\
7.9 & 1011 & 12.32 & 0.93 & 2256 & 47.11 & 1.08 & 15.9 & 3770 & 0.59 & 1.16 & 4871 & 0.45 & 1.22 \\
10.0 & 1314 & 9.82 & 1.19 & 2399 & 34.66 & 1.15 & 20.0 & 3928 & 1.02 & 1.22 & 4976 & 0.77 & 1.23 \\
12.6 & 1500 & 7.97 & 1.37 & 2417 & 26.99 & 1.17 & 25.1 & 3852 & 1.66 & 1.22 & 4702 & 1.27 & 1.18 \\
15.8 & 1546 & 6.43 & 1.47 & 2365 & 21.65 & 1.18 & 31.7 & 3614 & 2.65 & 1.18 & 4131 & 2.07 & 1.06 \\
19.9 & 1563 & 5.48 & 1.55 & 2299 & 18.23 & 1.19 & 39.9 & 3413 & 4.01 & 1.16 & 3661 & 3.19 & 0.97 \\
25.1 & 1624 & 4.97 & 1.70 & 2283 & 16.24 & 1.23 & 50.2 & 3400 & 5.64 & 1.21 & 3469 & 4.58 & 0.96 \\
31.6 & 1686 & 4.70 & 1.87 & 2299 & 15.16 & 1.31 & 63.2 & 3477 & 7.54 & 1.31 & 3466 & 6.20 & 1.01 \\
39.8 & 1710 & 4.58 & 2.04 & 2343 & 14.83 & 1.41 & 79.6 & 3630 & 9.56 & 1.44 & 3667 & 7.86 & 1.13 \\
50.1 & 1706 & 4.57 & 2.19 & 2364 & 14.97 & 1.51 & 100.2 & 3724 & 11.81 & 1.57 & 3806 & 9.69 & 1.23 \\
63.1 & 1674 & 4.66 & 2.32 & 2373 & 15.54 & 1.61 & 126.2 & 3747 & 14.28 & 1.68 & 3842 & 11.73 & 1.32 \\
79.5 & 1583 & 4.80 & 2.38 & 2355 & 16.50 & 1.71 & 159.0 & 3702 & 16.97 & 1.78 & 3851 & 13.88 & 1.43 \\
100.1 & 1399 & 4.86 & 2.31 & 2233 & 17.48 & 1.76 & 200.2 & 3491 & 20.23 & 1.82 & 3736 & 16.31 & 1.50 \\
126.0 & 1122 & 4.73 & 2.04 & 1995 & 18.30 & 1.71 & 252.1 & 3134 & 24.33 & 1.77 & 3522 & 18.99 & 1.53 \\
158.7 & 880 & 4.64 & 1.79 & 1705 & 18.93 & 1.61 & 317.4 & 2673 & 29.54 & 1.67 & 3142 & 22.42 & 1.50 \\
199.8 & 633 & 4.36 & 1.50 & 1356 & 19.10 & 1.45 & 399.7 & 2111 & 37.03 & 1.49 & 2547 & 27.54 & 1.37 \\
251.6 & 403 & 3.90 & 1.14 & 957 & 18.51 & 1.18 & 503.3 & 1473 & 48.59 & 1.20 & 1803 & 35.42 & 1.10 \\
316.9 & 246 & 3.59 & 0.83 & 593 & 17.34 & 0.85 & 633.7 & 896 & 65.65 & 0.85 & 1082 & 47.40 & 0.76 \\
399.0 & 145 & 3.44 & 0.55 & 320 & 16.31 & 0.53 & 798.0 & 475 & 88.23 & 0.52 & 544 & 63.26 & 0.43 \\
502.5 & 48 & 3.21 & 0.22 & 135 & 15.66 & 0.26 & 1004.9 & 184 & 117.07 & 0.23 & 202 & 81.93 & 0.19 \\
632.7 & 0 & 2.78 & 0.00 & 0 & 14.06 & 0.00 & 1265.4 & 0 & 170.68 & 0.00 & 0 & 117.63 & 0.00 \\
796.7 & 0 & 2.23 & 0.00 & 0 & 11.73 & 0.00 & 1593.5 & 0 & 264.52 & 0.00 & 0 & 176.88 & 0.00 \\
1003.3 & 0 & 2.21 & 0.00 & 0 & 11.26 & 0.00 & 2006.6 & 0 & 348.21 & 0.00 & 0 & 231.94 & 0.00 \\
1263.4 & 0 & 2.40 & 0.00 & 0 & 12.41 & 0.00 & 2526.7 & 0 & 399.60 & 0.00 & 0 & 260.34 & 0.00 \\
1590.9 & 0 & 2.75 & 0.00 & 0 & 14.43 & 0.00 & 3181.7 & 0 & 434.75 & 0.00 & 0 & 277.03 & 0.00 \\
2003.3 & 0 & 3.58 & 0.00 & 0 & 19.91 & 0.00 & 4006.5 & 0 & 384.76 & 0.00 & 0 & 227.67 & 0.00 \\
2522.6 & 0 & 5.00 & 0.00 & 0 & 28.15 & 0.00 & 5045.2 & 0 & 341.74 & 0.00 & 0 & 198.51 & 0.00 \\
3176.5 & 0 & 6.89 & 0.00 & 0 & 38.48 & 0.00 & 6353.1 & 0 & 323.69 & 0.00 & 0 & 195.09 & 0.00 \\
4000.0 & 0 & 9.71 & 0.00 & 0 & 54.20 & 0.00 & 8000.0 & 0 & 294.16 & 0.00 & 0 & 181.37 & 0.00 \\

\bottomrule
\end{tabular}
\caption{Same as Table \ref{tb:bb}, but showing the results for the $\tau \bar{\tau}$ channel.}
\label{tb:tautau}
\end{table*}
\end{center}

% #############################################
% Table mumu

\begin{center}
\begin{table*}
\scriptsize
\begin{tabular}{|r| c c c | c c c ||r| c c c |  c c c | }
\toprule
%1st
 \multicolumn{7}{|c||}{{\bf Annihilation}} & \multicolumn{7}{c|}{{\bf Decay}} \\
%2nd
  & \multicolumn{3}{c|}{{\bf NFW }} &  \multicolumn{3}{c||}{{\bf Burkert}} & &\multicolumn{3}{c|}{{\bf NFW}} &  \multicolumn{3}{c|}{\bf{Burkert}} \\
%3rd
{$m_\chi$} &  \multirow{2}{*}{$\hat{n}_s$} & {$\langle \sigma v \rangle^{90\%}_{u.l.} $} & \multirow{2}{*}{$z$-score} &  \multirow{2}{*}{$\hat{n}_s$} & {$\langle \sigma v \rangle^{90\%}_{u.l.}$}   & \multirow{2}{*}{$z$-score} & {$m_\chi$} &  \multirow{2}{*}{$\hat{n}_s$} & {$\tau^{90\%}_{l.l.}$} & \multirow{2}{*}{$z$-score} &  \multirow{2}{*}{$\hat{n}_s$} & {$\tau^{90\%}_{l.l.}$} & \multirow{2}{*}{$z$-score} \\ 
%4th
${\rm [GeV]}$ &   & {$10^{-24} {\rm \;[cm^3 s^{-1}]}$} &   &   & {$10^{-24} {\rm \;[cm^3 s^{-1}]}$} &  & ${\rm [GeV]}$ & &  {$10^{24} {\rm \;[s]}$} &   &    & {$10^{24} {\rm \;[s]}$} & \\
\midrule

5.0 & 549 & 18.92 & 0.56 & 994 & 60.80 & 0.54 & 10.0 & 1539 & 0.31 & 0.53 & 1748 & 0.25 & 0.50 \\
6.0 & 756 & 14.21 & 0.73 & 1607 & 50.53 & 0.81 & 12.0 & 2637 & 0.43 & 0.85 & 3238 & 0.34 & 0.86 \\
7.2 & 869 & 10.51 & 0.82 & 2085 & 41.13 & 1.01 & 14.4 & 3490 & 0.62 & 1.09 & 4503 & 0.47 & 1.14 \\
8.6 & 1086 & 8.61 & 1.00 & 2337 & 32.73 & 1.12 & 17.3 & 3865 & 0.93 & 1.20 & 5000 & 0.70 & 1.25 \\
10.4 & 1299 & 7.35 & 1.18 & 2407 & 26.33 & 1.16 & 20.8 & 3921 & 1.40 & 1.22 & 4978 & 1.05 & 1.24 \\
12.5 & 1465 & 6.40 & 1.34 & 2431 & 22.03 & 1.18 & 24.9 & 3878 & 2.02 & 1.23 & 4782 & 1.53 & 1.20 \\
15.0 & 1533 & 5.44 & 1.45 & 2396 & 18.48 & 1.19 & 29.9 & 3706 & 2.92 & 1.20 & 4343 & 2.26 & 1.11 \\
18.0 & 1542 & 4.73 & 1.50 & 2298 & 15.81 & 1.17 & 35.9 & 3440 & 4.15 & 1.16 & 3800 & 3.27 & 1.00 \\
21.6 & 1560 & 4.26 & 1.58 & 2251 & 14.09 & 1.19 & 43.1 & 3327 & 5.61 & 1.16 & 3503 & 4.50 & 0.95 \\
25.9 & 1607 & 3.99 & 1.70 & 2213 & 12.92 & 1.22 & 51.8 & 3288 & 7.31 & 1.19 & 3314 & 5.96 & 0.93 \\
31.1 & 1649 & 3.83 & 1.84 & 2204 & 12.24 & 1.27 & 62.2 & 3316 & 9.20 & 1.26 & 3270 & 7.59 & 0.96 \\
37.3 & 1660 & 3.74 & 1.97 & 2221 & 11.98 & 1.34 & 74.6 & 3406 & 11.16 & 1.36 & 3381 & 9.21 & 1.04 \\
44.8 & 1641 & 3.71 & 2.07 & 2222 & 11.97 & 1.41 & 89.6 & 3472 & 13.26 & 1.46 & 3503 & 10.90 & 1.14 \\
53.8 & 1608 & 3.73 & 2.17 & 2213 & 12.20 & 1.49 & 107.5 & 3490 & 15.52 & 1.55 & 3553 & 12.73 & 1.22 \\
64.5 & 1562 & 3.80 & 2.27 & 2198 & 12.65 & 1.57 & 129.1 & 3471 & 17.93 & 1.64 & 3554 & 14.71 & 1.29 \\
77.5 & 1477 & 3.89 & 2.31 & 2173 & 13.33 & 1.65 & 154.9 & 3421 & 20.45 & 1.72 & 3555 & 16.70 & 1.38 \\
93.0 & 1346 & 3.97 & 2.29 & 2089 & 14.05 & 1.71 & 186.0 & 3273 & 23.36 & 1.78 & 3476 & 18.89 & 1.45 \\
111.6 & 1138 & 3.91 & 2.13 & 1920 & 14.65 & 1.71 & 223.3 & 3006 & 26.97 & 1.76 & 3293 & 21.36 & 1.49 \\
134.0 & 906 & 3.77 & 1.87 & 1693 & 15.07 & 1.63 & 268.1 & 2641 & 31.42 & 1.69 & 3041 & 24.13 & 1.49 \\
160.9 & 709 & 3.66 & 1.62 & 1425 & 15.31 & 1.52 & 321.8 & 2229 & 37.07 & 1.57 & 2675 & 27.73 & 1.44 \\
193.1 & 519 & 3.44 & 1.37 & 1143 & 15.26 & 1.37 & 386.3 & 1777 & 44.81 & 1.41 & 2193 & 32.87 & 1.32 \\
231.9 & 351 & 3.12 & 1.10 & 849 & 14.87 & 1.16 & 463.7 & 1310 & 55.64 & 1.19 & 1645 & 40.09 & 1.12 \\
278.3 & 216 & 2.78 & 0.80 & 562 & 13.93 & 0.89 & 556.7 & 844 & 72.01 & 0.89 & 1067 & 51.15 & 0.82 \\
334.1 & 130 & 2.61 & 0.57 & 325 & 12.86 & 0.59 & 668.3 & 480 & 93.99 & 0.58 & 577 & 66.65 & 0.51 \\
401.1 & 67 & 2.48 & 0.34 & 149 & 11.86 & 0.31 & 802.2 & 210 & 123.04 & 0.29 & 216 & 87.16 & 0.22 \\
481.5 & 15 & 2.33 & 0.09 & 34 & 11.19 & 0.08 & 963.0 & 33 & 157.77 & 0.05 & 5 & 109.74 & 0.01 \\
578.0 & 0 & 2.09 & 0.00 & 0 & 10.39 & 0.00 & 1156.1 & 0 & 209.47 & 0.00 & 0 & 143.27 & 0.00 \\
693.9 & 0 & 1.80 & 0.00 & 0 & 9.25 & 0.00 & 1387.8 & 0 & 290.22 & 0.00 & 0 & 195.28 & 0.00 \\
833.0 & 0 & 1.55 & 0.00 & 0 & 8.13 & 0.00 & 1666.0 & 0 & 402.53 & 0.00 & 0 & 265.43 & 0.00 \\
1000.0 & 0 & 1.60 & 0.00 & 0 & 8.05 & 0.00 & 2000.0 & 0 & 487.75 & 0.00 & 0 & 323.02 & 0.00 \\

\bottomrule
\end{tabular}
\caption{Same as Table \ref{tb:bb}, but showing the results for the $\mu^+ \mu^-$ channel.}
\label{tb:mumu}
\end{table*}
\end{center}

% #############################################
% Table nuenue

\begin{center}
\begin{table*}
\scriptsize
\begin{tabular}{|r| c c c | c c c ||r| c c c |  c c c | }
\toprule
%1st
 \multicolumn{7}{|c||}{{\bf Annihilation}} & \multicolumn{7}{c|}{{\bf Decay}} \\
%2nd
  & \multicolumn{3}{c|}{{\bf NFW }} &  \multicolumn{3}{c||}{{\bf Burkert}} & &\multicolumn{3}{c|}{{\bf NFW}} &  \multicolumn{3}{c|}{\bf{Burkert}} \\
%3rd
{$m_\chi$} &  \multirow{2}{*}{$\hat{n}_s$} & {$\langle \sigma v \rangle^{90\%}_{u.l.} $} & \multirow{2}{*}{$z$-score} &  \multirow{2}{*}{$\hat{n}_s$} & {$\langle \sigma v \rangle^{90\%}_{u.l.}$}   & \multirow{2}{*}{$z$-score} & {$m_\chi$} &  \multirow{2}{*}{$\hat{n}_s$} & {$\tau^{90\%}_{l.l.}$} & \multirow{2}{*}{$z$-score} &  \multirow{2}{*}{$\hat{n}_s$} & {$\tau^{90\%}_{l.l.}$} & \multirow{2}{*}{$z$-score} \\ 
%4th
${\rm [GeV]}$ &   & {$10^{-24} {\rm \;[cm^3 s^{-1}]}$} &   &   & {$10^{-24} {\rm \;[cm^3 s^{-1}]}$} &  & ${\rm [GeV]}$ & &  {$10^{24} {\rm \;[s]}$} &   &    & {$10^{24} {\rm \;[s]}$} & \\
\midrule

5.0 & 422 & 1.26 & 0.54 & 2212 & 7.85 & 1.21 & 10.0 & 3546 & 2.32 & 1.31 & 4885 & 1.72 & 1.50 \\
5.7 & 842 & 1.32 & 1.08 & 2742 & 7.33 & 1.42 & 11.4 & 4187 & 2.90 & 1.48 & 5232 & 2.22 & 1.49 \\
6.4 & 236 & 0.90 & 0.24 & 1902 & 5.27 & 0.95 & 12.9 & 3034 & 4.48 & 1.03 & 4102 & 3.38 & 1.17 \\
7.3 & 908 & 1.12 & 0.95 & 2316 & 4.90 & 1.19 & 14.6 & 3377 & 5.57 & 1.14 & 4275 & 4.21 & 1.17 \\
8.3 & 1494 & 1.27 & 1.54 & 2735 & 4.65 & 1.44 & 16.6 & 4217 & 6.50 & 1.49 & 5105 & 5.04 & 1.46 \\
9.4 & 1174 & 1.00 & 1.16 & 1892 & 3.52 & 0.98 & 18.9 & 2937 & 9.65 & 1.01 & 3604 & 7.31 & 0.99 \\
10.7 & 1501 & 0.99 & 1.61 & 1954 & 3.15 & 1.09 & 21.5 & 2915 & 12.53 & 1.09 & 3163 & 9.98 & 0.94 \\
12.2 & 1299 & 0.85 & 1.42 & 2270 & 3.22 & 1.24 & 24.4 & 3450 & 13.75 & 1.27 & 3931 & 10.78 & 1.15 \\
13.8 & 947 & 0.67 & 1.12 & 1039 & 2.12 & 0.61 & 27.7 & 1141 & 26.00 & 0.45 & 537 & 22.70 & 0.17 \\
15.7 & 1242 & 0.77 & 1.42 & 1771 & 2.55 & 1.05 & 31.4 & 2517 & 22.81 & 1.00 & 2452 & 18.75 & 0.76 \\
17.8 & 1299 & 0.77 & 1.49 & 1782 & 2.48 & 1.09 & 35.7 & 2667 & 25.94 & 1.08 & 2780 & 20.73 & 0.88 \\
20.3 & 1232 & 0.71 & 1.52 & 1631 & 2.33 & 1.02 & 40.5 & 2399 & 31.43 & 0.99 & 2191 & 26.23 & 0.70 \\
23.0 & 1426 & 0.76 & 1.86 & 1493 & 2.17 & 1.02 & 46.0 & 2246 & 38.14 & 1.02 & 1898 & 32.92 & 0.68 \\
26.1 & 1177 & 0.70 & 1.58 & 1634 & 2.29 & 1.16 & 52.3 & 2433 & 41.19 & 1.15 & 2384 & 33.96 & 0.88 \\
29.7 & 1113 & 0.68 & 1.64 & 1588 & 2.28 & 1.20 & 59.3 & 2385 & 46.87 & 1.21 & 2425 & 38.16 & 0.96 \\
33.7 & 964 & 0.64 & 1.53 & 1507 & 2.32 & 1.18 & 67.4 & 2360 & 50.94 & 1.22 & 2482 & 40.45 & 0.99 \\
38.3 & 1110 & 0.73 & 1.80 & 1521 & 2.37 & 1.27 & 76.5 & 2474 & 55.62 & 1.38 & 2677 & 44.18 & 1.16 \\
43.5 & 1134 & 0.77 & 1.93 & 1699 & 2.70 & 1.44 & 86.9 & 2711 & 55.32 & 1.50 & 2879 & 44.26 & 1.23 \\
49.4 & 1062 & 0.74 & 2.12 & 1456 & 2.57 & 1.39 & 98.7 & 2273 & 68.12 & 1.45 & 2328 & 55.25 & 1.14 \\
56.1 & 1005 & 0.78 & 2.13 & 1492 & 2.72 & 1.56 & 112.1 & 2298 & 73.34 & 1.61 & 2414 & 59.34 & 1.31 \\
63.7 & 762 & 0.71 & 1.80 & 1490 & 2.87 & 1.71 & 127.3 & 2290 & 78.70 & 1.76 & 2649 & 60.39 & 1.58 \\
72.3 & 620 & 0.67 & 1.70 & 1230 & 2.85 & 1.51 & 144.6 & 1860 & 91.25 & 1.53 & 2036 & 71.88 & 1.29 \\
82.1 & 658 & 0.78 & 1.89 & 1050 & 2.84 & 1.39 & 164.2 & 1588 & 103.12 & 1.40 & 1674 & 81.98 & 1.13 \\
93.2 & 286 & 0.53 & 0.92 & 775 & 2.62 & 1.15 & 186.5 & 1147 & 128.61 & 1.14 & 1492 & 91.12 & 1.11 \\
105.9 & 146 & 0.45 & 0.51 & 620 & 2.58 & 1.01 & 211.8 & 1010 & 142.28 & 1.09 & 1556 & 92.73 & 1.25 \\
120.2 & 116 & 0.43 & 0.50 & 324 & 2.04 & 0.64 & 240.5 & 482 & 213.62 & 0.64 & 710 & 144.08 & 0.71 \\
136.6 & 234 & 0.61 & 1.08 & 394 & 2.37 & 0.89 & 273.1 & 595 & 205.97 & 0.90 & 737 & 150.71 & 0.86 \\
155.1 & 130 & 0.49 & 0.76 & 378 & 2.68 & 0.90 & 310.2 & 583 & 205.29 & 0.93 & 772 & 142.81 & 0.91 \\
176.1 & 71 & 0.43 & 0.48 & 168 & 2.21 & 0.45 & 352.2 & 203 & 301.23 & 0.37 & 168 & 222.95 & 0.22 \\
200.0 & 29 & 0.35 & 0.25 & 33 & 1.75 & 0.11 & 400.0 & 41 & 427.91 & 0.09 & 20 & 292.44 & 0.03 \\

\bottomrule
\end{tabular}
\caption{Same as Table \ref{tb:bb}, but showing the results for the $\nu_e \bar{\nu}_e$ channel.}
\label{tb:nuenue}
\end{table*}
\end{center}

% #############################################
% Table numunumu

\begin{center}
\begin{table*}
\scriptsize
\begin{tabular}{|r| c c c | c c c ||r| c c c |  c c c | }
\toprule
%1st
 \multicolumn{7}{|c||}{{\bf Annihilation}} & \multicolumn{7}{c|}{{\bf Decay}} \\
%2nd
  & \multicolumn{3}{c|}{{\bf NFW }} &  \multicolumn{3}{c||}{{\bf Burkert}} & &\multicolumn{3}{c|}{{\bf NFW}} &  \multicolumn{3}{c|}{\bf{Burkert}} \\
%3rd
{$m_\chi$} &  \multirow{2}{*}{$\hat{n}_s$} & {$\langle \sigma v \rangle^{90\%}_{u.l.} $} & \multirow{2}{*}{$z$-score} &  \multirow{2}{*}{$\hat{n}_s$} & {$\langle \sigma v \rangle^{90\%}_{u.l.}$}   & \multirow{2}{*}{$z$-score} & {$m_\chi$} &  \multirow{2}{*}{$\hat{n}_s$} & {$\tau^{90\%}_{l.l.}$} & \multirow{2}{*}{$z$-score} &  \multirow{2}{*}{$\hat{n}_s$} & {$\tau^{90\%}_{l.l.}$} & \multirow{2}{*}{$z$-score} \\ 
%4th
${\rm [GeV]}$ &   & {$10^{-24} {\rm \;[cm^3 s^{-1}]}$} &   &   & {$10^{-24} {\rm \;[cm^3 s^{-1}]}$} &  & ${\rm [GeV]}$ & &  {$10^{24} {\rm \;[s]}$} &   &    & {$10^{24} {\rm \;[s]}$} & \\
\midrule

5.0 & 1104 & 2.17 & 1.21 & 2458 & 8.77 & 1.30 & 10.0 & 3896 & 2.06 & 1.34 & 4694 & 1.63 & 1.33 \\
5.7 & 1049 & 1.70 & 1.16 & 2773 & 7.79 & 1.41 & 11.4 & 4694 & 2.55 & 1.58 & 6082 & 1.91 & 1.66 \\
6.4 & 492 & 1.12 & 0.51 & 1750 & 5.37 & 0.89 & 12.9 & 2728 & 4.39 & 0.91 & 3683 & 3.26 & 0.99 \\
7.3 & 1539 & 1.59 & 1.53 & 2174 & 5.02 & 1.13 & 14.6 & 3518 & 5.14 & 1.20 & 4311 & 3.93 & 1.17 \\
8.3 & 1400 & 1.30 & 1.46 & 2010 & 4.21 & 1.09 & 16.6 & 3181 & 7.10 & 1.14 & 3657 & 5.61 & 1.06 \\
9.4 & 961 & 0.95 & 1.03 & 1552 & 3.40 & 0.85 & 18.9 & 2214 & 10.33 & 0.80 & 2463 & 8.04 & 0.69 \\
10.7 & 1311 & 0.98 & 1.51 & 1924 & 3.35 & 1.10 & 21.5 & 2879 & 11.70 & 1.10 & 3151 & 9.19 & 0.93 \\
12.2 & 1196 & 0.86 & 1.43 & 2117 & 3.29 & 1.22 & 24.4 & 3118 & 13.56 & 1.19 & 3373 & 10.73 & 1.00 \\
13.8 & 912 & 0.71 & 1.12 & 1395 & 2.53 & 0.85 & 27.7 & 1764 & 21.15 & 0.71 & 1368 & 17.96 & 0.43 \\
15.7 & 1245 & 0.81 & 1.54 & 1800 & 2.70 & 1.15 & 31.4 & 2585 & 21.48 & 1.09 & 2521 & 17.66 & 0.82 \\
17.8 & 1162 & 0.73 & 1.56 & 1821 & 2.61 & 1.22 & 35.7 & 2668 & 25.06 & 1.19 & 2709 & 20.28 & 0.93 \\
20.3 & 1076 & 0.67 & 1.56 & 1726 & 2.50 & 1.19 & 40.5 & 2564 & 29.36 & 1.17 & 2449 & 24.28 & 0.86 \\
23.0 & 1296 & 0.75 & 1.95 & 1665 & 2.43 & 1.21 & 46.0 & 2623 & 33.35 & 1.27 & 2472 & 27.95 & 0.92 \\
26.1 & 1116 & 0.69 & 1.80 & 1681 & 2.45 & 1.29 & 52.3 & 2577 & 38.06 & 1.31 & 2516 & 31.49 & 0.99 \\
29.7 & 1211 & 0.73 & 2.04 & 1825 & 2.60 & 1.46 & 59.3 & 2891 & 40.00 & 1.54 & 2926 & 32.90 & 1.20 \\
33.7 & 1203 & 0.77 & 2.16 & 1779 & 2.68 & 1.49 & 67.4 & 2981 & 42.59 & 1.65 & 3222 & 33.96 & 1.36 \\
38.3 & 1166 & 0.78 & 2.23 & 1663 & 2.66 & 1.47 & 76.5 & 2778 & 49.07 & 1.64 & 2951 & 39.56 & 1.33 \\
43.5 & 1115 & 0.80 & 2.22 & 1760 & 2.92 & 1.60 & 86.9 & 2819 & 51.82 & 1.69 & 2892 & 42.33 & 1.33 \\
49.4 & 978 & 0.76 & 2.10 & 1472 & 2.79 & 1.44 & 98.7 & 2292 & 62.55 & 1.48 & 2263 & 51.78 & 1.11 \\
56.1 & 962 & 0.83 & 2.19 & 1503 & 3.00 & 1.56 & 112.1 & 2332 & 66.44 & 1.61 & 2328 & 55.18 & 1.24 \\
63.7 & 839 & 0.85 & 1.96 & 1463 & 3.19 & 1.60 & 127.3 & 2290 & 70.29 & 1.67 & 2443 & 56.83 & 1.38 \\
72.3 & 704 & 0.81 & 1.86 & 1292 & 3.25 & 1.52 & 144.6 & 1966 & 79.60 & 1.54 & 2095 & 63.59 & 1.26 \\
82.1 & 583 & 0.80 & 1.69 & 1201 & 3.39 & 1.53 & 164.2 & 1897 & 85.10 & 1.60 & 2220 & 65.07 & 1.45 \\
93.2 & 328 & 0.64 & 1.02 & 953 & 3.29 & 1.31 & 186.5 & 1452 & 101.70 & 1.31 & 1844 & 74.01 & 1.28 \\
105.9 & 206 & 0.58 & 0.69 & 779 & 3.26 & 1.15 & 211.8 & 1243 & 113.27 & 1.20 & 1736 & 78.40 & 1.27 \\
120.2 & 244 & 0.67 & 0.93 & 611 & 3.17 & 1.01 & 240.5 & 959 & 134.10 & 1.04 & 1266 & 95.68 & 1.04 \\
136.6 & 230 & 0.73 & 0.91 & 507 & 3.06 & 0.97 & 273.1 & 776 & 158.18 & 0.98 & 959 & 117.69 & 0.94 \\
155.1 & 126 & 0.58 & 0.64 & 410 & 3.17 & 0.85 & 310.2 & 618 & 175.22 & 0.85 & 784 & 126.87 & 0.81 \\
176.1 & 43 & 0.45 & 0.27 & 251 & 2.88 & 0.57 & 352.2 & 338 & 227.72 & 0.52 & 428 & 160.43 & 0.48 \\
200.0 & 54 & 0.51 & 0.35 & 145 & 2.67 & 0.38 & 400.0 & 212 & 270.12 & 0.37 & 230 & 193.88 & 0.29 \\

\bottomrule
\end{tabular}
\caption{Same as Table \ref{tb:bb}, but showing the results for the $\nu_\mu \bar{\nu}_\mu$ channel.}
\label{tb:numunumu}
\end{table*}
\end{center}

% #############################################
% Table nutaunutau

\begin{center}
\begin{table*}
\scriptsize
\begin{tabular}{|r| c c c | c c c ||r| c c c |  c c c | }
\toprule
%1st
 \multicolumn{7}{|c||}{{\bf Annihilation}} & \multicolumn{7}{c|}{{\bf Decay}} \\
%2nd
  & \multicolumn{3}{c|}{{\bf NFW }} &  \multicolumn{3}{c||}{{\bf Burkert}} & &\multicolumn{3}{c|}{{\bf NFW}} &  \multicolumn{3}{c|}{\bf{Burkert}} \\
%3rd
{$m_\chi$} &  \multirow{2}{*}{$\hat{n}_s$} & {$\langle \sigma v \rangle^{90\%}_{u.l.} $} & \multirow{2}{*}{$z$-score} &  \multirow{2}{*}{$\hat{n}_s$} & {$\langle \sigma v \rangle^{90\%}_{u.l.}$}   & \multirow{2}{*}{$z$-score} & {$m_\chi$} &  \multirow{2}{*}{$\hat{n}_s$} & {$\tau^{90\%}_{l.l.}$} & \multirow{2}{*}{$z$-score} &  \multirow{2}{*}{$\hat{n}_s$} & {$\tau^{90\%}_{l.l.}$} & \multirow{2}{*}{$z$-score} \\ 
%4th
${\rm [GeV]}$ &   & {$10^{-24} {\rm \;[cm^3 s^{-1}]}$} &   &   & {$10^{-24} {\rm \;[cm^3 s^{-1}]}$} &  & ${\rm [GeV]}$ & &  {$10^{24} {\rm \;[s]}$} &   &    & {$10^{24} {\rm \;[s]}$} & \\
\midrule

5.0 & 919 & 1.94 & 1.03 & 2439 & 8.64 & 1.29 & 10.0 & 3922 & 2.08 & 1.36 & 4962 & 1.60 & 1.41 \\
5.7 & 1025 & 1.63 & 1.16 & 2809 & 7.76 & 1.42 & 11.4 & 4656 & 2.59 & 1.57 & 6011 & 1.95 & 1.63 \\
6.4 & 434 & 1.08 & 0.44 & 1824 & 5.41 & 0.91 & 12.9 & 2893 & 4.33 & 0.95 & 3938 & 3.22 & 1.06 \\
7.3 & 1387 & 1.48 & 1.38 & 2241 & 5.03 & 1.16 & 14.6 & 3541 & 5.19 & 1.19 & 4400 & 3.94 & 1.18 \\
8.3 & 1467 & 1.32 & 1.51 & 2237 & 4.38 & 1.20 & 16.6 & 3530 & 6.83 & 1.25 & 4147 & 5.35 & 1.18 \\
9.4 & 1044 & 0.98 & 1.08 & 1663 & 3.46 & 0.89 & 18.9 & 2451 & 10.01 & 0.86 & 2835 & 7.71 & 0.78 \\
10.7 & 1404 & 1.00 & 1.56 & 1956 & 3.32 & 1.10 & 21.5 & 2935 & 11.79 & 1.10 & 3210 & 9.28 & 0.94 \\
12.2 & 1252 & 0.87 & 1.44 & 2184 & 3.30 & 1.23 & 24.4 & 3256 & 13.47 & 1.23 & 3590 & 10.61 & 1.05 \\
13.8 & 947 & 0.71 & 1.14 & 1310 & 2.42 & 0.79 & 27.7 & 1606 & 22.21 & 0.64 & 1145 & 19.00 & 0.36 \\
15.7 & 1271 & 0.81 & 1.52 & 1811 & 2.67 & 1.13 & 31.4 & 2597 & 21.66 & 1.07 & 2534 & 17.81 & 0.81 \\
17.8 & 1222 & 0.75 & 1.55 & 1834 & 2.59 & 1.19 & 35.7 & 2706 & 25.05 & 1.17 & 2770 & 20.20 & 0.92 \\
20.3 & 1140 & 0.69 & 1.57 & 1723 & 2.47 & 1.15 & 40.5 & 2558 & 29.63 & 1.13 & 2415 & 24.57 & 0.82 \\
23.0 & 1361 & 0.77 & 1.94 & 1637 & 2.38 & 1.16 & 46.0 & 2554 & 34.29 & 1.21 & 2343 & 28.98 & 0.86 \\
26.1 & 1157 & 0.70 & 1.76 & 1684 & 2.42 & 1.26 & 52.3 & 2565 & 38.61 & 1.27 & 2508 & 31.93 & 0.96 \\
29.7 & 1215 & 0.73 & 1.95 & 1778 & 2.52 & 1.40 & 59.3 & 2785 & 41.41 & 1.45 & 2818 & 34.00 & 1.14 \\
33.7 & 1168 & 0.74 & 2.01 & 1720 & 2.59 & 1.40 & 67.4 & 2841 & 44.39 & 1.54 & 3054 & 35.33 & 1.26 \\
38.3 & 1173 & 0.77 & 2.13 & 1638 & 2.59 & 1.42 & 76.5 & 2723 & 50.41 & 1.57 & 2911 & 40.42 & 1.29 \\
43.5 & 1140 & 0.80 & 2.16 & 1758 & 2.87 & 1.56 & 86.9 & 2811 & 52.34 & 1.65 & 2924 & 42.49 & 1.30 \\
49.4 & 1013 & 0.77 & 2.11 & 1478 & 2.74 & 1.43 & 98.7 & 2307 & 63.66 & 1.48 & 2305 & 52.35 & 1.13 \\
56.1 & 989 & 0.82 & 2.19 & 1508 & 2.93 & 1.57 & 112.1 & 2338 & 67.99 & 1.61 & 2373 & 55.98 & 1.27 \\
63.7 & 825 & 0.81 & 1.92 & 1480 & 3.11 & 1.64 & 127.3 & 2307 & 72.23 & 1.70 & 2530 & 57.40 & 1.45 \\
72.3 & 687 & 0.77 & 1.82 & 1281 & 3.14 & 1.52 & 144.6 & 1949 & 82.34 & 1.54 & 2097 & 65.43 & 1.27 \\
82.1 & 610 & 0.80 & 1.76 & 1162 & 3.23 & 1.49 & 164.2 & 1816 & 89.47 & 1.55 & 2073 & 69.04 & 1.36 \\
93.2 & 320 & 0.61 & 1.00 & 902 & 3.09 & 1.26 & 186.5 & 1361 & 108.31 & 1.27 & 1752 & 78.05 & 1.23 \\
105.9 & 190 & 0.54 & 0.64 & 733 & 3.06 & 1.11 & 211.8 & 1180 & 120.31 & 1.18 & 1703 & 81.56 & 1.27 \\
120.2 & 205 & 0.60 & 0.80 & 512 & 2.79 & 0.89 & 240.5 & 794 & 153.42 & 0.92 & 1083 & 107.29 & 0.94 \\
136.6 & 237 & 0.70 & 0.98 & 474 & 2.85 & 0.95 & 273.1 & 724 & 169.86 & 0.96 & 897 & 125.51 & 0.92 \\
155.1 & 130 & 0.56 & 0.68 & 404 & 3.04 & 0.87 & 310.2 & 614 & 182.01 & 0.88 & 792 & 130.06 & 0.85 \\
176.1 & 53 & 0.45 & 0.34 & 225 & 2.68 & 0.54 & 352.2 & 296 & 245.65 & 0.47 & 345 & 174.87 & 0.40 \\
200.0 & 46 & 0.46 & 0.32 & 107 & 2.37 & 0.29 & 400.0 & 151 & 307.78 & 0.28 & 158 & 216.74 & 0.21 \\

\bottomrule
\end{tabular}
\caption{Same as Table \ref{tb:bb}, but showing the results for the $\nu_\tau \bar{\nu}_\tau$ channel.}
\label{tb:nutaunutau}
\end{table*}
\end{center}

% \begin{center}
% \begin{table*}
% \begin{tabular}{|r| c c c | c c c |}
% \toprule
% %1st
%  \multicolumn{7}{|c|}{{\bf Annihilation}} \\
% %2nd
%   & \multicolumn{3}{c|}{{\bf NFW }} &  \multicolumn{3}{c|}{{\bf Burkert}}\\
% %3rd
% {$m_\chi$} &  \multirow{2}{*}{$\hat{n}_s$} & {$\langle \sigma v \rangle^{90\%}_{u.l.} $} & \multirow{2}{*}{$z$-score} &  \multirow{2}{*}{$\hat{n}_s$} & {$\langle \sigma v \rangle^{90\%}_{u.l.}$}   & \multirow{2}{*}{$z$-score} \\ 
% %4th
% ${\rm [GeV]}$ &   & {$10^{-24} {\rm \;[cm^2]}$} &   &   & {$10^{-24} {\rm \;[cm^2]}$} \\
% \midrule
% 15.0 & 1171.43 & 147.32 & 1.04 & 2299.23 & 526.48 & 1.08 \\

% \bottomrule
% \end{tabular}
% \caption{Table with the results for the final state channel $b\bar{b}$ for the annihilation mode and for both the NFW and Burkert profile. The best fit value on the number of signal evens $\hat{n}_s$ is shown together with the resulting upper limit in $\langle \sigma v \rangle^{90\%}_{u.l.}$ and lower limit on $\tau^{90\%}_{l.l.}$ along with the significance given in number of sigmas, $z$-score.}
% \label{tb:nuenue}
% \end{table*}
% \end{center}

\end{document}